\documentclass[longauth]{aa} 

\usepackage[varg]{txfonts}
\usepackage{graphicx}
\usepackage{natbib}
\usepackage[colorlinks = true, linkcolor = blue, citecolor = blue]{hyperref}
\bibpunct{(}{)}{;}{a}{}{,} 

\begin{document}

\title{RefPlanets\thanks{Based on observations made with ESO Telescopes at the La Silla Paranal Observatory under programme IDs: 095.C-0312(B), 096.C-0326(A), 097.C-0524(A), 097.C-0524(B), 098.C-0197(A), 099.C-0127(A), 099.C-0127(B), 0102.C-0435(A)}: Search for reflected light from extra-solar planets with SPHERE / ZIMPOL}

\subtitle{}

    \author{S.~Hunziker\inst{\ref{instch1}}    
    \and H.M.~Schmid\inst{\ref{instch1}}
   \and D.~Mouillet\inst{\ref{instf1},\ref{instf2}} 
   \and J.~Milli\inst{\ref{insteso2}}
   \and A.~Zurlo\inst{\ref{instcl2},\ref{instcl3},\ref{instf3}}
   \and P.~Delorme\inst{\ref{instf1}}
   \and L.~Abe\inst{\ref{instf5}}
   \and H.~Avenhaus\inst{\ref{instd1},\ref{instch1}}
   \and A.~Baruffolo\inst{\ref{insti1}}
   \and A.~Bazzon\inst{\ref{instch1}}   
   \and A.~Boccaletti\inst{\ref{instf4}}
   \and P.~Baudoz\inst{\ref{instf4}}      
   \and J.L.~Beuzit\inst{\ref{instf3}}
   \and M.~Carbillet\inst{\ref{instf5}}
   \and G.~Chauvin\inst{\ref{instf1},\ref{instcl1}} 
   \and R.~Claudi\inst{\ref{insti1}} 
   \and A.~Costille\inst{\ref{instf3}} 
   \and J.-B.~Daban\inst{\ref{instf5}}
   \and S.~Desidera\inst{\ref{insti1}}
   \and K.~Dohlen\inst{\ref{instf3}}     
   \and C.~Dominik\inst{\ref{instnl2}}
   \and M.~Downing\inst{\ref{insteso1}}
   \and N.~Engler\inst{\ref{instch1}}     
   \and M.~Feldt\inst{\ref{instd1}}      
   \and T.~Fusco\inst{\ref{instf3},\ref{instf6}}     
   \and C.~Ginski\inst{\ref{instnl3}}  
   \and D.~Gisler\inst{\ref{instch3},\ref{instd2}}
   \and J.H.~Girard\inst{\ref{insteso2}}      
   \and R.~Gratton\inst{\ref{insti1}}
   \and Th.~Henning\inst{\ref{instd1}}      
   \and N.~Hubin\inst{\ref{insteso1}}  
   \and M.~Kasper\inst{\ref{insteso1}}   
   \and C.U.~Keller\inst{\ref{instnl3}}      
   \and M.~Langlois\inst{\ref{instf7},\ref{instf3}}      
   \and E.~Lagadec\inst{\ref{instf5}}
   \and P.~Martinez\inst{\ref{instf5}}
   \and A.L.~Maire\inst{\ref{instd1},\ref{instb1}}
   \and F.~Menard\inst{\ref{instf1},\ref{instf2}}
   \and M.R.~Meyer\inst{\ref{instusa2}} 
   \and A.~Pavlov\inst{\ref{instd1}}
   \and J.~Pragt\inst{\ref{instnl1}}
   \and P.~Puget\inst{\ref{instf1}} 
   \and S.P.~Quanz\inst{\ref{instch1}}
   \and E.~Rickman\inst{\ref{instch2}}         
   \and R.~Roelfsema\inst{\ref{instnl1}}
   \and B.~Salasnich\inst{\ref{insti1}}
   \and J.-F.~Sauvage\inst{\ref{instf3},\ref{instf6}}        
   \and R.~Siebenmorgen\inst{\ref{insteso1}}
   \and E.~Sissa\inst{\ref{insti1}}
   \and F.~Snik\inst{\ref{instnl3}}      
   \and M.~Suarez\inst{\ref{insteso1}}
   \and J.~Szul\'agyi\inst{\ref{instch4}}
   \and Ch.~Thalmann\inst{\ref{instch1}}  
   \and M.~Turatto\inst{\ref{insti1}}   
   \and S.~Udry\inst{\ref{instch2}}
   \and R.G.~van Holstein\inst{\ref{instnl3}}
   \and A.~Vigan\inst{\ref{instf3}}  
   \and F.~Wildi\inst{\ref{instch2}}
          }

\institute{
ETH Zurich, Institute for Particle Physics and Astrophysics, 
Wolfgang-Pauli-Strasse 27, 
CH-8093 Zurich, Switzerland\label{instch1}
\and
NOVA Optical Infrared Instrumentation Group at ASTRON, Oude
Hoogeveensedijk 4, 7991 PD Dwingeloo, The Netherlands\label{instnl1}
\and
Universit\'{e} Grenoble Alpes, IPAG, 38000 Grenoble, France\label{instf1}
\and
CNRS, IPAG, 38000 Grenoble, France\label{instf2}
\and
European Southern Observatory, Alonso de Cordova 3107, Casilla
19001 Vitacura, Santiago 19, Chile\label{insteso2}
\and
Institute for Computational Science, University of Z\"urich, Winterthurerstrasse 190, 8057 Z\"urich, Switzerland\label{instch4}
\and
Istituto Ricerche Solari Locarno, Via Patocchi 57,
6605 Locarno Monti, Switzerland\label{instch3}
\and
Kiepenheuer-Institut f\"{u}r Sonnenphysik, Schneckstr. 6, D-79104
Freiburg, Germany\label{instd2}
\and
Anton Pannekoek Institute for Astronomy, University of Amsterdam,
PO Box 94249, 1090 GE Amsterdam, The Netherlands\label{instnl2}
\and
LESIA, CNRS, Observatoire de Paris, Universit\'{e} Paris Diderot,
UPMC, 5 place J. Janssen, 92190 Meudon, France\label{instf4}
\and
Leiden Observatory, Leiden University, P.O. Box 9513, 2300 RA
Leiden, The Netherlands\label{instnl3}
\and
Laboratoire Lagrange, UMR7293, Universit\'{e} de Nice Sophia-Antipolis, 
CNRS, Observatoire de la C\^{o}te d'Azur, Boulevard de l'Observatoire, 
06304 Nice, Cedex 4, France\label{instf5}
\and
INAF - Osservatorio Astronomico di Roma, via Frascati 33, 
I-00087 Monte Porzio Catone, Italy\label{insti3}
\and
Max-Planck-Institut f\"{u}r Astronomie, K\"{o}nigstuhl 17, 69117
Heidelberg, Germany\label{instd1}
\and
INAF – Osservatorio Astronomico di Padova, Vicolo
dell’Osservatorio 5, 35122 Padova, Italy\label{insti1}
\and
Aix Marseille Universit\'{e}, CNRS, CNES, LAM (Laboratoire
d’Astrophysique de Marseille) UMR 7326, 13388, Marseille,
France\label{instf3}
\and
Unidad Mixta International Franco-Chilena de Astronomia, CNRS/INSU
UMI 3386 and Departemento de Astronomia, Universidad de Chile, Casilla 36-D,
Santiago, Chile\label{instcl1}
\and
Nucleo de Astronomia, Facultad de Ingenieria y Ciencias, Universidad Diego Portales, Av. Ejercito 441, Santiago, Chile\label{instcl2}
\and
Escuela de Ingenieria Industrial, Facultad de Ingenieria y Ciencias, Universidad Diego Portales, Av. Ejercito 441, Santiago, Chile\label{instcl3}
\and
European Southern Observatory, Karl Schwarzschild St, 2, 85748
Garching, Germany\label{insteso1}
\and
ONERA, The French Aerospace Lab BP72, 29 avenue de la
Division Leclerc, 92322 Ch\^{a}tillon Cedex, France\label{instf6}
\and
Centre de Recherche Astrophysique de Lyon, CNRS/ENSL
Universit\'{e} Lyon 1, 9 av. Ch. Andr\'{e}, 69561 Saint-Genis-Laval,
France\label{instf7}
\and
INAF - Osservatorio Astrofisico di Arcetri, Largo E. Fermi 5, 
I-50125 Firenze, Italy\label{insti4}
\and
Geneva Observatory, University of Geneva, Chemin des Mailettes
51, 1290 Versoix, Switzerland\label{instch2}
\and
STAR Institute, Universit\'{e} de Li\`{e}ge, All\'{e}e du Six Ao\^{u}t 19c, 4000
Li\`{e}ge, Belgium\label{instb1}
\and
Department of Astronomy, University of Michigan, Ann Arbor, MI 48109, USA\label{instusa2}
            }

\date{Received --- ; accepted --- }

\abstract{}
{RefPlanets is a guaranteed time observation (GTO) programme that uses the Zurich IMaging POLarimeter (ZIMPOL) of SPHERE/VLT for a blind search for exoplanets in wavelengths from 600-900~nm. The goals of this study are the characterization of the unprecedented high polarimetic contrast and polarimetric precision capabilities of ZIMPOL for bright targets, the search for polarized reflected light around some of the closest bright stars to the Sun and potentially the direct detection of an evolved cold exoplanet for the first time.}
{For our observations of $\alpha$~Cen~A and B, Sirius~A, Altair, $\epsilon$~Eri and $\tau$~Ceti we used the polarimetric differential imaging (PDI) mode of ZIMPOL which removes the speckle noise down to the photon noise limit for angular separations $\gtrapprox$0.6\arcsec. We describe some of the instrumental effects that dominate the noise for smaller separations and explain how to remove these additional noise effects in post-processing. We then combine PDI with angular differential imaging (ADI) as a final layer of post-processing to further improve the contrast limits of our data at these separations.}
{For good observing conditions we achieve polarimetric contrast limits of 15.0--16.3~mag at the effective inner working angle of $\sim$0.13\arcsec, 16.3--18.3~mag at 0.5\arcsec and 18.8--20.4~mag at 1.5\arcsec. The contrast limits closer in ($\lessapprox$0.6\arcsec) depend significantly on the observing conditions, while in the photon noise dominated regime ($\gtrapprox$0.6\arcsec), the limits mainly depend on the brightness of the star and the total integration time. We compare our results with contrast limits from other surveys and review the exoplanet detection limits obtained with different detection methods. For all our targets we achieve unprecedented contrast limits. Despite the high polarimetric contrasts we are not able to find any additional companions or extended polarized light sources in the data that has been taken so far.}
{}


\authorrunning{Hunziker et al.}
\maketitle

\section{Introduction}

High-contrast imaging is a key technique for the search and classification of extra-solar planets which is one of the primary goals in modern astronomy. However, the technical requirements are very challenging and up to now only about a dozen young, giant planets have been directly imaged \citep[e.g.][]{Macintosh15, Bowler16, Schmidt16, Chauvin17, Keppler18}. Young, self-contracting giant planets are hot with temperatures of $T\approx 1000-2000~{\rm K}$ \citep[e.g.][]{Baraffe03, Spiegel12}, therefore they are bright in the near-infrared (NIR) and the required contrast $C=F_{\rm pl}/F_{\rm star}\approx 10^{-5\pm 1}$ is within reach of modern extreme adaptive optics (AO) systems, like SPHERE at the VLT \citep{Beuzit08}, GPI at Gemini \citep{Macintosh14}, the NGS AO system at Keck \citep{vanDam04} or SCExAO at Subaru  \citep{Jovanovic15}. Unfortunately, young stars with planets are rare in the solar neighbourhood. Furthermore, for the young stars in the nearest star forming regions at $d\approx 150$~pc the expected angular separations of planets tend to already be quite small and hence they are observationally challenging to detect.

Most old planets, including all habitable planets, are cold and therefore produce only scattered light in the visual to NIR (<$2~{\rm \mu m}$) wavelength range \citep{Sudarsky03}. Light-scattering by the planets' atmosphere produces a polarization signal which can be distinguished from the unpolarized light of the much brighter central star \citep{Seager00, Stam04, Buenzli09}. The contrast of this reflected light from extra-solar planets with respect to the brightness of their host stars is very challenging $(C\lessapprox10^{-7})$, but polarimetric differential imaging (PDI) has been shown to be a very effective technique to reveal faint reflected light signals. For these reasons the SPHERE "planet finder" instrument includes the Zurich IMaging POLarimeter \citep[ZIMPOL,][]{Schmid18} which was designed for the search of light from reflecting planets in the visual wavelength range using innovative polarimetric techniques \citep{Schmid06a, Thalmann08}. 

We investigate in this paper the achievable contrast of SPHERE/ZIMPOL for a first series of deep observations of promising targets obtained within the RefPlanets project, which is a part of the guaranteed time observation (GTO) program of the SPHERE consortium. An important goal of this work is a better understanding of the limitations of this instrument in order to optimize the SPHERE/ZIMPOL observing strategy for high-contrast targets, and possibly to conceive upgrades for this instrument or improve concepts for future instruments, for example for the Extremely Large Telescope \citep[ELT,][]{Kasper10, Keller10}. Pushing the limits of high-contrast imaging polarimetry should be useful for the future investigation of many types of planets around the nearest stars, including Earth twins. 

The following subsections describe the expected polarization signal from reflecting planets and the search strategy using SPHERE/ZIMPOL. The GTO observations are presented in Section 2, and Section 3 discusses our standard data reduction procedures for ZIMPOL polarimetry. Section 4 provides the description of the angular differential imaging method that we applied to our data and the metric for the assessment of the point-source contrast. Section 5 shows our detailed search results for $\alpha$~Cen~A. Section 6 discusses in more detail the physical meaning of the contrast limits and Section 7 presents our conclusions. In Appendix~\ref{sec:Advanced data reduction steps} and \ref{Appendix: instrmental polarization calculation} we present the advanced data reduction steps necessary to reach the best possible polarimetric contrast limits with ZIMPOL and in Appendix~\ref{Appendix: Results and discussion for the additional targets} we present and discuss the detection limits for all other targets of our survey.

\subsection{The polarization of the reflected light from planets}

\begin{figure*}
\centering
\begin{tabular}{ll}
(a) & (b) \\
\includegraphics[totalheight=3.6in]{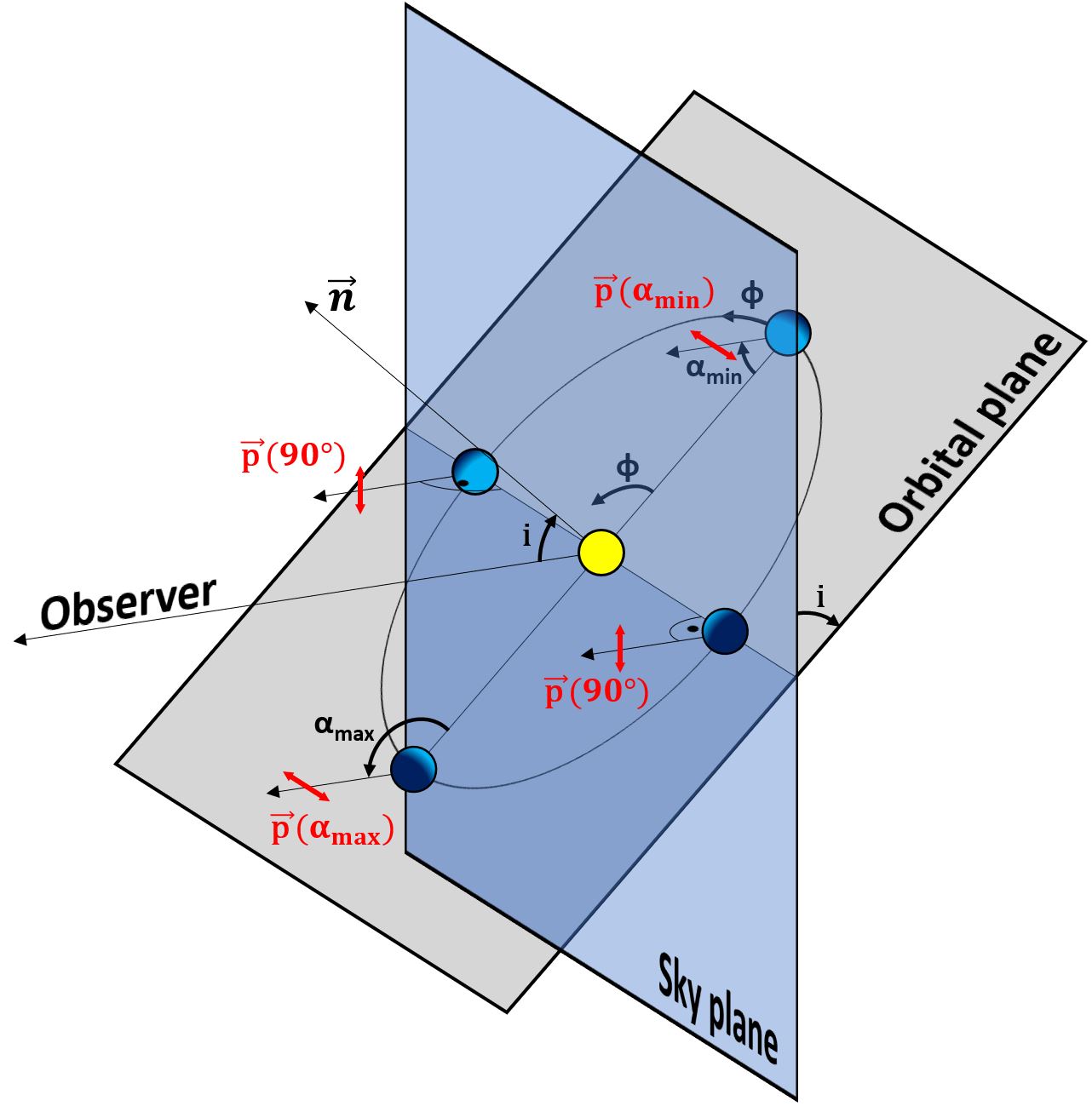} &
\includegraphics[totalheight=3.4in]{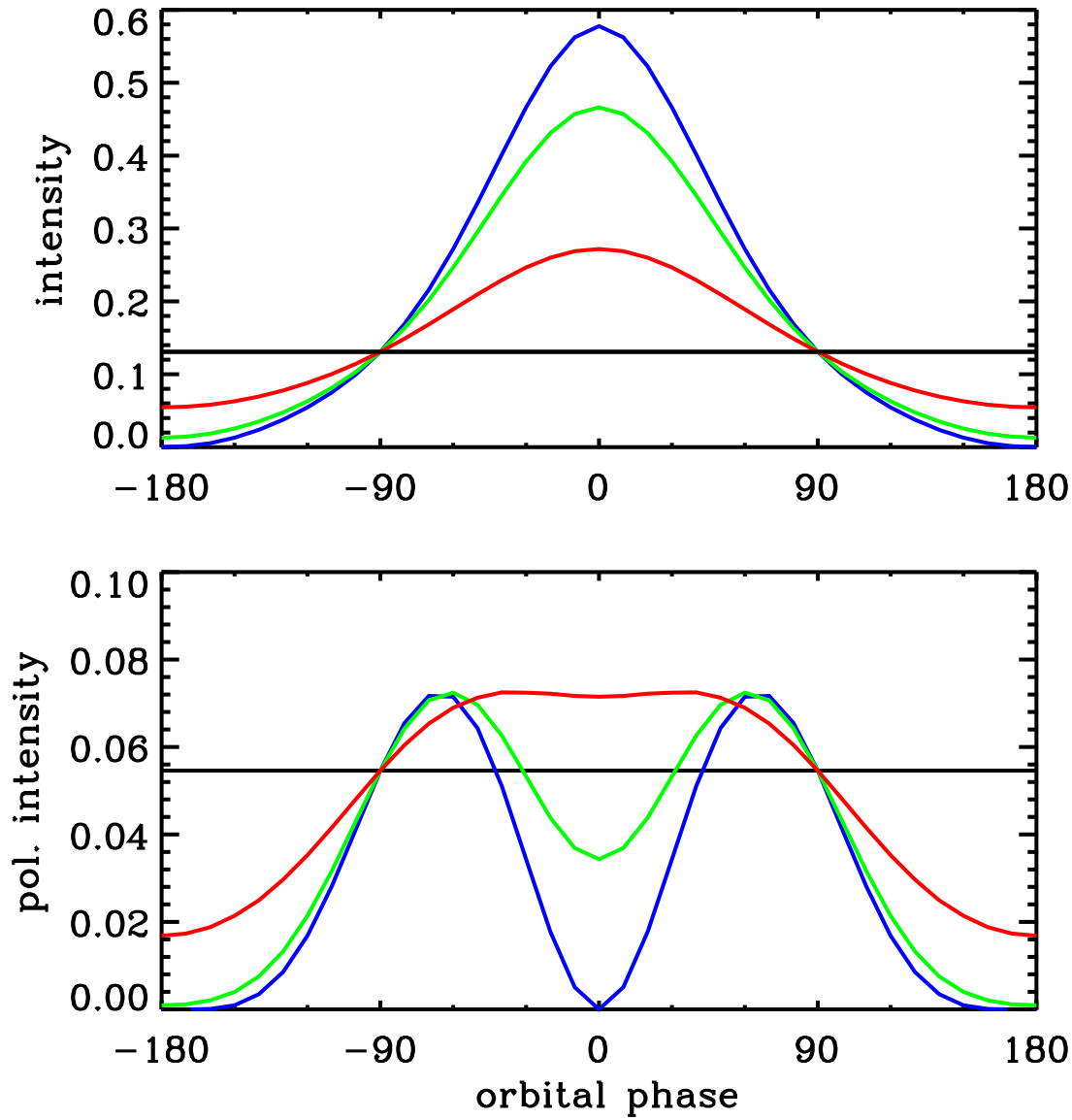} \\
\end{tabular}
\caption{(a) Diagram showing the essential planes and angles needed to characterize the reflected light intensity: the orbital phase of the planet $\phi$, the inclination of the orbital plane with respect to the sky plane $i$ and the scattering angle $\alpha$. The scattering angle $\alpha$ is measured along the scattering plane and our definition of the direction of a positive polarization $p(\alpha)$ is perpendicular to this plane. In special cases the polarization of the reflected light could be negative, this would correspond to a direction of $p(\alpha)$ perpendicular to the red arrows. (b) Normalized intensity and polarized intensity of the reflected light as function of the orbital phase $\phi$, calculated with the reference planet atmosphere model used in this paper: a Rayleigh scattering atmosphere from \citep{Buenzli09} with an optical depth of $\tau_{\rm sc}=2$, a single scattering albedo of $\omega=0.90$, and a ground surface (= cloud) albedo of $A_S=1$. The different colors show the phase functions of planets on circular orbits seen at four different inclinations: $0^\circ$ (black), $30^\circ$ (red), $60^\circ$ (green), $90^\circ$ (blue).}
  \label{PhaseCurvesIQ}
\end{figure*}

The expected polarization signal from reflecting planets has been described with simple models \citep{Seager00}, with detailed calculations for e.g. Jupiter and Earth-like planets \citep{Stam04, Stam08}, or for a parameter grid of planets with Rayleigh scattering atmospheres \citep{Buenzli09, Bailey18}. The intensity and polarized intensity phase functions depending on the orbital phase angle $\phi$ and the planet-star-observer scattering angle $\alpha$ for one such model is illustrated in Fig.~\ref{PhaseCurvesIQ}. These models are guided by polarimetric observations of Solar System objects, for which the typical fractional polarization is quite high $p(\alpha)>10~\%$ for visible wavelengths and scattering angles in the range $\alpha~\approx~60^\circ-120^\circ$ \citep[e.g.][]{Schmid06a}.

Observations of individual objects have shown that for Rayleigh scattering atmospheres like Uranus and Neptune \citep{Schmid06b} the fractional polarization can be substantially higher than this value ($p(90^\circ)> 20~\%$). For mostly haze scattering atmospheres as found on Titan \citep{Tomasko82, Bazzon14} or in the polar regions of Jupiter \citep{Smith84, Schmid11, McLean17} the fractional polarization can even reach values up to $p(90^\circ)\approx 50~\%$. On the other hand, the Mie scattering process in the clouds that dominate the atmospheres of Venus, Saturn or the equatorial regions of Jupiter produces a lower polarization in the visual wavelengths $<10~\%$ \citep{Smith84, Hansen74}. And for larger objects without any significant atmosphere like Mercury, Moon, Mars and other rocky bodies \citep[e.g.][]{Dollfus85} the polarization of the reflected light is somewhere in between $p(90^\circ)\approx 5-20~\%$. Finally, for the polarization of Earth \citet{Bazzon13} determined fractional polarizations of about 19~\% in V-band and 13~\% in R-band mainly caused by Rayleigh scattering in the atmosphere.

For Rayleigh scattering, haze scattering, and the reflection from solid planet surfaces, the resulting polarization for $\alpha~\approx~30^\circ-150^\circ$ is perpendicular to the scattering plane, just like illustrated in Fig.~\ref{PhaseCurvesIQ}(a). This means that for extra-solar planets the polarization is usually positive in perpendicular direction to the line connecting star and planet as projected onto the sky. The polarization, however, can be negative for the reflection from clouds as observed for Venus \citep{Hansen74}, or for reflections with small scattering angles ($\alpha\lessapprox 25^\circ$) on rocky or icy surfaces \citep{Dollfus85}.

\subsection{The signal from extra-solar planets}
\label{sec:The signal from extra-solar planets}

The signal of a reflecting planet depends on the surface properties, which define the reflectivity $I(\alpha)$ and the fractional polarization $p(\alpha)$ of the planet, as well as the planet size and its separation from the central star. The reflectivity and polarization depend on the scattering angle $\alpha$ given by the orbital phase $\phi$ and orbit inclination $i$ as sketched in Fig.~\ref{PhaseCurvesIQ}(a). We set the phase $\phi=0$ in conjunction, when the planet illumination as seen by the observer is maximal. For circular orbits the dependence is

\begin{equation}
\alpha = {\rm arccos}(\sin i \cdot \cos \phi)
\label{equ: angle dependence}
\end{equation}

and the scattering angle varies between a minimum and maximum value $\alpha_{\rm min}$ and $\alpha_{\rm max}$ as indicated in Fig.~\ref{PhaseCurvesIQ}(a). For edge-on orbits ($i=90^\circ$), Eq.~(\ref{equ: angle dependence}) simplifies to $\alpha=\left|\phi\right|$ for $\phi=-180^\circ$ to $180^\circ$, and for pole-on systems ($i=0^\circ$) we see one single scattering angle $\alpha=90^\circ$ during the whole orbit. Small and large scattering angles $\alpha\lessapprox 30^\circ$ and $\alpha\gtrapprox 150^\circ$ are only observable for strongly inclined orbits $i>60^\circ$, but at the corresponding phase angles, planets are typically faint in polarized flux (see Fig.~\ref{PhaseCurvesIQ}(b)), in addition, the angular separation is small and therefore a successful detection will be particularly difficult \citep[e.g.][]{Schworer15}.


For Rayleigh-like scattering the fractional polarization $p(\alpha)$ is highest around $\alpha\approx 90^\circ$ while the reflectivity $I(\alpha)$ is increasing for $\alpha\rightarrow 0^\circ$. Therefore the maximum polarized intensity $p(\alpha)\,I(\alpha)$ is expected for a scattering angle $\alpha\approx 60^\circ$. The full dependence of the normalized intensity $I(\phi)$ and polarized intensity $p(\phi)\,I(\phi)$ as function of orbital phase for a Rayleigh scattering planet is illustrated in Fig.~\ref{PhaseCurvesIQ}(b). The figure shows simulated phase functions for planets on circular orbits with inclinations of $i=0^\circ,\, 30^\circ,\, 60^\circ$, and $90^\circ$.





The model in Fig.~\ref{PhaseCurvesIQ}(b) was selected from the model grid of Rayleigh scattering atmospheres derived in \citet{Buenzli09}. We use it as a reference case for the reflected intensity and polarization of a planetary atmosphere. The model is similar to Uranus and Neptune which are quite favourable cases for a polarimetric search for planets. A giant planet might have a thinner Rayleigh scattering layer $\tau_{\rm sc} < 1$ on top of a cloud layer, resulting in a lower polarization fraction \citep[see][]{Buenzli09}. This is because the reflection from a cloud layer produces significantly less polarization than the reflection from a thick ($\tau_{\rm sc} \gtrapprox 1$) Rayleigh scattering layer. The model shown in Fig.~\ref{PhaseCurvesIQ}(b) has an optical depth of $\tau_{\rm sc}=2$ for the Rayleigh scattering layer, with a single scattering albedo of $\omega=0.95$ above a cloud layer approximated by a Lambertian surface with an albedo of $A_S=1$. This model yields for quadrature phase $\alpha=90^\circ$ a reflectivity of $I(90^\circ)=0.131$ and a corresponding polarized signal amplitude of $p(90^\circ)\cdot I(90^\circ)=0.055$. The parameter $p(90^\circ)$ is a good way of characterizing the polarization of an extra-solar planet because planets at all inclinations will pass through this phase at least twice. In this phase , the fractional polarization is expected to be close to the maximum and the apparent separation from the star is maximized for planets on circular orbits (see Fig.~\ref{PhaseCurvesIQ}(b), \ref{PhaseCurves2D}).

The polarization $p(\alpha)$ refers to the amplitude of the polarization but our raw data consists of independent measurements of the Stokes $Q$ and $U$ parameters. Since we can assume that the reflected light from a planet is polarized along the axis perpendicular to the connecting line between star and planet, we use the transformation into polar coordinates from \citet{Schmid06b} to derive $Q_{\phi}$ and $U_{\phi}$. In the dominating single scattering scenario, the tangential polarization $Q_{\phi}$ should contain all the polarized intensity of the reflected light, while $U_{\phi}$ should be zero everywhere. Because of this relationship we will refer to $Q_{\phi}$ as the polarized intensity throughout this work.

 
The key parameter for the polarimetric search of reflecting extra-solar planets is the polarization contrast $C_{\rm pol}$, this is the polarized flux $Q_{\phi}$ from the planet relative to the total flux from the central star:

\begin{equation}
C_{\rm pol}(\alpha) = p(\alpha)\cdot C_{\rm flux}(\alpha) = 
p(\alpha)\cdot I(\alpha) \frac{R_p^2}{d_p^2}\,,
\label{equ: pol contrast}
\end{equation}

where $R_p$ is the radius of the planet, $d_p$ the physical separation between planet and star and $I(\alpha)$ and $p(\alpha)$ the reflectivity and fractional scattering polarization for a given scattering angle, respectively. In this notation, the reflectivity $I(\alpha=0)$ is equivalent to the geometric albedo $A_g$ of a planet. The ratio $R_p^2/d_p^2$ for a Jupiter-sized planet with radius $R_J$ at a separation $d_p=1$~AU is $R_J^2/{\rm AU}^2=2.3\cdot 10^{-7}$ and the total polarization contrast of a planet with our reference model with $p(90^\circ)\cdot I(90^\circ)=0.055$ would be of order $C_{\rm pol}\approx 10^{-8}$. A Neptune-sized planet would have to be located at about 0.5~AU to produce the same polarization contrast. With increasing physical separation $d_p$ the contrast decreases rapidly with $1/d_p^{2}$ (see Eq. (\ref{equ: pol contrast})). With increasing distance to the star, the angular separation $\rho$ of a planet at a constant $d_p$ also decreases. Thus moving it closer to the star where high contrasts cannot be maintained. The combination of both effects limits the sample of possible targets for a search of reflected light to the most nearby stars. In addition to that, the sample is limited to only the brightest stars because photon noise increases like $1/\sqrt{F}$ with the lower photon flux $F$ of stars that are fainter in the visible wavelengths.

\subsection{Targets for the search of extra-solar planets}


The detection space for our SPHERE/ZIMPOL high-contrast observations starts at about $\rho \approx 0.1\arcsec$, and the current polarimetric contrast limits after post-processing are of the order $10^{-7}$ for $\rho<0.5\arcsec$ and $10^{-8}$ for $\rho>0.5\arcsec$. Therefore only the nearest stars within about 5~pc can have a bright enough reflecting planet with $R_p\approx R_J$ and a contrast of $C_{\rm pol}\gtrapprox 10^{-8}$ with a sufficiently large angular separation $\rho>0.1\arcsec$ for a successful detection. Based on these criteria, some of the best stellar systems for the search of a Jupiter-sized planet in reflected light with SPHERE/ZIMPOL are $\alpha$~Cen~A and B, Sirius~A, $\epsilon$~Eri, $\tau$~Cet, Altair and a few others as determined by \citet{Thalmann08}.  


No extra-solar planet is known to exist around these high priority stars which would fulfil the above detection limit criteria. There is strong evidence from radial velocity and astrometric studies for the presence of a giant planet in $\epsilon$ Eri \citep[e.g.][]{Hatzes00, Mawet19}, but the derived separation is 3~AU and therefore the expected signal is at the level of only $C_{\rm pol}\approx 10^{-9}$. For $\tau$ Cet, the presence of planets has been proposed based on radial velocity data \citep{Feng17}, but none is expected to produce a contrast $C_{\rm pol}\gtrapprox 10^{-9}$. The radial velocity constraints for the A-stars Sirius~A and Altair are very loose because their spectra are not well suited for sensitive radial velocity searches, and undetected giant planets at 1~AU may be present. The radial velocity limits for planets are very stringent for $\alpha$ Cen B \citep{Zhao18}, but less well constrained for $\alpha$ Cen A \citep{Zhao18}. However, the simple calculation of the reflected light contrast does not consider the possibility that a planet could be exceptionally bright due to certain reasons, e.g. an extensive ring system surrounding the planet \citep[e.g.][]{Arnold04}. Because of the absence of obvious targets, we decided to carry out an exploratory blind search for ``unexpectedly'' bright companions, with the additional aim to investigate the detection limits of this instrument and to define the best observing strategies for possible future searches. 

\begin{figure}
\resizebox{\hsize}{!}{\includegraphics{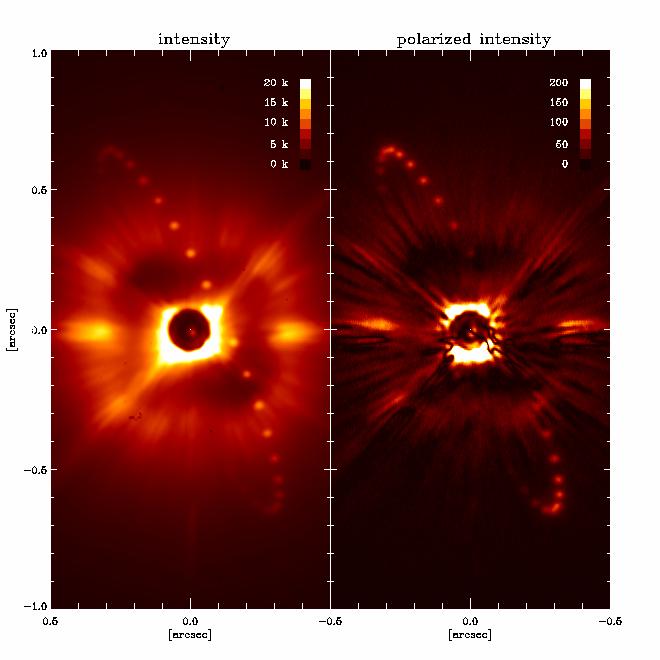}}
\caption{Apparent positions of a model planet on a circular $80^\circ$ inclined orbit with $r=1$~AU around $\alpha$ Cen A in a typical coronagraphic intensity (left) and polarization frame (right) at 10~day intervals. The brightness of the planet signal with respect to $\alpha$~Cen~A is exaggerated by a factor of $10^4$ for the intensity and $10^3$ for the polarization. The relative brightness of the point for different phases is according to the model presented in Fig.~\ref{PhaseCurvesIQ}.}
\label{PhaseCurves2D}
\end{figure}

For such a survey, one needs to consider that planets around the nearest stars are moving fast through our field-of-view (FOV). This is illustrated in Fig.~\ref{PhaseCurves2D}, which simulates the orbit of a planet with a circular orbit with a separation of 1~AU around $\alpha$ Cen A on top of single coronagraphic intensity or polarimetric frames. The individual points are the orbital positions of this model planet separated by 10~days. The relative brightness of the points are calculated for an orbit inclination of $i=80^\circ$ coplanar with the $\alpha$~Cen binary \citep{Kervella16} and using the same Rayleigh scattering atmosphere model as in Fig.~\ref{PhaseCurvesIQ}, but with the brightness upscaled by a factor of $10^4$ for the intensity and $10^3$ for the polarization to make the dots visible on top of a single coronagraphic observation. Of course, the angular motion depends on the orbital parameters and the distance of the systems and our example $\alpha$ Cen A system would show for a planet the fastest angular orbital motion for a given orbital separation because of its proximity. 

Without going into details, already the $\alpha$ Cen A example in Fig.~\ref{PhaseCurves2D} illustrates, that planets on inclined orbits have phases with large separation when they are relatively bright and easy to detect, and phases where they are close to the star and faint and challenging to detect. Therefore, a blind search provides only planet detection limits valid for that observing date. One should also notice that data taken during different nights cannot simply be coadded for the search of extra-solar planets due to the expected short orbital periods. Instead it would be necessary to use a tool like K-Stacker \citep{Nowak18} that combines the results from multiple epochs while considering the orbital motion of a planet.

\section{Observations}

\subsection{The SPHERE/ZIMPOL instrument}

The polarimetric survey for extra-solar planets was carried out
with the SPHERE "Planet Finder" instrument \citep{Beuzit08, Beuzit19}
on VLT Unit Telescope 3 (UT3) of the European Southern 
Observatory. SPHERE is an extreme adaptive optics system
with a fast tip-tilt mirror and a fast high-order 
deformable mirror with 41x41 actuators and a Shack-Hartman
wave-front sensor \citep[e.g.][]{Fusco06}. The system includes an image de-rotator, atmospheric dispersion correctors, 
calibration components and the IRDIS \citep{Dohlen08}, IFS \citep{Claudi08} and
ZIMPOL focal plane instruments for high-contrast imaging. 

This program was carried out with ZIMPOL 
which was specifically designed for the polarimetric search of 
reflected light from extra-solar planets around the nearest, bright stars
in the spectral range 500-900~nm. The SPHERE/ZIMPOL system is  
described in detail in \citet{Schmid18} and we highlight here some of the important
properties for high-contrast imaging of reflected light from planets:
\begin{itemize}
\item the polarimetric mode is based on a fast modulation - demodulation technique which reaches a polarimetric sensitivity\footnote{Degree of suppression of the light by the polarimetry} of $\Delta p<10^{-4}$ \citep{Schmid18} in the light halo of a bright star. This is possible because the used modulation frequency of 968~Hz is faster than the seeing variations and therefore the speckle noise suppression for PDI is particularly good as long as the coherence time $\tau_0$ is greater than about $2~{\rm ms}$. This condition was usually satisfied during the RefPlanets observations (see Table~\ref{table: Refplanets data}).
\item ZIMPOL polarimetry can be combined with coronagraphy for the suppression of the diffraction limited 
PSF peak of the bright star, for a sensitive search of faint point-sources in the light halo of a bright star. 
\item ZIMPOL has a small pixel scale of 3.6 mas/pix, a detector mode with a high pixel gain of 10.5 e$^-$~ADU$^{-1}$ and a full well capacity of 640~ke$^-$~pix$^{-1}$. This allows to search for very faint
polarized signals in coronagraphic images of very bright stars $m_R<4^m$ with broad-band filters by "just" pushing the photon noise limit thanks to the photon collecting power of the VLT telescope.  
\end{itemize}

The combination of high-contrast imaging using AO and coronagraphy provides for point-sources a raw contrast at a level $10^{-4}-10^{-5}$, while polarimetry in combination with angular differential imaging (ADI) yields a further contrast improvement for polarimetric differential imaging of about $10^{-3}$, so that a total contrast
of $C_{\rm pol}\approx 10^{-8}$ is reachable with sufficiently long integrations. 

\subsection{Observations}
In Table \ref{table: Refplanets data} we list all observations which 
were carried out so far for the RefPlanets GTO program. We observed six of the most favourable targets in the solar neighbourhood identified by \citet{Thalmann08} as ideal targets for the search of planets in reflected light.

All data were taken with the fast modulation 
polarimetry mode, which is the mode of choice for high flux 
applications. The first observations in 2015 were made with different filters in 
camera 1 and camera 2 of ZIMPOL. But it was noticed that some disturbing polarimetric 
residuals can be corrected if the simultaneous
camera 1 and camera 2 frames are taken with the same filter passband, because
the residuals have opposite signs and compensate when camera 1 and camera 2 frames taken with the same filter are combined. Of course, the contrast also improves with the combination of data from both cameras because of the lower photon noise limit.
From 2016 onwards we took for each hour of 
coronagraphic observations one or two short polarimetric cycles  
with the star offset from the focal plane mask for the calibration of the flux, the point-spread function (PSF), 
and the polarimetric beam shift \citep{Schmid18}. These PSFs were taken with neutral density (ND) filters to avoid detector saturation.

The main criterion for the filter selection is a high photon throughput. Filters with broader passbands provide more photons and stars with $m_R>1^m$ were observed usually in the VBB filter ($\lambda_{c,\rm VBB}=735$~nm, $\Delta\lambda_{\rm VBB}=291$~nm). For Sirius~A, $\alpha$~Cen~A and Altair we used filters with smaller band widths to avoid detector saturation with the minimum detector integration time of 1.1~s available for ZIMPOL, namely, the R\_PRIM ($\lambda_{c,\rm R\_PRIM}=626$~nm, $\Delta\lambda_{\rm R\_PRIM}=149$~nm), N\_R ($\lambda_{c,\rm N\_R}=646$~nm, $\Delta\lambda_{\rm N\_R}=57$~nm) and N\_I ($\lambda_{c,\rm N\_I}=817$~nm, $\Delta\lambda_{\rm N\_I}=81$~nm) filters. Only for $\tau$~Ceti we deviated from this strategy and chose the R\_PRIM filter instead of the VBB filter because we noticed that certain disturbing wavelength dependent instrumental effects (instrumental polarization, beam shift) are easier to correct during the data reduction for data narrower passbands.

Almost all objects were observed with SPHERE/ZIMPOL in P1-mode, in which the image de-rotator is fixed. In this mode the sky rotates as a function of the telescope parallactic angle and altitude allowing for ADI \citep{Marois08} in combination with PDI because most of the strong aberrations -- mainly caused by the deformable mirror (DM) -- are fixed with respect to the detector. The P1-mode stabilizes the instrument polarization after the HWP2-switch, but does not stabilize the telescope pupil, which still rotates with the telescope altitude. Therefore, speckles related to the telescope pupil cannot be suppressed with ADI. We observed only $\epsilon$~Eri in the field-stabilized polarimetric P2-mode to make use of the improved capability of the instrument to detect weak extended scattering polarization from circumstellar dust which could be detectable with our FOV of $3.6 \arcsec \times 3.6 \arcsec$ \citep[e.g.][]{Backman09, Greaves14}.

For all observations we used the medium sized classical Lyot coronagraph CLC-MT-WF with a dark focal plane mask spot deposited on a plate with a radius corresponding to 77.5~mas \citep{Schmid18}, however, the effective inner working angle (IWA) of the reduced data is generally larger and depends on the star centering accuracy and stability. The spot in this coronagraphic mask has a transparency of 
about 0.1~\% \citep{Schmid18} and during good conditions and with good centering the star is visible behind the coronagraph so that an accurate centering of the frames in possible.

Our usual observing strategy for deep coronagraphic observations consists of one-hour blocks with about five to ten polarimetric cycles. Each cycle consists of observations with all four half-wave plate orientations (Q$^+$, Q$^-$, U$^+$, U$^-$). Between these blocks we took short non-coronagraphic cycles with a neutral density filter, by offsetting the star from the coronagraphic mask, to acquire samples of the unsaturated PSF for image quality assessments, flux calibrations, and the measurement of the beam shift effect.

\begin{table*}
\tiny
\caption{Summary of all RefPlanets observations completed until the end of 2018. For each observation we also list observing conditions (seeing in arc seconds and coherence time $\tau_0$ in ms) and the total field rotation (relevant for the efficiency of angular differential imaging).}
\label{table: Refplanets data}
\centering
\begin{tabular}{l l l l l l l l l l l l l}
\hline\hline
Date (UT) & Object & m$_R$ & \multicolumn{2}{c}{Filters} & DIT & \# of & $t_{\rm exp}$\tablefootmark{a} & Seeing & $\tau_0$ & Air mass & Field\\
 & & & & & (sec) & pol. & & $\left(\arcsec\right)$ & (ms) & & rotation \\
 & & (mag) & cam1 & cam2 & & cycles & & & & & $\left(^{\circ}\right)$ \\
 
\hline                       
    2015/05/01 & $\alpha$ Cen A & -0.5 & N\_I & N\_R & 1.2  & 66 & 2h 38.4min & 0.6--1.0 & 1.7--2.5 & 1.24--1.47 & 90.4\\
    2015/05/02 & $\alpha$ Cen B & 1.0 & VBB & R\_PRIM & 1.2  & 90 & 3h 36min & 0.6--1.1 & 1.5--2.2 & 1.24--1.46 & 97.1\\
	2016/02/17 & Sirius~A & -1.5 & N\_I & N\_I & 1.2  & 34 & 1h 21.6min & 0.7--2.0 & 1.6--2.5 & 1.01--1.13 & 101.1\\	
	\textbf{2016/02/20} & \textbf{Sirius~A} & \textbf{-1.5} & \textbf{N\_I} & \textbf{N\_I} & \textbf{1.2}  & \textbf{73} & \textbf{2h 55.2min} & \textbf{1.0--2.0} & \textbf{1.8--3.5} & \textbf{1.01--1.39} & \textbf{113.2}\\
    2016/04/18 & $\alpha$ Cen A & -0.5 & N\_R & N\_R & 1.2  & 40 & 1h 36min & 1.1--1.5 & 1.8--2.4 & 1.26--1.48 & 46.7\\
    2016/04/21 & $\alpha$ Cen A & -0.5 & N\_R & N\_R & 1.2  & 80 & 3h 12min & 0.7--1.7 & 2.0--4.0 & 1.24--1.71 & 107.9\\
    2016/06/22 & $\alpha$ Cen B & 1.0 & VBB & VBB & 1.1  & 74 & 2h 42.8min & 0.3--0.8 & 3.5--6.0 & 1.24--1.60 & 106.6\\
    \textbf{2016/07/21} & \textbf{Altair} & \textbf{0.6} & \textbf{R\_PRIM} & \textbf{R\_PRIM} & \textbf{1.2}  & \textbf{63} & \textbf{2h 31.2min} & \textbf{0.4--0.8} & \textbf{4.5--7.0} & \textbf{1.20--1.44} & \textbf{67.2}\\
    2016/07/22 & Altair & 0.6 & R\_PRIM & R\_PRIM & 1.2  & 30 & 1h 12min & 0.4--0.7 & 3.0--5.0 & 1.20--1.32 & 27.6\\
	2016/10/10 & $\epsilon$ Eri & 3.0 & VBB & VBB & 3.0  & 42 & 2h 48min & 0.5--0.8 & 4.5--6.7 & 1.05--1.37 & 0\tablefootmark{b}\\
	\textbf{2016/10/11} & \textbf{{\boldmath$\epsilon$} Eri} & \textbf{3.0} & \textbf{VBB} & \textbf{VBB} & \textbf{3.0}  & \textbf{48} & \textbf{3h 12min} & \textbf{0.5--0.8} & \textbf{3.3--6.5} & \textbf{1.04--1.18} & \textbf{0\tablefootmark{b}}\\
	2016/10/12 & $\epsilon$ Eri & 3.0 & VBB & VBB & 5.0  & 15 & 1h 40min & 1.0--1.8 & 1.8--2.1 & 1.04--1.21 & 0\tablefootmark{b}\\
    \textbf{2017/04/30} & \textbf{{\boldmath$\alpha$} Cen A} & \textbf{-0.5} & \textbf{N\_R} & \textbf{N\_R} & \textbf{1.2}  & \textbf{84} & \textbf{3h 21.6min} & \textbf{0.5--0.7} & \textbf{3.0--4.4} & \textbf{1.24--2.21} & \textbf{121.5}\\
    2017/05/01 & $\alpha$ Cen A & -0.5 & N\_R & N\_R & 1.2  & 39 & 1h 33.6min & 0.8--1.3 & 2.0--2.5 & 1.37--1.84 & 47.8\\
    \textbf{2017/06/19} & \textbf{{\boldmath$\alpha$} Cen B} & \textbf{1.0} & \textbf{VBB} & \textbf{VBB} & \textbf{1.1}  & \textbf{141} & \textbf{3h 26.8min} & \textbf{0.3--1.0} & \textbf{4.5--9.5} & \textbf{1.24--1.52} & \textbf{115.6}\\
    \textbf{2018/10/14} & \textbf{{\boldmath$\tau$} Ceti} & \textbf{2.9} & \textbf{R\_PRIM} & \textbf{R\_PRIM} & \textbf{14} & \textbf{30} & \textbf{2h 48min} & \textbf{0.4--0.7} & \textbf{5.0--10} & \textbf{1.01--1.12} & \textbf{130.5}\\
    2018/10/15 & $\tau$ Ceti & 2.9 & R\_PRIM & R\_PRIM & 14 & 24 & 2h 14.4min & 0.6--1.6 & 2.0--4.0 & 1.01--1.12 & 112.1\\
    2018/10/16 & $\tau$ Ceti & 2.9 & R\_PRIM & R\_PRIM & 14 & 32 & 2h 59min & 0.6--1.0 & 2.5--3.7 & 1.01--1.28 & 104.9\\
    2018/10/19 & $\tau$ Ceti & 2.9 & R\_PRIM & R\_PRIM & 14 & 29 & 2h 42.4min & 0.6--1.4 & 2.6--5.3 & 1.01--1.60 & 107.5\\
\hline                        
\end{tabular}
\tablefoot{The datasets with the deepest limits for each target are marked in bold font. 
\tablefoottext{a}{The total exposure time per camera.}
\tablefoottext{b}{$\epsilon$ Eri was observed in the field stabilized ZIMPOL P2-polarimetry mode.}
}
\end{table*}

\section{Basic data reduction}
The data reduction is mainly carried out with the IDL-based sz-software (SPHERE/ZIMPOL) pipeline developed at ETH Zurich. Basic data preprocessing, reduction and calibration steps are essentially identical to the ESO Data Reduction and Handling (DRH) software package developed for SPHERE \citep{Pavlov08}. The basic steps are described briefly in this subsection and more technical information is available in \cite{Schmid12, Schmid18}. In addition to that, we describe in the appendix the more advanced sz-pipeline routines and additional data reduction procedures required especially for high-contrast imaging and polarimetry.

The fast modulation and on-chip demodulation imaging polarimetry of ZIMPOL produces raw frames where the simultaneous $I_\perp$ and $I_\parallel$ polarization signals are registered on alternating rows of the CCD detectors. Basically, the ZIMPOL raw polarization signal $Q^Z$ is the difference of the ``even-row'' $I_\perp$ and the ``odd-row'' $I_\parallel$ subframes $Q^Z=I_\perp-I_\parallel$. The raw intensity signal is derived from adding the two subframes $I^Z=I_\perp+I_\parallel$.

Just like for any other CCD detector data, the basic data reduction steps include image extraction, frame flips for the correct image orientation, a first bias subtraction based on the pre- and overscan pixel level, bias frame subtraction for fixed pattern noise removal, and flat-fielding. Special steps for the ZIMPOL-system are the differential polarimetric combination of the subframes, taking into account the alternating modulation phases for the CCD pixel charge trap correction \citep{Gisler04, Schmid12}, and 
calibrating the polarimetric efficiency $\epsilon_{\rm pol}$ or modulation-demodulation efficiency. The polarimetric combination of the frames of a polarimetric cycle $Q^+$, $Q^-$, $U^+$, $U^-$ taken with the four half-wave plate orientations is again done in a standard way. For non-field stabilized observations, the data combination must also consider the image rotation. As basic data product of one polarimetric cycle one obtains four frames $I_Q$, $Q$, $I_U$, and $U$, which can be combined with the frames from many other cycles for higher signal-to-noise ratio (SNR) results.

The basic PDI data reduction steps listed above are not sufficient for reaching the very high polarimetric contrast required for the search of reflecting planets. Especially at smaller separations $\lessapprox0.6\arcsec$ the noise is still dominated by residuals of order $10^{-6}$ in terms of contrast compared to the brightness of the star (see Fig.~\ref{fig:data_redu_steps_single_img}). This is why we additionally apply more advanced calibration steps described in Appendix~\ref{sec:Advanced data reduction steps} and \ref{Appendix: instrmental polarization calculation}. The steps include:
\begin{itemize}
\item Frame transfer smearing correction
\item Telescope polarization correction
\item Correction of the differential polarimetric beam shift
\end{itemize}

\section{Post-processing and the determination of the contrast limits}
\begin{figure}
\resizebox{\hsize}{!}{\includegraphics{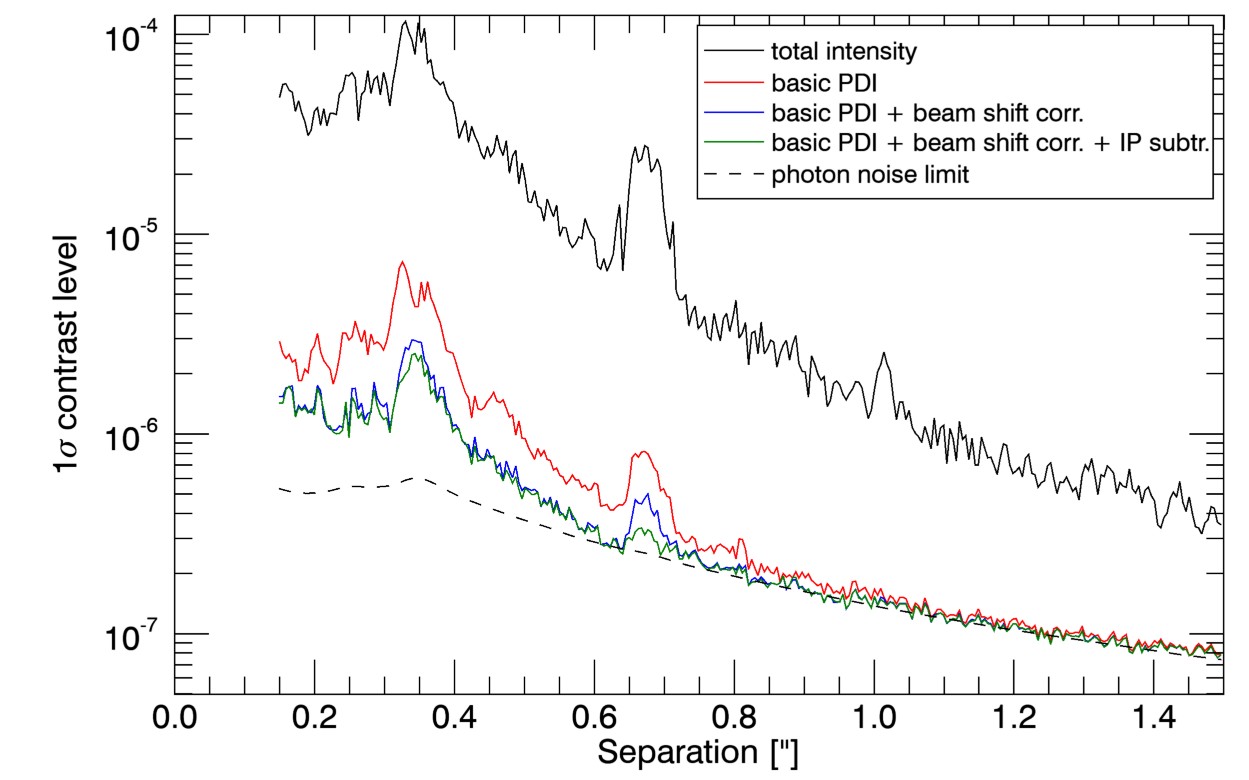}}
\caption{The 1$\sigma$ radial contrast levels for the data shown after the different major data reduction steps -- basic PDI by ZIMPOL, correction of the polarimetric beam shift and subtraction of instrument polarization (IP) -- for the polarized intensity $Q$ and for the corresponding intensity $I$ of one single combined zero-phase and $\pi$-phase ($2 \times 1.2$~s) exposure of $\alpha$~Cen~A in the N\_R filter.}
\label{fig:data_redu_steps_single_img}
\end{figure}

There is still a landscape of residual noise visible after the basic data reduction, beam shift and frame transfer smearing correction, and the subtraction of the residual instrument polarization. This can be seen for example in bottom panel in Fig.~\ref{fig:bs_corr_exmpl_ACenA30Apr}. We show this quantitatively with 1$\sigma$ noise levels for a series of short 2.4~s exposures measured after the different reduction steps in Fig.~\ref{fig:data_redu_steps_single_img}. After the full data reduction, the residual noise at small separations $<0.6\arcsec$ still dominates the photon noise by a factor of about 2--5 in this particular example. For larger separations $> 0.6 \arcsec$ the residual noise is close to the photon noise limit. In a effort to further reduce the noise at small separations, we used a principle component analysis (PCA) \citep{Amara12} based ADI algorithm to model the fixed and slowly varying residual noise features.

\subsection{PCA based ADI}
\begin{figure}
\resizebox{\hsize}{!}{\includegraphics{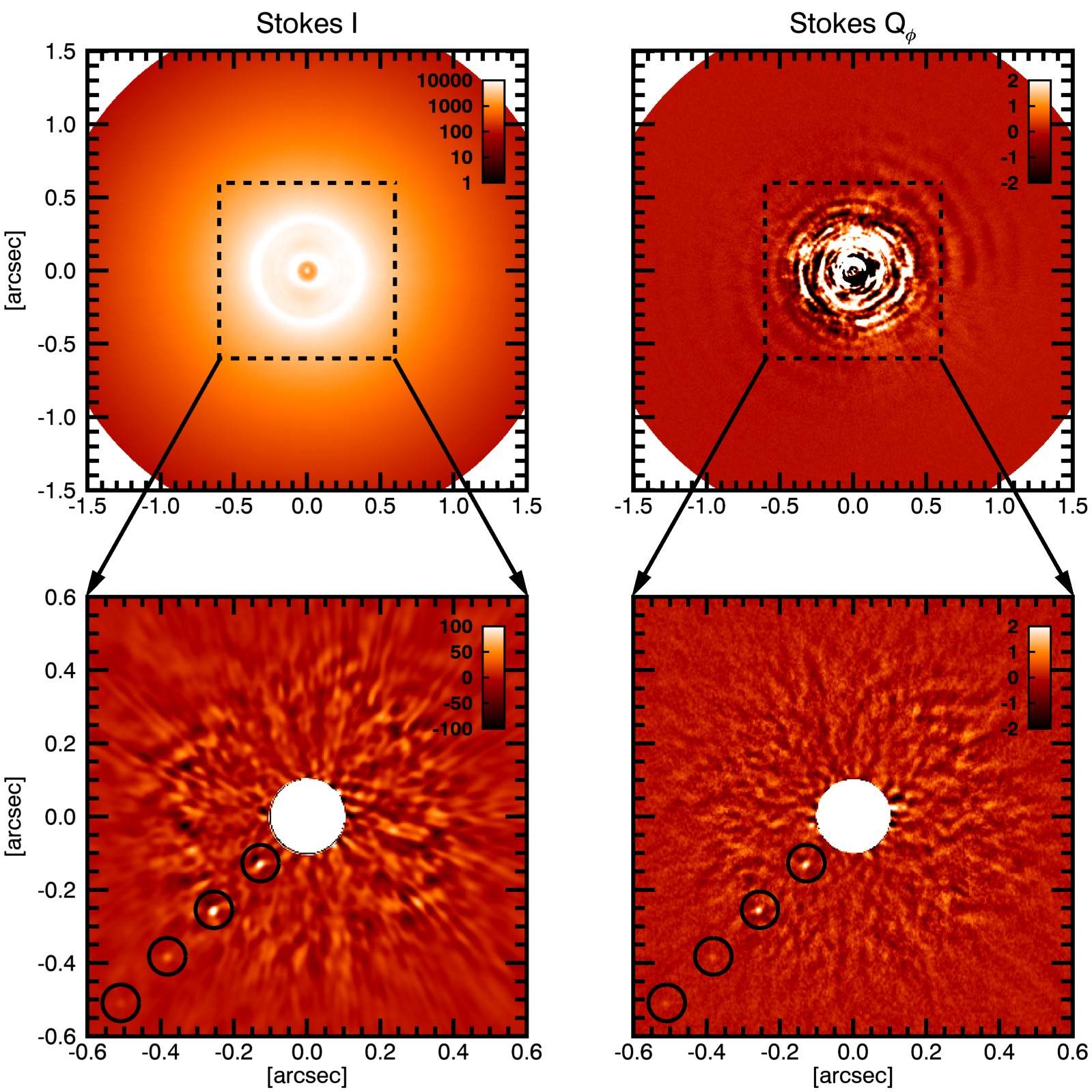}}
\caption{Total intensity (Stokes I) and polarized intensity (Stokes Q$_{\phi}$) for the complete dataset of $\alpha$~Cen~A in the N\_R filter. The frames in the bottom row show a closer look at the speckle-dominated region closer to the star after injecting artificial point-sources (black circles) and applying PCA-ADI with 20~PCs. }
\label{fig:5sigm_fkplnts_20PCs}
\end{figure}

The top row in Fig.~\ref{fig:5sigm_fkplnts_20PCs} shows the coronagraphic total intensity and polarized intensity data of $\alpha$~Cen~A from June 2017 after de-rotating and combining all frames. The intensity exhibits a PSF speckle halo with a strong radial gradient over two orders of magnitude. The differential polarization shows arc-like patterns in the de-rotated and combined image which originate from the de-rotated fixed residual noise pattern. ADI can be used to efficiently model and subtract such large scale patterns before de-rotating and combining the images. The PCA-ADI approach was used successfully before by \citet{vanHolstein17} to improve the contrast limits of SPHERE/IRDIS polarimetry data.

We used a customized version of the core code from the PynPoint pipeline \citep{Amara12, Stolker19} for the ADI process. The complete speckle subtraction process was applied to the stacks of Q$^+$, Q$^-$, U$^+$, U$^-$ and intensity frames separately after preprocessing and centering the frames. For the polarized intensity frames we applied PCA in an annulus around the star from 0.1\arcsec to 1\arcsec in order to cover the speckle-dominated region. For the total intensity frames we increased the outer radius to 1.8\arcsec since the whole FOV is dominated by speckles and other fixed pattern noise. We used a fixed number of 20 principle components (PCs), or 10~PCs in the case of the $\tau$~Ceti polarimetry, to model and subtract the residual noise patterns because this seemed to be the sweet-spot that produced deep contrast limits at most separations. In the bottom row of Fig.~\ref{fig:5sigm_fkplnts_20PCs} we show an example for the result after removing 20~PCs from the intensity and polarized intensity frames. In both cases the contrast improved significantly. Typical contrast limit improvements for PCA-ADI were between a factor of 5--10 for the total intensity and up to a factor of 3 for the polarized intensity. In Fig.~\ref{fig:5sigm_fkplnts_20PCs}, the resulting images after PCA-ADI also contain a number of artificial planets with SNR$\approx$5, they were introduced for estimating the contrast limits after PCA-ADI.

For the de-rotation and combination of the frames, we applied the noise-weighted algorithm as described in \citet{Bottom17}. This algorithm is simple to implement in a direct imaging data reduction pipeline and it often improved the SNR of the artificial planets significantly, with typical SNR gains of about 8\% and a maximum gain up to 26\% for our $\alpha$~Cen~A test dataset. 

\subsection{Polarimetric point-source contrast}
\begin{figure}
\resizebox{\hsize}{!}{\includegraphics{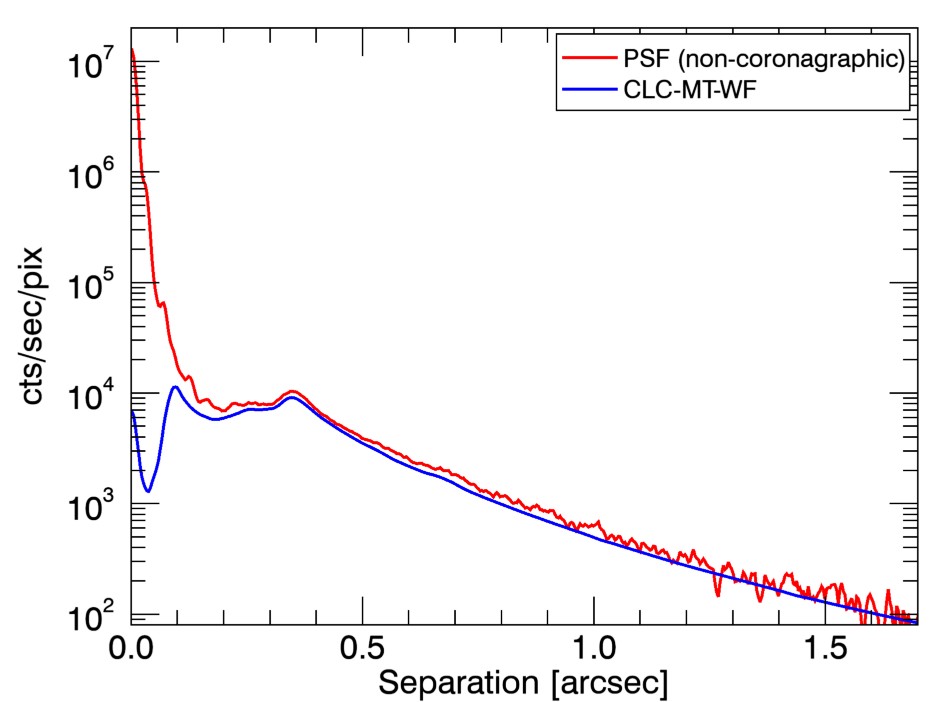}}
\caption{The coronagraphic PSF with the CLC-MT-WF coronagraph (blue) compared to the non-coronagraphic PSF of $\alpha$~Cen~A in the N\_R filter (red). The coronagraphic PSF was upscaled by a factor of $\sim 8\cdot10^{3}$ to account for the use of a neutral density filter during the measurement.}
\label{fig:psf_profiles}
\end{figure}

The contrast limits after the PCA-ADI step were calculated using artificial point-sources arranged in a spiral pattern around the star which we tried to recover with SNR=5. We determined the SNR according to the methods derived in \citet{2014ApJ...792...97M}, including the correction for small sample statistics. The artificial planet PSF was simulated with a non-coronagraphic PSF from one of the beam shift measurements, upscaled with the mean value of the transmission curve for the neutral density filter that was used to avoid saturation. We visually selected the non-coronagraphic PSF that best fits the shape of the coronagraphic PSF of the combined intensity image at separations $>0.3\arcsec$. This ensures that we do not severely over- or underestimate the aperture flux that a point-source would have in our data, and therefore ensures accurate contrast limit estimations. For the $\alpha$~Cen~A data, we show the radial profiles of both PSFs in Fig.~\ref{fig:psf_profiles}, normalized to the number of counts on the detector per second and pixel.

The aperture radius $r_{ap}$ used for the contrast estimation and SNR calculation was optimized for high SNR under the assumption that the searched point-source is weak compared to the PSF of the central star and read-out noise is negligible. With increasing $r_{ap}$, the number of counts from a faint source increases, however, a larger aperture also has an increased background noise $\sigma_{bck} \propto r_{ap}$. We derived an optimized $r_{ap}\approx\lambda/D$, corresponding to about 4-6 pixels (14-22~mas) depending on the observed wavelength. The flux in each aperture was background subtracted individually using the mean value of the pixels in a two pixel wide concentric annulus around the aperture, because the point-source contrast should not be affected by residual, non-axisymmetric, large scale structures in the image (e.g. stray light from $\alpha$~Cen~A in the observations of $\alpha$~Cen~B). 

We also calculated raw contrast curves for both Stokes I and Q$_{\phi}$ without PCA-ADI to investigate how the other advanced data reduction steps improve the contrast limits at different separations. The calculated raw contrast curves do not require the insertion of fake signals and are independent on the field rotation. Therefore, the raw contrast is more suitable for assessing the quality of small subsets of the data or even single exposures.

For the raw contrast we also used methods derived in \citet{2014ApJ...792...97M} to calculate the noise at different separations to the star and turn this into the signal aperture flux required for a detection. The detection threshold was set to a constant false positive fraction (FPF) corresponding to the FPF of an N$\sigma$ detection with Gaussian distributed noise. The required aperture flux was then turned into a contrast limit estimation by dividing it through the aperture flux of the unsaturated stellar PSF.

In order to apply the signal detection method described in \citet{2014ApJ...792...97M}, the underlying distribution of noise aperture fluxes has to be approximately Gaussian. We applied a Shapiro-Wilk test and found that this condition is satisfied for all separations.

\section{Results for $\alpha$ Cen A}
\label{sec:Results for alpha Cen A}
We present a detailed analysis of the results from the deepest observations of $\alpha$~Cen~A. We derive contrast limits and analyse the properties of the noise at different separations and for different total DITs. This detailed analysis shows the outstanding performance of ZIMPOL in terms of speckle suppression in PDI mode. In addition to that, we present and discuss the best results for all other targets of our survey in Appendix~\ref{Appendix: Results and discussion for the additional targets}.

\subsection{Total intensity and polarized intensity}
The deepest observations of $\alpha$~Cen~A were carried out in the N\_R filter and in P1 polarimetry mode during a single half-night with good observing conditions (see Table~\ref{table: Refplanets data}). For the full data reduction we used the best 84 out of 88 polarimetric cycles with a total $t_{\rm exp}$ of 201.6~min for each camera. This is our longest exposure time with a narrow-band filter. We combined the results of both cameras to improve the photon noise limit by an additional factor of around $\sqrt{2}$. The resulting de-rotated and combined images are shown in Fig. \ref{fig:5sigm_fkplnts_20PCs}. Just as described in Sec.~\ref{sec:The signal from extra-solar planets}, we transformed the polarized intensity frames $Q$, $U$ into the $Q_{\phi}$, $U_{\phi}$ basis. We expect a positive Q$_{\phi}$ and no U$_{\phi}$ signal from the reflected light of a companion. The bottom panels in Fig. \ref{fig:5sigm_fkplnts_20PCs} show the inner, speckle-dominated region after inserting four artificial point-sources in the lower left corner and subsequently removing 20 PCs modes. The final image for $Q_{\phi}$ is clean and shows no disturbing residuals except for a few very close to the coronagraph. However, the total intensity shows some strong disturbing features that are extended in the radial direction. These features are residuals from the diffraction pattern of the rotating telescope spiders. The residuals are unpolarized and hence mostly cancelled in the $Q_{\phi}$ result.


\subsection{Contrast curve}
\label{sec:Contrast curve}
\begin{figure*}
\centering
\begin{tabular}{ll}
(a) & (b) \\
\includegraphics[totalheight=2.8in]{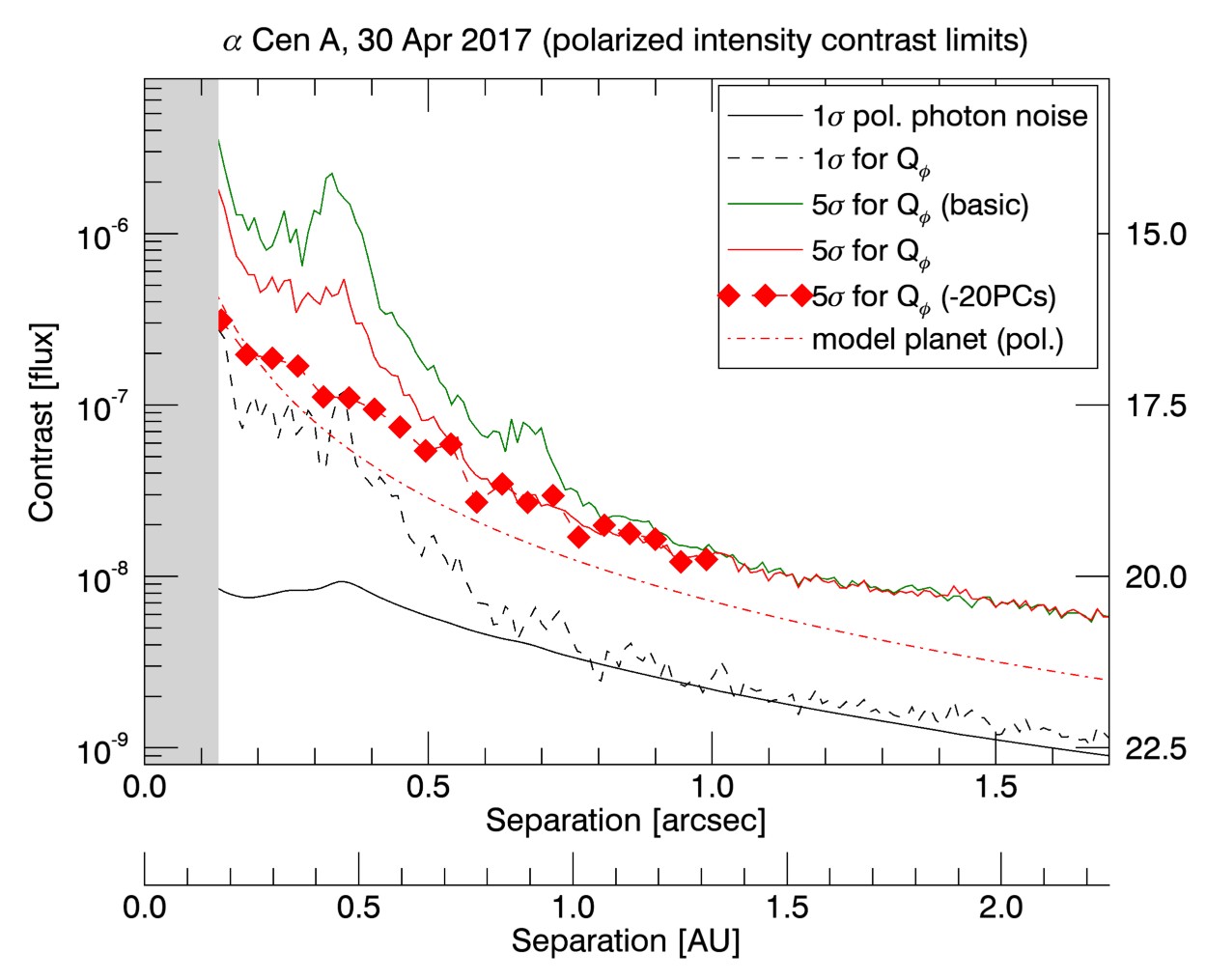} & \includegraphics[totalheight=2.8in]{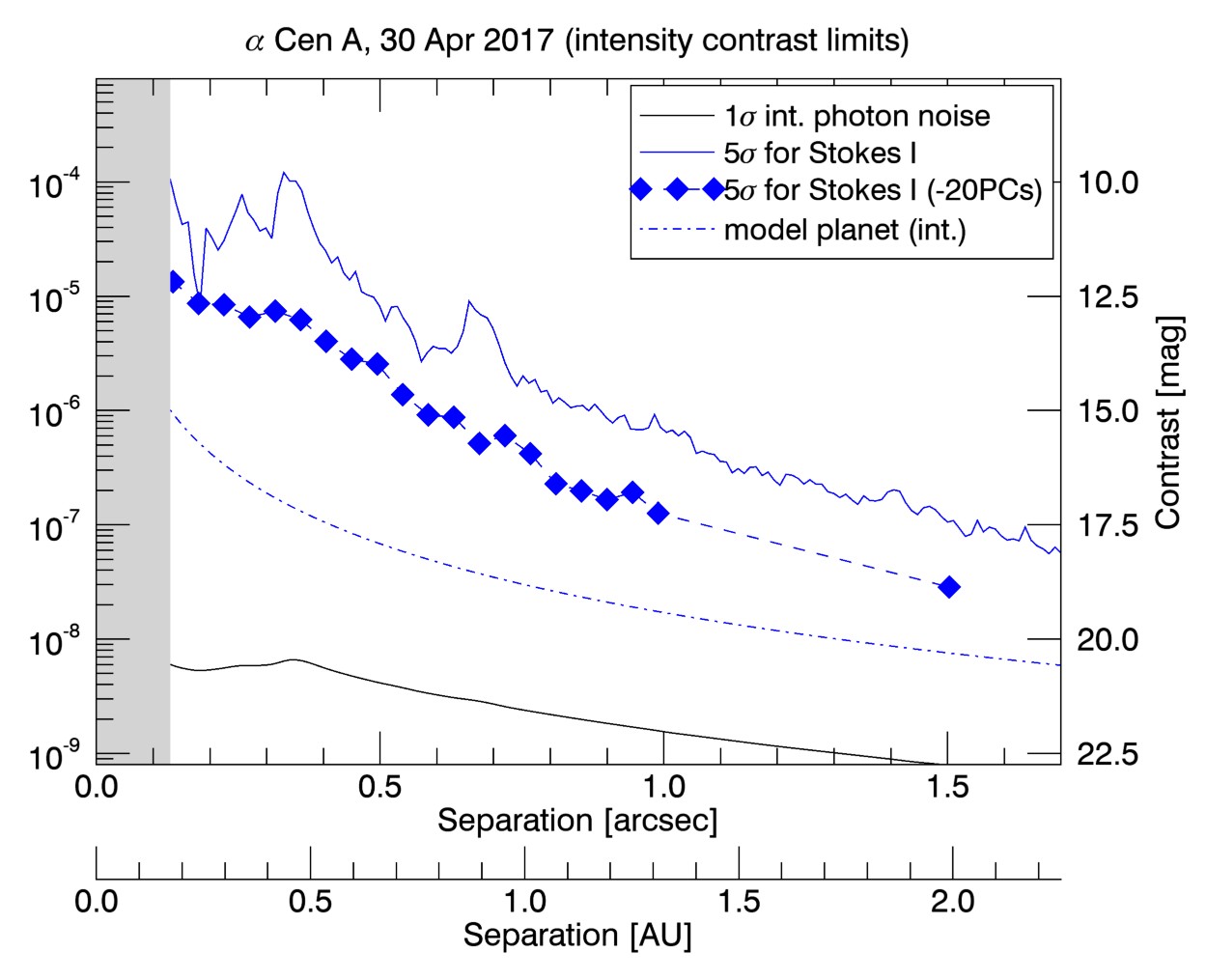} \\
\end{tabular}
\caption{Radial contrast limits as a function of separation for the deepest ZIMPOL high-contrast dataset of $\alpha$~Cen~A in the N\_R filter. (a) The plot shows 1$\sigma$ and 5$\sigma$ limits for the polarized intensity as well as the 1$\sigma$ photon noise limit for the polarized intensity. The basic data reduction (green line) was done without beam shift correction and residual instrument polarization subtraction, the complete reduction includes both steps. The diamond symbols show the improvement of the contrast limits after applying PCA-ADI. (b) The plot shows 5$\sigma$ limits for the intensity as well as the 1$\sigma$ photon noise limit for the intensity. For both contrast limits -- polarized intensity and intensity -- we also show the corresponding contrast of our reference model planet.}
  \label{fig:contrast_methods_ACenA30Apr}
\end{figure*}

Figure~\ref{fig:contrast_methods_ACenA30Apr}(a) shows the 1$\sigma$ and 5$\sigma$ contrast limits for polarized intensity $Q_{\phi}$ together with the 1$\sigma$ photon noise limit. The contrast is limited by speckle noise when the photon noise is lower than the measured 1$\sigma$ point-source contrast, which is the case for separations $\lessapprox$~$0.6\arcsec$, corresponding to $\lessapprox$~$1$~AU for $\alpha$~Cen~A. The solid green line shows the 5$\sigma$ contrast limits after applying the basic data reduction steps without beam shift correction and residual instrument polarization subtraction, the solid red line includes both additional corrections. The symbols show the corresponding contrast improvements after additional PCA speckle subtraction.

The additional corrections -- including PCA-ADI -- improve the contrast limits mostly in the speckle noise dominated region close to the star at separations $\lessapprox$~$0.5 \arcsec$. The contrast can be improved to about 2--5 times the fundamental limit due to photon noise for these separations. For separations $\gtrapprox$~$0.5 \arcsec$ the improvement for the polarized intensity is zero but the limits are already close to the photon noise and ADI could only make it worse. This is why we have chosen to apply ADI in combination with PDI only in an annulus instead of applying it to the whole frame. 

The solid blue line in Fig.~\ref{fig:contrast_methods_ACenA30Apr}(b) is the contrast limit for the total intensity. The corresponding photon noise limit for the intensity is a factor of $\sqrt{2}$ lower than the photon noise limit for the polarization shown in Fig. \ref{fig:contrast_methods_ACenA30Apr}(a) because only 50\% of the photons contribute to the polarized signal $Q_{\phi}$. For the total intensity contrast we also applied PCA-ADI and calculated the resulting contrast limits inside 1\arcsec and at 1.5\arcsec. The results show that speckle noise dominates at all separations. The PCA-ADI procedure can be used to improve the limits but they still exceed the photon noise limit by factors of about 100-1000. However, the detection limits for the total intensity could be further improved with the ZIMPOL pupil stabilized imaging mode without polarimetry. This should produce better contrast limits for the same exposure time.

\subsection{Companion size limit}
\begin{figure}
\resizebox{\hsize}{!}{\includegraphics{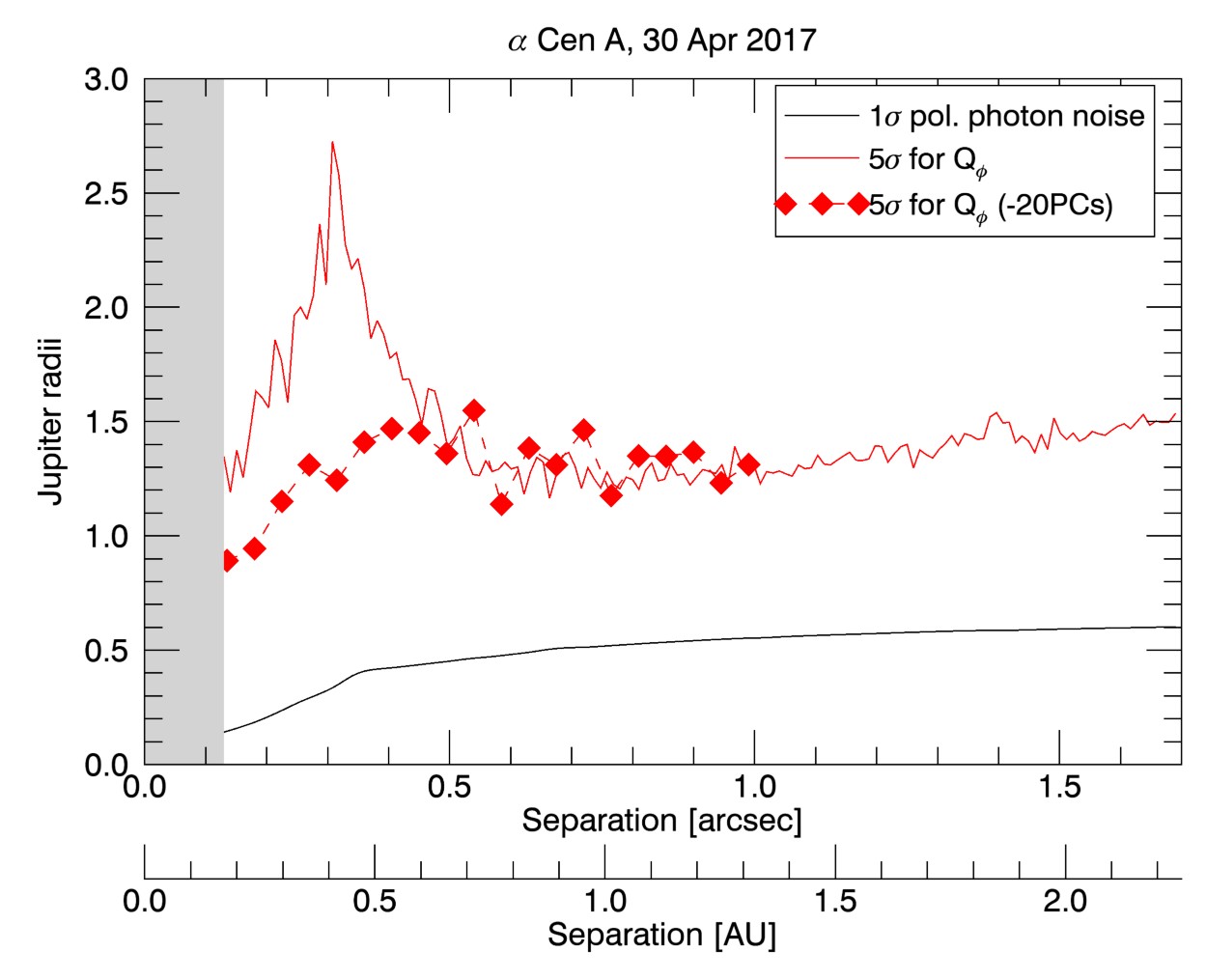}}
\caption{Polarized intensity contrast limits for $\alpha$~Cen~A, turned into the minimum size of a planet that could be observed at each apparent separation. We assume that planets are at maximum apparent separation ($\alpha = 90^{\circ}$) and we adopt our reference model from Sec.~\ref{sec:The signal from extra-solar planets} for the reflective properties of the light.}
\label{fig:how_many_jupiter}
\end{figure}

The detection limits can be turned into size upper limits for a planet with some assumptions about its reflective properties and orbital phase. We adopt again the reference model from Sec.~\ref{sec:The signal from extra-solar planets} with $Q(90^\circ)=p(90^\circ)\cdot I(90^\circ) = 0.055$ and use the contrast curve from Sec.~\ref{sec:Contrast curve} to calculate the upper radius limits for a companion that would still be detectable with polarimetry. The 5$\sigma$ limits shown in Fig.~\ref{fig:how_many_jupiter} result in sizes smaller than 1~$R_{\rm J}$ for small separations $\sim$~$0.2$~AU (0.15\arcsec) and stay between 1--1.5~$R_{\rm J}$ within the whole FOV. The sensitivity improves considerably towards smaller separations (short period planets) because the brightness of the planet scales with $d_P^{-2}$. Companions larger than the calculated limits should be detectable with an average SNR of at least 5. For comparison we also show what size the $1\sigma$ photon noise limit corresponds to. The radius limits in Fig.~\ref{fig:how_many_jupiter} are proportional to $(p(\alpha) \cdot I(\alpha))^{-1/2}$, therefore improving for planets with higher reflectivity and fractional polarization.


\subsection{Contrast gain through longer integration}
\label{sec:Contrast gain through longer integration}
\begin{figure}
\resizebox{\hsize}{!}{\includegraphics{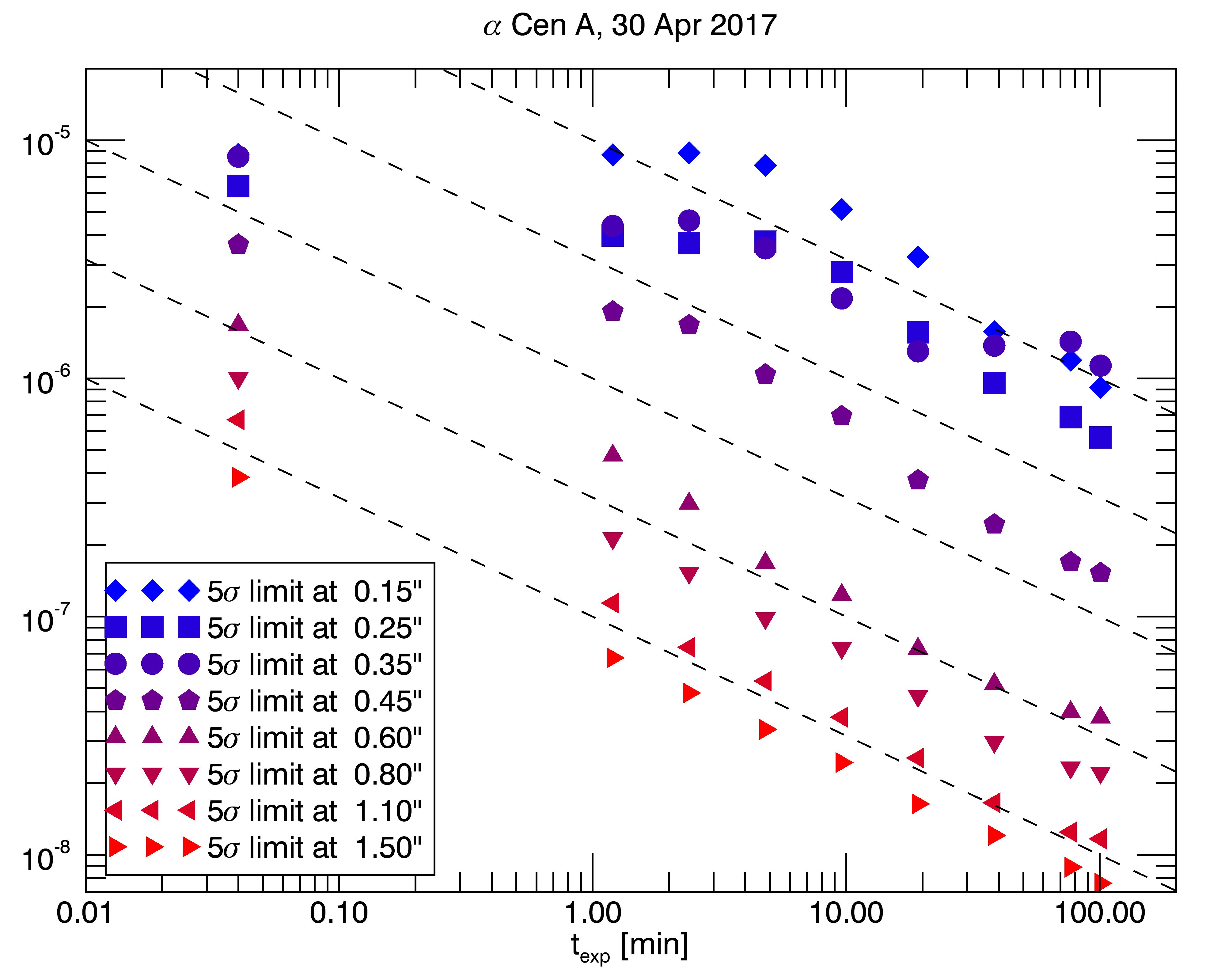}}
\caption{Contrast limits at different $t_{\rm exp}$, calculated for a range of different separations from the star (indicated by different symbols). The dashed lines are proportional to $t_{\rm exp}^{-1/2}$, therefore emphasising the expected behaviour of the noise in the photon noise limited case.}
\label{fig:contrast_develop}
\end{figure}

One simple way of improving the achievable contrast limits is through longer integration $t_{\rm exp}$. Especially if photon noise dominates, the detections limits should be proportional to $t_{exp}^{-1/2}$. At small separations from the star the noise is dominated by the noise residuals that were not eliminated perfectly in the PDI step. This can be seen for example in the bottom frame of Fig.~\ref{fig:ins_pol_corr_exmpl_ACenA30Apr}. Some of the aberrations -- especially the ones to the right and left of the coronagraph caused by the deformable mirror (DM) -- are quasi-static throughout the observation \citep{Cantalloube19}. This changes the statistics of the noise for smaller separations and can ultimately prevent the detection of a point-source signal with a reasonable $t_{\rm exp}$. 

In Fig. \ref{fig:contrast_develop} we show how the polarimetric contrast evolves at different separations if we combine more and more polarimetric cycles. The points at $t_{\rm exp} = 0.04~{\rm min}$ show the contrast in a single zero-phase and $\pi$-phase combined $2 \times 1.2$ second exposure just like the bottom frame of Fig. \ref{fig:ins_pol_corr_exmpl_ACenA30Apr}. All other points show the polarimetric contrast in Stokes Q from one single camera after combining the exposures of multiple polarimetric cycles. PCA-ADI was not applied because the procedure requires a certain amount of field rotation to be effective, and therefore would make it difficult to directly compare the results for different total exposure times.

The data show that the noise for separations $\gtrapprox 0.6 \arcsec$ is proportional to $t_{exp}^{-1/2}$, just as expected in the photon noise dominated regime, all the way from the shortest to the longest $t_{\rm exp}$, totally in agreement with what we see in the corresponding contrast curve (Fig. \ref{fig:contrast_methods_ACenA30Apr}). This indicates for these separations that longer integrations would certainly improve the achievable contrast to a deeper level.

For small separations $\lessapprox 0.6 \arcsec$ and short integration times, the contrast first barely improves with increasing $t_{\rm exp}$. Towards longer integration times, however, it also changes to a $t_{\rm exp}^{-1/2}$ scaling. The transition from a flat curve to a square-root scaling happens later for smaller separations. This can be explained because at small separations the noise is dominated by quasi-static aberrations, therefore angular averaging by the field rotation increases the SNR of a point-source, however the efficiency of this process depends on the separation and the speed of the field rotation. This explanation is supported by Fig.~\ref{fig:contrast_develop} where the transitions for the four separations $\rho = 0.15\arcsec...0.45\arcsec$ happen when the field rotation leads to an azimuthal shift of about $1.5-2.5 \lambda /D$ at the corresponding separation. This corresponds to about the characteristic size of a speckle. As a reference: In the data used for this study, the speed of the field rotation during the relevant time period is $\sim$26~mas/10~min or $1.6~\lambda/D$/10~min at the observed wavelength.

\subsection{Detection limits}
The $\alpha$~Cen A/B system is a close binary with semimajor axis of 23.5~AU, which restricts the range of stable planetary orbits around the individual components. \citet{Wiegert97} and \citet{Quarles16} found that orbits around $\alpha$~Cen A are stable for semimajor axes up to $\sim$3~AU. Stable orbits would preferably be coplanar to the binary orbital plane with inclination $i = 79.2^\circ$ but deviations up to $\pm 45^\circ$ are not unlikely from a stability point of view. There are also reports of other massive planets around one component in close binary systems with a separation smaller than 25~AU (e.g. HD~196885 \citep{Correia07}, Gliese~86 \citep{Queloz00, Lagrange06}, $\gamma$~Cep \citep{Hatzes03, Neuhauser07} and HD~41004~A \citep{Zucker04}).

The radial velocity limits for $\alpha$~Cen~A \citep[e.g.][]{Zhao18} exclude the presence of massive planets with $M~sin(i)~>~53~M_{\textrm{Earth}}$ for the classically defined habitable zone from about 1 to 2~AU with even more stringent limits for smaller separations. This evidence is not in favour of a planet around $\alpha$~Cen~A with a mass larger or comparable to Jupiter but planets up to almost $100~M_{\textrm{Earth}}$ cannot be excluded. Depending on the exact composition, formation history and age, gas giants with masses like that could already be close to Jupiter sized \citep[e.g.][]{Swift12}.

With our radius limits in Fig.~\ref{fig:how_many_jupiter} we show that we might not be far from being able to detect a planet of this size around $\alpha$~Cen A. An important unknown factor in the radius limits are the reflective properties of the planet. For the limits in Fig.~\ref{fig:how_many_jupiter} we assumed a model with $Q(90^\circ)= 0.055$ for the polarized reflectivity of the reflected light. This is optimistic for the reflection of stellar light by the atmosphere of a giant planet, however, some models predict even larger values. The combination of reflection and polarization could also be larger due to other reasons. Calculations from \citet{Arnold04} have shown that a planet with a Saturn-like unresolved ring could have an exceptionally high brightness in reflected light.

It is not unreasonable to assume that $\alpha$~Cen A could harbour a still undetected companion that could be observed with SPHERE/ZIMPOL in reflected visible light. Our best detection limits based on one single half-night show no evidence for a Jupiter sized planet with exceptionally high fraction of polarized reflectivity. However, there is a temporal aspect to the detection limits because of the strong dependence of the reflected light intensity and polarization fraction on the phase angle $\alpha$ (see Fig.~\ref{PhaseCurves2D}), even a Jupiter sized planet with exceptionally high reflection and polarization would be faint for a large range of phase angles. Therefore, only a series of multiple observations could verify the absence of such a planet. Alternatively, one can carry out a detailed combined analysis of the detection limits and possible companion orbits for an estimate on the likelihood of observing a companion. We did an investigation like this for $\alpha$~Cen A and discuss the procedure and the results in Sec.~\ref{sec:Interpreting the contrast limits}.

As far as we know, there has not been a direct imaging search comparable to our study for planetary companions in reflected light around $\alpha$~Cen A. \citet{Kervella06} performed an extensive direct imaging search for faint comoving companions around $\alpha$~Cen~A/B with NACO at the VLT in $J$-,$H$- and $K$-band observations. But their results are difficult to compare to ours because the IWA of their contrast limits is larger than the FOV of ZIMPOL. \citet{Schroeder00} conducted a survey for low mass stellar and sub-stellar companions with the Hubble Space Telescope (HST) for some of the brightest stars closest to the Sun. Their contrast limits for $\alpha$~Cen A in a range of separations 0.5\arcsec-1.5\arcsec are about 7.5-8.5~mag at a wavelength of $\sim$1.02~$\mu$m. Our much deeper contrast limits in intensity are about 13.7-17.4~mag and in polarized intensity 18.3-20.4~mag but with an effective IWA of only $\sim$0.13\arcsec for the R-band.

\section{Discussion}
\begin{table*}
\tiny
\caption{Summary of 5$\sigma$ contrast limits for the intensity $C_{\rm flux}$ and polarized intensity $C_{\rm pol}$ at some key separations for each target}
\label{table: Refplanets contrast}
\centering
\begin{tabular}{l l l l|l|l|l|l|l|l|l|l}
\hline\hline
 Object & m$_R$ & Filters & $t_{\rm exp}$\tablefootmark{a} & \multicolumn{4}{|c|}{$C_{\rm pol}$ (mag)} & \multicolumn{4}{|c}{$C_{\rm flux}$ (mag)} \\
 & & & & \multicolumn{2}{|c|}{Inside AO contr. rad.\tablefootmark{b}} & & & \multicolumn{2}{|c|}{Inside AO contr. rad.\tablefootmark{b}} & & \\
 & & & & $\lessapprox 0.35 \arcsec$ & $\lessapprox 0.45 \arcsec$ & $0.5\arcsec$ & $1.5\arcsec$ & $\lessapprox 0.35 \arcsec$ & $\lessapprox 0.45 \arcsec$ & $0.5\arcsec$ & $1.5\arcsec$ \\
\hline                       
Sirius~A & -1.5 & N\_I & 2h 55.2min & & 15.0 & 15.8 & 18.8 & & 11.0 & 11.8 & 15.7\\
Altair & 0.6 & R\_PRIM & 2h 31.2min & 16.8 & & 17.9 & 20.4 & 12.7 & & 14.6 & 19.5\\
$\epsilon$ Eri & 3.0 & VBB & 3h 12min & 15.8 & & 16.2 & 19.6 & 10.4 & & 11.3 & 16.3\\
$\alpha$ Cen A & -0.5 & N\_R & 3h 21.6min & 16.8 & & 18.3 & 20.4 & 12.8 & & 14.0 & 18.9\\
$\alpha$ Cen B & 1.0 & VBB & 3h 26.8min & 17.1 & & 18.5 & 20.4 & 13.0 & & 14.1 & 18.5\\
$\tau$ Ceti & 2.9 & R\_PRIM & 2h 48min & 15.8 & & 16.7 & 18.8 & 12.4 & & 13.7 & 18.2\\
\hline                        
\end{tabular}
\tablefoot{\tablefoottext{a}{The combined total exposure time} \tablefoottext{b}{Average value for the contrast limit at separations inside the AO control radius ($\lessapprox 20\lambda/D$)}
}
\end{table*}

We have shown the exceptional capability of SPHERE/ZIMPOL polarimetry for the search of reflected light from extra-solar planets on our prime target $\alpha$~Cen~A in Sec.~\ref{sec:Results for alpha Cen A} and our additional targets in Appendix~\ref{Appendix: Results and discussion for the additional targets}. The combination of high resolution and polarimetric sensitivity of our observations is far beyond of any other instrument. For $\alpha$~Cen A, B and Altair we derive polarimetric contrast limits better than 20~mag at separations >1\arcsec. Even at the effective coronagraphic IWA of 0.13\arcsec the polarimetric contrast limits can be around 16~mag. The same performance would also be possible for Sirius~A during better observing conditions. A summary of the resulting 5$\sigma$ contrast limits for all targets can be found in Table~\ref{table: Refplanets contrast}. For the less bright objects $\epsilon$~Eri and $\tau$~Ceti we still see polarimetric contrast limits better than 18.9~mag and 18.2~mag at separations >1\arcsec, respectively, with 16~mag close to the effective IWA for $\tau$~Ceti. Photon noise limited polarimetric contrasts can be achieved already at separations as small as 0.6\arcsec.

\subsection{Comparison to thermal infrared imaging}
\label{sec:Comparison to thermal infrared imaging}
Only a small number of other high-contrast direct imaging searches for planetary companions are published for our targets \citep{Schroeder00, Kervella06, Thalmann11, Vigan15, Mizuki16, Boehle19, Mawet19}. The observations were typically carried out with available near-IR high-contrast imagers and the aim was usually a search for thermal light from brown dwarfs or very massive, self-luminous planets. The detection of such objects around the nearest stars would have been possible, but is quite unexpected. For these near-IR observations, the expected signal for the reflected light from a planet is far out of reach, but the obtained results represent the best limits achieved so far. The most sensitive limits were obtained with a combination of SDI and ADI for Sirius~A with the SPHERE/IRDIFS mode \citep{Vigan15}. Our observation of this object suffered from bad observing conditions, however, for PDI and ADI observations of similarly bright targets, our reported contrast limits show an improvement of 2--3~mag at all separations up to 1.7\arcsec. However, much improved sensitivity is severely needed to detect a planet in reflected light. The only targets where a detection seems to be possible in a single night are $\alpha$~Cen~A and B. The lower brightness of the other targets decreases the sensitivity at a given angular separation and the larger distance to them increases the contrast of companions for the same angular separations.

The physical meaning of the contrast limits for the reflected light is different compared to the limits from IR-surveys for the thermal emission from the planet. The contrast limits in the infrared probe the intrinsic luminosity and surface temperature and can be transformed into upper limits for the planet mass with models for planet formation and evolution \citep[e.g.][]{Baraffe03, Spiegel12} if the age of the system is known and if the irradiation from the star can be neglected. Evolved planets are usually close to or at equilibrium temperature and emit for separations of $\sim$1~AU or larger at longer wavelengths ($\sim$10~$\mu$m) where reaching high-contrast is difficult with current ground-based observations. The intrinsic flux of planets drops off exponentially towards visible wavelengths. For example, assuming perfect black body spectra and a solar-like host star, even a self-luminous 800~K Jupiter-sized planet would only have a contrast of order $3 \cdot 10^{-11}$ in the visual I-band. While the contrast of the reflected light would be around $3 \cdot 10^{-8}$ for $d_p<1$~AU and $I(90^\circ)=0.131$ (for a discussion with wavelength dependent reflectivity see \citet{Sudarsky03}). Planets with d$_p \approx 1$~AU around $\alpha$~Cen~A/B would have to be at temperatures above $\sim$1000~K to be brighter in thermal emission compared to reflected light at visible wavelengths. This is the reason why we can only probe reflected stellar light in the visible wavelengths for all our targets and we do not expect any contribution from thermal emission.

The recently launched NEAR \citep{Kasper17, Kaufl18} survey using the VISIR instrument at the VLT aims to achieve high contrasts at 10~$\mu$m for $\alpha$~Cen~A/B and possibly detect evolved planets in the habitable zone of this binary system. The results of the NEAR campaign will be especially interesting for our survey since it will at least provide exceptionally deep detection limits in the IR that can be directly compared to our own limits for $\alpha$~Cen~A/B in reflected light at visible wavelengths.

\subsection{Interpreting the contrast limits}
\label{sec:Interpreting the contrast limits}
\begin{figure*}
\centering
\includegraphics[width=\linewidth]{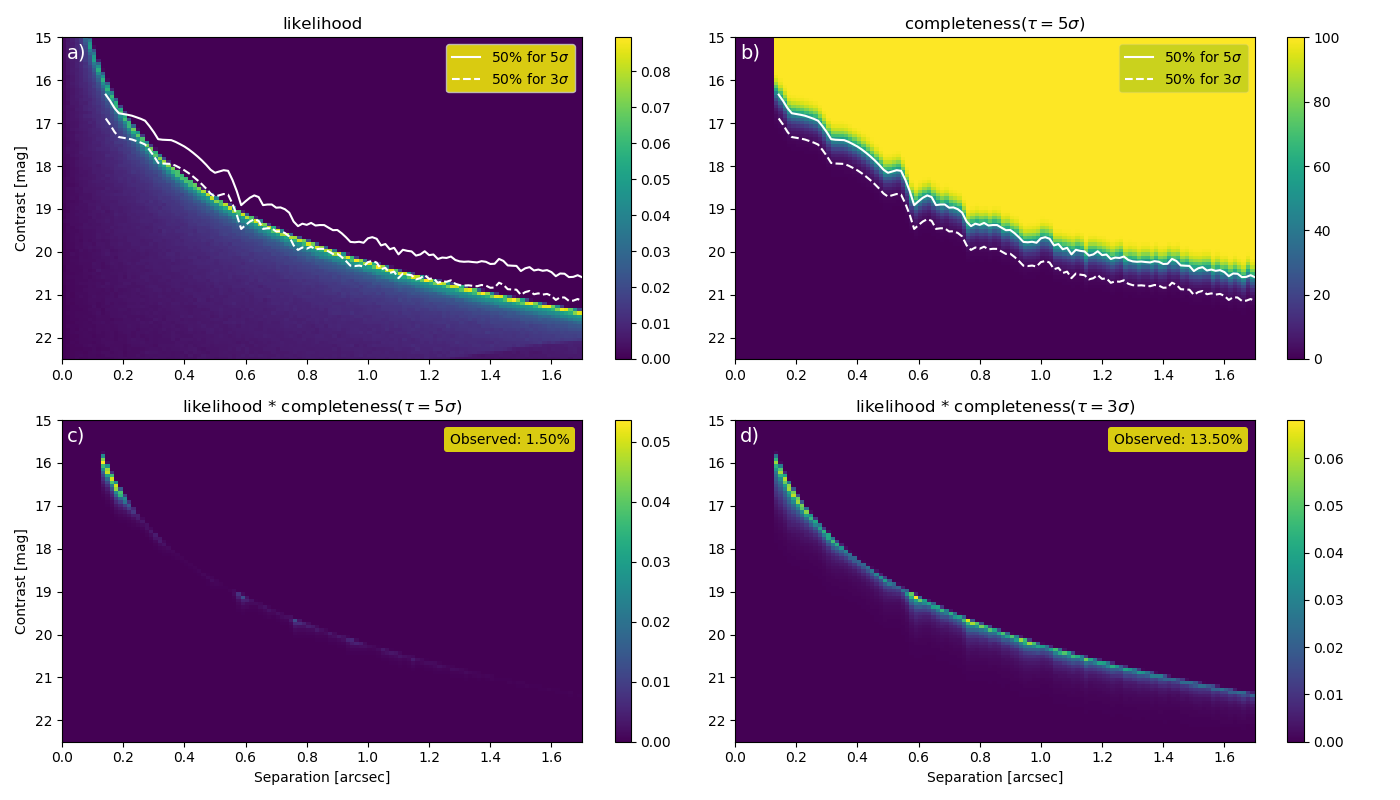}
\caption{The results of the Monte-Carlo sampling of 5\,000\,000 Jupiter sized planets on circular orbits around $\alpha$~Cen~A. (a) The fraction of samples (in percent) that end up at a certain apparent separation and polarized intensity contrast at all times. (b) The full performance map for a certain detection threshold $N\tau$. It shows the probability to detect an existing planet for all separations and contrasts with an SNR of $N$. (c) and (d) The combination of the likelihood and the performance map gives the fraction of planets at each separation and contrast that are detected as either SNR=5 or SNR=3 signals depending on the detection threshold.}
  \label{fig:ACenA30Apr_completeness}
\end{figure*}

In contrast to the thermal emission, the polarized intensity of reflected stellar light depends strongly on the planet radius $R_P$, the planet-star separation $d_p$, the reflective properties of the atmosphere and the phase of the planet (see Eq.~(\ref{equ: pol contrast})). Therefore, contrast limits yield -- for a given physical separation, orbital phase and reflective properties -- an upper limit for the planet radius. This means that the upper limits for the size of a companion as presented for the $\alpha$~Cen~A data come with a set of critical assumptions. The contrast limits are determined for the apparent separation between star and planet $\rho$. The reflected light brightness of the planet, however, depends on the physical separation $d_p$ and planets located at apparent separation $\rho$ can have any physical separation $d_p \geq \rho$. This introduces a degeneracy into the calculation of physical parameters that cannot be lifted without further assumptions. Because of this, we assumed for the radius upper limits in Fig.~\ref{fig:how_many_jupiter} that the physical separation corresponds to the apparent separation, in addition to fixing the scattering model. This assumption can be justified for a blind search for planets with a Monte-Carlo (MC) simulation of apparent separations and contrasts for a random sample of planets. We simulated 5\,000\,000 Jupiter sized planets on circular orbits around $\alpha$~Cen~A with randomly distributed semi-major axes and inclinations and the Rayleigh scattering atmosphere model discussed in Sec.~\ref{sec:The signal from extra-solar planets}. We used a flat prior distribution for the orbital phase angles in the interval [0, 2$\pi$] and for the semi-major axes in the interval [0.01, 3]~AU. The inner boundary for the semi-major axis has a negligible effect on the final result as long as it is smaller than 0.18~AU (the effective IWA of our data). Planets with larger semi-major axes would be unstable due to the close binary. For the inclination we assumed a Gaussian prior with a standard deviation of 45$^\circ$, centred on the inclination of the binary orbit. Large mutual inclinations of binary and planetary orbits are unlikely due to stability reasons \citep{Quarles16}. We chose $\alpha$~Cen~A as our example because it has some of the best detection limits. Panel~(a) in Fig.~\ref{fig:ACenA30Apr_completeness} shows the likelihood of one of the simulated planets having a certain apparent separation and contrast. The likelihood was calculated by dividing the number of MC-samples in each contrast-separation bin by the total number of sampled planets. The likelihood drops to zero towards the upper right corner because planets at large separations have an upper limit for their reflected light intensity determined by their size and reflective properties. The dividing line with the strongly increased likelihood in the center is $\propto \rho^{-2}$, representing planets at maximum elongation, corresponding to orbital phase angles close to 90$^\circ$ and 270$^\circ$. It is more likely for a planet to be located around this line independent from the inclination of its orbit. For orbits close to edge-on the apparent movement of the planet is slower at these phase angles, this naturally increases the likelihood of it being observed during this phase. For orbits closer to face-on the apparent separation of the planet will not change much during the orbit, this also increases the likelihood of the planet being observed during maximum elongation. Around 66\% of all sampled planets end up inside the parameter space shown in Fig. \ref{fig:ACenA30Apr_completeness} and 24\% end up in close proximity ($\Delta {\rm m} \approx 0.4$~mag) to the line with maximum separation and contrast. This is a large fraction considering that we did not assume any prior knowledge about the orbital phase of the sample planets. We compare the likelihood to the completeness or the performance map \citep[see][]{JensenClem18} of our observation in panel~(b) of Fig.~\ref{fig:ACenA30Apr_completeness}, adopting the previously shown contrast curve for $\alpha$~Cen~A (Fig.~\ref{fig:contrast_methods_ACenA30Apr}(b)) and assuming a Gaussian noise distribution. The full performance map in panel~(b) is drawn for a detection threshold $\tau=5\sigma$, additionally we show the 50\% completeness contour for $\tau=3\sigma$. The completeness can be understood as the fraction of true positives given $\tau=3\sigma$ or 5$\sigma$. We multiply the performance map and likelihood in panels (c) and (d) in Fig. \ref{fig:ACenA30Apr_completeness} to show the expected fraction of detectable planets for both detection thresholds and calculate the total integrated fraction of observed planets. Only about 1.5\% of the samples would produce a signal with SNR=5 in our data but the number increases by almost a factor of 10 to about 13.5\% for signals with SNR=3. This happens because the shape of the contrast curve resembles the $\propto \rho^{-2}$ shape of the parameter space where the likelihood is strongly increased. If both curves are on a similar level in terms of contrast, just like in our case with $\alpha$~Cen~A, a small contrast improvement can considerably increase the possibility of a detection. The same happens if we lower the detection threshold but this simultaneously increases the probability for a false detection (false-alarm probability) significantly. For Gaussian distributed noise and a 1024$\times$1024~px$^2$ detector the expected number of random events exceeding >5$\sigma$ is smaller than one, but the number increases to $\sim1000$ for >3$\sigma$. Therefore, the 5$\sigma$ threshold should definitely be respected in a blind search. However, a detection between 3-5$\sigma$ could be enough if there were multiple independent such detections with ZIMPOL that could be combined into one single, more significant detection. 

The MC-simulation shows that the reflected light from a Jupiter sized planet around $\alpha$~Cen~A could be detected as a 5$\sigma$ signal in a single half-night, when it is located relatively close (0.13\arcsec-0.3\arcsec) to the star. It should also be possible to detect Jupiter sized planets at any other separation as 3$\sigma$ signals and multiple 3$\sigma$ detections could be combined to a 5$\sigma$ result. Alternatively, a less significant detection could be considered sufficient if the position of the planet is known from another high-contrast detection or from the astrometric reflex motion of the star.

\subsection{Improving the contrast limits}
There are multiple ways to improve the detection limits with ZIMPOL with future observations. Different strategies are required for blind searches when compared to follow up observations of already known planets. For blind searches the most effective way is to just increase the total integration time. We have shown in Sec.~\ref{sec:Contrast gain through longer integration} that the contrast improves with the square-root of the integration time. The observations should be done in P1 polarimetry mode to enable ADI for improving the contrast at smaller separations. For longer total integration times it will be necessary to combine the data from multiple observing nights. This is not straight forward for our targets because the apparent orbital motion is large. The most extreme case is $\alpha$~Cen~A for which a planet on a face-on circular orbit would move 40~mas or $\sim2\lambda/$D per day at the IWA of 0.13\arcsec and 10~mas or $\sim0.5\lambda/$D at 1.7\arcsec. For the combination of data from different, even consecutive nights it will be necessary to consider the Keplerian motion of planets. This is possible with data analysis tools like K-Stacker \citep{Nowak18}. K-Stacker was developed especially for finding weak planet signals in a time series of images when they move on Keplerian orbits. For a time series spanning weeks it would also be necessary to additionally consider the change of the reflected polarized intensity as function of the orbital phase (Fig.~\ref{PhaseCurves2D}). The orbital motion of planets around nearby stars could also be used as an advantage to further improve the contrast limits. \citet{Males15} developed the concept of Orbital Differential Imaging (ODI) that exploits the orbital motion of a planet in multi-epoch data to remove the stellar PSF, while minimizing the subtraction of the planet signal.

Follow-up observations of a known planet would have major advantages over a blind search because the prior knowledge of orbital phase or orbit location from RV or astrometric measurements can be exploited for optimizing the observing strategy and simplify the analysis of the data. Currently, the best planets for a successful follow up with ZIMPOL are the giant planets $\epsilon$~Eri~b and GJ~876~b and the terrestrial planet Proxima~Centauri~b. The planet around $\epsilon$~Eri can be observed at the favourable photon noise limited apparent separation of $\sim$0.8\arcsec with ZIMPOL. However, it is expected to be rather faint in reflected light because of its large semi-major axis of $\sim$3~AU. The polarimetric contrasts of both Proxima~Centauri~b and GJ~876~b are expected to be less demanding but the expected maximum separation of only $\sim$0.04\arcsec, corresponding to $\sim$2$\lambda$/D in the visible, is very challenging. This requires a specialized instrumental setup for SPHERE/ZIMPOL for example an optimized pupil mask developed to suppress the first Airy ring at 2$\lambda$/D as proposed by \citet{Patapis18}.

For companions with known separation the selection of the ZIMPOL instrument mode can also be optimized. The P1 polarimetry mode should be used for companions close to or inside the AO control ring <0.7\arcsec because it allows the use of ADI for additional speckle noise suppression. ADI also helps to reduce static noise induced by the instrument itself. However, for larger separations ADI is not necessary and the field stabilized P2 polarimetry mode could be used. This would allow to use longer DIT without diluting the planet signal due to the field rotation during exposures.

For planets with well known orbital parameters like semi-major axis, inclination and orbital phase, it would be possible to plan observations to be executed at the right time when the reflected intensity and apparent separation are optimal. And finally, if also the position angle of the orbit is known, it would be possible to align the polarimetric Q-direction of ZIMPOL with the expected orientation of the polarized signal from the planet. This would allow to only observe in a rotated Q polarization coordinate system without spending half of the time observing U, which is expected to be zero. The observation time would be cut in half for the same detection sensitivity or the contrast limit would be improved by a factor of $\sqrt{2}$ in the same amount of telescope time.

There are certainly other ways to improve the detection limits which were not sufficiently investigated yet. The use of narrow-band versus broad-band filters could be beneficial because instrumental effects like beam shift and instrumental polarization are wavelength dependent and the post-processing cannot fully account for this. Therefore, the applied corrections are not optimal for observations taken with broad-band filters and would provide better results for narrow-band filters. Another way to improve the detection limits is frame selection. The gain both of the mentioned techniques is difficult to quantify because we did not find any point-sources in our data. Adding more data, even data of bad quality, generally improved the calculated detection limits because it decreased the noise level of the data. However, data with bad quality also lowers the signal of a point-source but this effect can only be studied properly if a real signal is present in the data because for deep coronagraphic observations we do not know the exact PSF shape for each image.

\section{Conclusion}
We have observed $\alpha$~Cen~A and B, Sirius~A, $\epsilon$~Eri and $\tau$~Cet using SPHERE/ZIMPOL in polarimetry mode. The target list for the search of reflected light from extra-solar planets with direct imaging is short and the targets were selected for achieving deep detection limits within a few hours of observation. We were not able to detect a polarized intensity signal above the detection threshold from any of our targets, however, our data provide some of the deepest contrast limits for direct imaging to date. The achieved limits for our brightest targets show that the detection of polarized reflected light from a 1~R$_{\textrm{J}}$ sized object would be possible in a single night under good observing conditions (Seeing $\lessapprox0.8\arcsec$, $\tau_0 \gtrapprox 4~{\rm ms}$) for our nearest neighbours $\alpha$~Cen~A/B with a realistic model for a reflecting atmosphere. Unfortunately, our null result is not constraining for the occurrence rate of giant planets because of the strong time dependence of the reflected light intensity and given the low frequency of gas giants with 1-10 Jupiter masses between 0.3-3 AU is expected to be only about 4\% \citep{Cumming08, Fernandes19}, slightly higher for A-stars \citep{Johnson10}, but lower for intermediate separation binaries \citep{Kraus16}.

Our results show the capability of ZIMPOL to remove the unpolarized stellar PSF and they deliver the deepest contrast limits for direct imaging at visible wavelengths from 600-900~nm. The performance is close to the photon noise limit and this allows to scale the contrast limits for different total integration times and for targets with different brightnesses. This will be useful in the future for planning further observations in particular for larger programs with deeper observations of the surroundings of the nearest stars by combining the results of many nights. Due to the strong phase dependence the search for reflected light is especially well suited as potential follow up observation of targets with known orbital phases, already determined with different methods (e.g. RV, astrometry). Another use of the highly sensitive polarimetry with ZIMPOL could be the determination of the linear polarization of the thermal light of low mass companions. This measurement has been tried before for a few different targets at infrared wavelengths \citep[e.g.][]{JensenClem16, vanHolstein17}. The main difficulty with brown dwarf companions is that the linear polarization degree for the thermal light is expected to be <1\% \citep{Stolker17}. Another problem for ZIMPOL polarimetry is the low luminosity of L and T dwarfs in the visible wavelengths.

Another important aspect of this work are our investigations on the limitations of SPHERE/ZIMPOL at the VLT. We have investigated and corrected the residual instrument polarization and most importantly the polarimetric beam shift effect. The beam shift effect is well known in optics but ZIMPOL is the first astronomical instrument where this effect is apparent in the data because of its high spatial resolution and polarimetric sensitivity. Despite all the calibrations and corrections applied in this work, there remain substantial speckle residuals in the differential polarimetry which currently limit the contrast performance at small separations (<0.6\arcsec). However, the achieved polarimetric contrast $C_{\rm pol}$ is in most cases more than 10 times deeper for separations <1\arcsec and up to $\sim$50 times deeper at the smallest separations compared to the imaging contrast $C_{\rm flux}$ achieved with classical PCA-ADI processing. Over a larger range of separations, deep polarimetric contrast limits and even photon noise limited performance is achieved without additional PCA-ADI. Therefore, polarimetry is a most attractive method to push the detection limits for reflecting planets with future high-contrast instruments \citep{Kasper10}. In particular, the speckle suppression can be further improved by better avoiding color dependent disturbing effects (see Appendix~\ref{sec:Advanced data reduction steps}) or by taking advantage of the much improved light gathering power of the upcoming generation of 30-40 meter ELTs. With larger telescopes it will be possible to achieve the same contrast limits for fainter stars at higher resolution, significantly increasing the sample size of nearby targets and possibly allow imaging the planets in the habitable zone of nearby M dwarfs like Proxima Centauri. The experience gained with the SPHERE/ZIMPOL RefPlanets survey described in this work should therefore be helpful for the trade-off studies and description of the design of such instruments.

\begin{acknowledgements}
SH and HMS acknowledge the financial support by the Swiss National Science Foundation through grant 200020\_162630/1. SPHERE is an instrument designed and built by a consortium consisting of IPAG (Grenoble, France), MPIA (Heidelberg, Germany), LAM (Marseille, France), LESIA (Paris, France), Laboratoire Lagrange (Nice, France), INAF–Osservatorio di Padova (Italy), Observatoire de Genève (Switzerland), ETH Zurich (Switzerland), NOVA (Netherlands), ONERA (France) and ASTRON (Netherlands) in collaboration with ESO. SPHERE was funded by ESO, with additional contributions from CNRS (France), MPIA (Germany), INAF (Italy), FINES (Switzerland) and NOVA (Netherlands).  SPHERE also received funding from the European Commission Sixth and Seventh Framework Programmes as part of the Optical Infrared Coordination Network for Astronomy (OPTICON) under grant number RII3-Ct-2004-001566 for FP6 (2004–2008), grant number 226604 for FP7 (2009–2012) and grant number 312430 for FP7 (2013–2016). We also acknowledge financial support from the Programme National de Planétologie (PNP) and the Programme National de Physique Stellaire (PNPS) of CNRS-INSU in France. This work has also been supported by a grant from the French Labex OSUG@2020 (Investissements d’avenir – ANR10 LABX56). The project is supported by CNRS, by the Agence Nationale de la Recherche (ANR-14-CE33-0018). It has also been carried out within the frame of the National Centre for Competence in Research PlanetS supported by the Swiss National Science Foundation (SNSF). MRM and SPQ are pleased to acknowledge this financial support of the SNSF. Finally, this work has made use of the the SPHERE Data Centre, jointly operated by OSUG/IPAG (Grenoble), PYTHEAS/LAM/CESAM (Marseille), OCA/Lagrange (Nice), Observatoire de Paris/LESIA (Paris), and Observatoire de Lyon, also supported by a grant from Labex  OSUG@2020 (Investissements d’avenir – ANR10 LABX56). We thank P. Delorme and E. Lagadec (SPHERE Data Centre) for their efficient help during the data reduction process. This research has made use of the SIMBAD database, operated at CDS, Strasbourg, France. This work has been partially supported by the project PRIN-INAF 2016 The Cradle of Life- GENESIS-SKA (General Conditions in Early Planetary Systems for the rise of life with SKA). We also acknowledge support from INAF/Frontiera (Fostering high ResolutiON Technology and Innovation for Exoplanets and Research in Astrophysics) through the "Progetti Premiali" funding scheme of the Italian Ministry of Education, University, and Research. T.H. acknowledges support from the European Research Council under the Horizon 2020 Framework Program via the ERC Advanced Grant Origins 83 24 28.
\end{acknowledgements}

\bibliographystyle{aa} 
\bibliography{paper_ZIMPOL_contrast_v00} 

\begin{thebibliography}{95}
\expandafter\ifx\csname natexlab\endcsname\relax\def\natexlab#1{#1}\fi

\bibitem[{{Amara} \& {Quanz}(2012)}]{Amara12}
{Amara}, A. \& {Quanz}, S.~P. 2012, \mnras, 427, 948

\bibitem[{{Arnold} \& {Schneider}(2004)}]{Arnold04}
{Arnold}, L. \& {Schneider}, J. 2004, \aap, 420, 1153

\bibitem[{{Backman} {et~al.}(2009){Backman}, {Marengo}, {Stapelfeldt}, {Su},
  {Wilner}, {Dowell}, {Watson}, {Stansberry}, {Rieke}, {Megeath}, {Fazio}, \&
  {Werner}}]{Backman09}
{Backman}, D., {Marengo}, M., {Stapelfeldt}, K., {et~al.} 2009, \apj, 690, 1522

\bibitem[{{Bailey} {et~al.}(2018){Bailey}, {Kedziora-Chudczer}, \&
  {Bott}}]{Bailey18}
{Bailey}, J., {Kedziora-Chudczer}, L., \& {Bott}, K. 2018, \mnras, 480, 1613

\bibitem[{{Bailey} {et~al.}(2010){Bailey}, {Lucas}, \& {Hough}}]{Bailey10}
{Bailey}, J., {Lucas}, P.~W., \& {Hough}, J.~H. 2010, \mnras, 405, 2570

\bibitem[{{Baraffe} {et~al.}(2003){Baraffe}, {Chabrier}, {Barman}, {Allard}, \&
  {Hauschildt}}]{Baraffe03}
{Baraffe}, I., {Chabrier}, G., {Barman}, T.~S., {Allard}, F., \& {Hauschildt},
  P.~H. 2003, \aap, 402, 701

\bibitem[{{Bazzon} {et~al.}(2012){Bazzon}, {Gisler}, {Roelfsema}, {Schmid},
  {Pragt}, {Elswijk}, {de Haan}, {Downing}, {Salasnich}, {Pavlov}, {Beuzit},
  {Dohlen}, {Mouillet}, \& {Wildi}}]{Bazzon12}
{Bazzon}, A., {Gisler}, D., {Roelfsema}, R., {et~al.} 2012, in \procspie, Vol.
  8446, Ground-based and Airborne Instrumentation for Astronomy IV, 844693

\bibitem[{{Bazzon} {et~al.}(2014){Bazzon}, {Schmid}, \& {Buenzli}}]{Bazzon14}
{Bazzon}, A., {Schmid}, H.~M., \& {Buenzli}, E. 2014, \aap, 572, A6

\bibitem[{{Bazzon} {et~al.}(2013){Bazzon}, {Schmid}, \& {Gisler}}]{Bazzon13}
{Bazzon}, A., {Schmid}, H.~M., \& {Gisler}, D. 2013, \aap, 556, A117

\bibitem[{{Beuzit} {et~al.}(2008){Beuzit}, {Feldt}, {Dohlen}, {Mouillet},
  {Puget}, {Wildi}, {Abe}, {Antichi}, {Baruffolo}, {Baudoz}, {Boccaletti},
  {Carbillet}, {Charton}, {Claudi}, {Downing}, {Fabron}, {Feautrier},
  {Fedrigo}, {Fusco}, {Gach}, {Gratton}, {Henning}, {Hubin}, {Joos}, {Kasper},
  {Langlois}, {Lenzen}, {Moutou}, {Pavlov}, {Petit}, {Pragt}, {Rabou}, {Rigal},
  {Roelfsema}, {Rousset}, {Saisse}, {Schmid}, {Stadler}, {Thalmann}, {Turatto},
  {Udry}, {Vakili}, \& {Waters}}]{Beuzit08}
{Beuzit}, J.-L., {Feldt}, M., {Dohlen}, K., {et~al.} 2008, in Society of
  Photo-Optical Instrumentation Engineers (SPIE) Conference Series, Vol. 7014,
  Society of Photo-Optical Instrumentation Engineers (SPIE) Conference Series,
  18

\bibitem[{{Beuzit} {et~al.}(2019){Beuzit}, {Vigan}, {Mouillet}, {Dohlen},
  {Gratton}, {Boccaletti}, {Sauvage}, {Schmid}, {Langlois}, {Petit},
  {Baruffolo}, {Feldt}, {Milli}, {Wahhaj}, {Abe}, {Anselmi}, {Antichi},
  {Barette}, {Baudrand}, {Baudoz}, {Bazzon}, {Bernardi}, {Blanchard}, {Brast},
  {Bruno}, {Buey}, {Carbillet}, {Carle}, {Cascone}, {Chapron}, {Chauvin},
  {Charton}, {Claudi}, {Costille}, {De Caprio}, {Delboulb{\'e}}, {Desidera},
  {Dominik}, {Downing}, {Dupuis}, {Fabron}, {Fantinel}, {Farisato},
  {Feautrier}, {Fedrigo}, {Fusco}, {Gigan}, {Ginski}, {Girard}, {Giro},
  {Gisler}, {Gluck}, {Gry}, {Henning}, {Hubin}, {Hugot}, {Incorvaia}, {Jaquet},
  {Kasper}, {Lagadec}, {Lagrange}, {Le Coroller}, {Le Mignant}, {Le Ruyet},
  {Lessio}, {Lizon}, {Llored}, {Lundin}, {Madec}, {Magnard}, {Marteaud},
  {Martinez}, {Maurel}, {M{\'e}nard}, {Mesa}, {M{\"o}ller-Nilsson}, {Moulin},
  {Moutou}, {Orign{\'e}}, {Parisot}, {Pavlov}, {Perret}, {Pragt}, {Puget},
  {Rabou}, {Ramos}, {Reess}, {Rigal}, {Rochat}, {Roelfsema}, {Rousset}, {Roux},
  {Saisse}, {Salasnich}, {Santambrogio}, {Scuderi}, {Segransan}, {Sevin},
  {Siebenmorgen}, {Soenke}, {Stadler}, {Suarez}, {Tiph{\`e}ne}, {Turatto},
  {Udry}, {Vakili}, {Waters}, {Weber}, {Wildi}, {Zins}, \& {Zurlo}}]{Beuzit19}
{Beuzit}, J.~L., {Vigan}, A., {Mouillet}, D., {et~al.} 2019, arXiv e-prints,
  arXiv:1902.04080

\bibitem[{{Boehle} {et~al.}(2019){Boehle}, {Quanz}, {Lovis}, {S{\`e}gransan},
  {Udry}, \& {Apai}}]{Boehle19}
{Boehle}, A., {Quanz}, S.~P., {Lovis}, C., {et~al.} 2019, arXiv e-prints,
  arXiv:1907.04334

\bibitem[{{Bond} {et~al.}(2017){Bond}, {Schaefer}, {Gilliland}, {Holberg},
  {Mason}, {Lindenblad}, {Seitz-McLeese}, {Arnett}, {Demarque}, {Spada},
  {Young}, {Barstow}, {Burleigh}, \& {Gudehus}}]{Bond17}
{Bond}, H.~E., {Schaefer}, G.~H., {Gilliland}, R.~L., {et~al.} 2017, \apj, 840,
  70

\bibitem[{{Booth} {et~al.}(2017){Booth}, {Dent}, {Jord{\'a}n}, {Lestrade},
  {Hales}, {Wyatt}, {Casassus}, {Ertel}, {Greaves}, {Kennedy}, {Matr{\`a}},
  {Augereau}, \& {Villard}}]{Booth17}
{Booth}, M., {Dent}, W.~R.~F., {Jord{\'a}n}, A., {et~al.} 2017, \mnras, 469,
  3200

\bibitem[{{Bottom} {et~al.}(2017){Bottom}, {Ruane}, \& {Mawet}}]{Bottom17}
{Bottom}, M., {Ruane}, G., \& {Mawet}, D. 2017, Research Notes of the American
  Astronomical Society, 1, 30

\bibitem[{{Bowler}(2016)}]{Bowler16}
{Bowler}, B.~P. 2016, \pasp, 128, 102001

\bibitem[{{Buenzli} \& {Schmid}(2009)}]{Buenzli09}
{Buenzli}, E. \& {Schmid}, H.~M. 2009, \aap, 504, 259

\bibitem[{{Cantalloube} {et~al.}(2019){Cantalloube}, {Dohlen}, {Milli},
  {Brandner}, \& {Vigan}}]{Cantalloube19}
{Cantalloube}, F., {Dohlen}, K., {Milli}, J., {Brandner}, W., \& {Vigan}, A.
  2019, The Messenger, 176, 25

\bibitem[{{Chauvin} {et~al.}(2017){Chauvin}, {Desidera}, {Lagrange}, {Vigan},
  {Gratton}, {Langlois}, {Bonnefoy}, {Beuzit}, {Feldt}, {Mouillet}, {Meyer},
  {Cheetham}, {Biller}, {Boccaletti}, {D'Orazi}, {Galicher}, {Hagelberg},
  {Maire}, {Mesa}, {Olofsson}, {Samland}, {Schmidt}, {Sissa}, {Bonavita},
  {Charnay}, {Cudel}, {Daemgen}, {Delorme}, {Janin-Potiron}, {Janson},
  {Keppler}, {Le Coroller}, {Ligi}, {Marleau}, {Messina}, {Molli{\`e}re},
  {Mordasini}, {M{\"u}ller}, {Peretti}, {Perrot}, {Rodet}, {Rouan}, {Zurlo},
  {Dominik}, {Henning}, {Menard}, {Schmid}, {Turatto}, {Udry}, {Vakili}, {Abe},
  {Antichi}, {Baruffolo}, {Baudoz}, {Baudrand}, {Blanchard}, {Bazzon}, {Buey},
  {Carbillet}, {Carle}, {Charton}, {Cascone}, {Claudi}, {Costille}, {Deboulbe},
  {De Caprio}, {Dohlen}, {Fantinel}, {Feautrier}, {Fusco}, {Gigan}, {Giro},
  {Gisler}, {Gluck}, {Hubin}, {Hugot}, {Jaquet}, {Kasper}, {Madec}, {Magnard},
  {Martinez}, {Maurel}, {Le Mignant}, {M{\"o}ller-Nilsson}, {Llored}, {Moulin},
  {Orign{\'e}}, {Pavlov}, {Perret}, {Petit}, {Pragt}, {Puget}, {Rabou},
  {Ramos}, {Rigal}, {Rochat}, {Roelfsema}, {Rousset}, {Roux}, {Salasnich},
  {Sauvage}, {Sevin}, {Soenke}, {Stadler}, {Suarez}, {Weber}, {Wildi},
  {Antoniucci}, {Augereau}, {Baudino}, {Brandner}, {Engler}, {Girard}, {Gry},
  {Kral}, {Kopytova}, {Lagadec}, {Milli}, {Moutou}, {Schlieder},
  {Szul{\'a}gyi}, {Thalmann}, \& {Wahhaj}}]{Chauvin17}
{Chauvin}, G., {Desidera}, S., {Lagrange}, A.~M., {et~al.} 2017, \aap, 605, L9

\bibitem[{{Claudi} {et~al.}(2008){Claudi}, {Turatto}, {Gratton}, {Antichi},
  {Bonavita}, {Bruno}, {Cascone}, {De Caprio}, {Desidera}, {Giro}, {Mesa},
  {Scuderi}, {Dohlen}, {Beuzit}, \& {Puget}}]{Claudi08}
{Claudi}, R.~U., {Turatto}, M., {Gratton}, R.~G., {et~al.} 2008, in Society of
  Photo-Optical Instrumentation Engineers (SPIE) Conference Series, Vol. 7014,
  Ground-based and Airborne Instrumentation for Astronomy II, 70143E

\bibitem[{{Correia} {et~al.}(2007){Correia}, {Udry}, {Mayor}, {Eggenberger},
  {Naef}, {Beuzit}, {Perrier}, {Queloz}, {Sivan}, {Pepe}, {Santos}, \&
  {S{\'e}gransan}}]{Correia07}
{Correia}, A.~C.~M., {Udry}, S., {Mayor}, M., {et~al.} 2007, arXiv e-prints
  [\eprint[arXiv]{0711.3343}]

\bibitem[{{Cotton} {et~al.}(2016){Cotton}, {Bailey}, {Kedziora-Chudczer},
  {Bott}, {Lucas}, {Hough}, \& {Marshall}}]{Cotton16}
{Cotton}, D.~V., {Bailey}, J., {Kedziora-Chudczer}, L., {et~al.} 2016, \mnras,
  455, 1607

\bibitem[{{Cumming} {et~al.}(2008){Cumming}, {Butler}, {Marcy}, {Vogt},
  {Wright}, \& {Fischer}}]{Cumming08}
{Cumming}, A., {Butler}, R.~P., {Marcy}, G.~W., {et~al.} 2008, \pasp, 120, 531

\bibitem[{{Dohlen} {et~al.}(2008){Dohlen}, {Langlois}, {Saisse}, {Hill},
  {Origne}, {Jacquet}, {Fabron}, {Blanc}, {Llored}, {Carle}, {Moutou}, {Vigan},
  {Boccaletti}, {Carbillet}, {Mouillet}, \& {Beuzit}}]{Dohlen08}
{Dohlen}, K., {Langlois}, M., {Saisse}, M., {et~al.} 2008, in Society of
  Photo-Optical Instrumentation Engineers (SPIE) Conference Series, Vol. 7014,
  Ground-based and Airborne Instrumentation for Astronomy II, 70143L

\bibitem[{{Dollfus}(1985)}]{Dollfus85}
{Dollfus}, A. 1985, Advances in Space Research, 5, 47

\bibitem[{{Feng} {et~al.}(2017){Feng}, {Tuomi}, {Jones}, {Barnes},
  {Anglada-Escud{\'e}}, {Vogt}, \& {Butler}}]{Feng17}
{Feng}, F., {Tuomi}, M., {Jones}, H.~R.~A., {et~al.} 2017, \aj, 154, 135

\bibitem[{{Fernandes} {et~al.}(2019){Fernandes}, {Mulders}, {Pascucci},
  {Mordasini}, \& {Emsenhuber}}]{Fernandes19}
{Fernandes}, R.~B., {Mulders}, G.~D., {Pascucci}, I., {Mordasini}, C., \&
  {Emsenhuber}, A. 2019, \apj, 874, 81

\bibitem[{{Fusco} {et~al.}(2006){Fusco}, {Petit}, {Rousset}, {Sauvage},
  {Dohlen}, {Mouillet}, {Charton}, {Baudoz}, {Kasper}, {Fedrigo}, {Rabou},
  {Feautrier}, {Downing}, {Gigan}, {Conan}, {Beuzit}, {Hubin}, {Wildi}, \&
  {Puget}}]{Fusco06}
{Fusco}, T., {Petit}, C., {Rousset}, G., {et~al.} 2006, in Society of
  Photo-Optical Instrumentation Engineers (SPIE) Conference Series, Vol. 6272,
  62720K

\bibitem[{{Gisler} {et~al.}(2004){Gisler}, {Schmid}, {Thalmann}, {Povel},
  {Stenflo}, {Joos}, {Feldt}, {Lenzen}, {Tinbergen}, {Gratton}, {Stuik},
  {Stam}, {Brand ner}, {Hippler}, {Turatto}, {Neuhauser}, {Dominik}, {Hatzes},
  {Henning}, {Lima}, {Quirrenbach}, {Waters}, {Wuchterl}, \&
  {Zinnecker}}]{Gisler04}
{Gisler}, D., {Schmid}, H.~M., {Thalmann}, C., {et~al.} 2004, in Society of
  Photo-Optical Instrumentation Engineers (SPIE) Conference Series, Vol. 5492,
  Ground-based Instrumentation for Astronomy, ed. A.~F.~M. {Moorwood} \&
  M.~{Iye}, 463--474

\bibitem[{{Greaves} {et~al.}(2014){Greaves}, {Sibthorpe}, {Acke}, {Pantin},
  {Vandenbussche}, {Olofsson}, {Dominik}, {Barlow}, {Bendo}, {Blommaert},
  {Brandeker}, {de Vries}, {Dent}, {Di Francesco}, {Fridlund}, {Gear},
  {Harvey}, {Hogerheijde}, {Holland}, {Ivison}, {Liseau}, {Matthews},
  {Pilbratt}, {Walker}, \& {Waelkens}}]{Greaves14}
{Greaves}, J.~S., {Sibthorpe}, B., {Acke}, B., {et~al.} 2014, \apjl, 791, L11

\bibitem[{{Greaves} {et~al.}(2004){Greaves}, {Wyatt}, {Holland}, \&
  {Dent}}]{Greaves04}
{Greaves}, J.~S., {Wyatt}, M.~C., {Holland}, W.~S., \& {Dent}, W.~R.~F. 2004,
  \mnras, 351, L54

\bibitem[{{Hansen} \& {Hovenier}(1974)}]{Hansen74}
{Hansen}, J.~E. \& {Hovenier}, J.~W. 1974, Journal of Atmospheric Sciences, 31,
  1137

\bibitem[{{Hatzes} {et~al.}(2003){Hatzes}, {Cochran}, {Endl}, {McArthur},
  {Paulson}, {Walker}, {Campbell}, \& {Yang}}]{Hatzes03}
{Hatzes}, A.~P., {Cochran}, W.~D., {Endl}, M., {et~al.} 2003, \apj, 599, 1383

\bibitem[{{Hatzes} {et~al.}(2000){Hatzes}, {Cochran}, {McArthur}, {Baliunas},
  {Walker}, {Campbell}, {Irwin}, {Yang}, {K{\"u}rster}, {Endl}, {Els},
  {Butler}, \& {Marcy}}]{Hatzes00}
{Hatzes}, A.~P., {Cochran}, W.~D., {McArthur}, B., {et~al.} 2000, \apjl, 544,
  L145

\bibitem[{{Holman} \& {Wiegert}(1999)}]{Holman99}
{Holman}, M.~J. \& {Wiegert}, P.~A. 1999, \aj, 117, 621

\bibitem[{{Jensen-Clem} {et~al.}(2018){Jensen-Clem}, {Mawet}, {Gomez Gonzalez},
  {Absil}, {Belikov}, {Currie}, {Kenworthy}, {Marois}, {Mazoyer}, {Ruane},
  {Tanner}, \& {Cantalloube}}]{JensenClem18}
{Jensen-Clem}, R., {Mawet}, D., {Gomez Gonzalez}, C.~A., {et~al.} 2018, \aj,
  155, 19

\bibitem[{{Jensen-Clem} {et~al.}(2016){Jensen-Clem}, {Millar-Blanchaer},
  {Mawet}, {Graham}, {Wallace}, {Macintosh}, {Hinkley}, {Wiktorowicz},
  {Perrin}, {Marley}, {Fitzgerald}, {Oppenheimer}, {Ammons}, {Rantakyr{\"o}},
  \& {Marchis}}]{JensenClem16}
{Jensen-Clem}, R., {Millar-Blanchaer}, M., {Mawet}, D., {et~al.} 2016, \apj,
  820, 111

\bibitem[{{Johnson} {et~al.}(2010){Johnson}, {Aller}, {Howard}, \&
  {Crepp}}]{Johnson10}
{Johnson}, J.~A., {Aller}, K.~M., {Howard}, A.~W., \& {Crepp}, J.~R. 2010,
  \pasp, 122, 905

\bibitem[{{Jovanovic} {et~al.}(2015){Jovanovic}, {Martinache}, {Guyon},
  {Clergeon}, {Singh}, {Kudo}, {Garrel}, {Newman}, {Doughty}, {Lozi}, {Males},
  {Minowa}, {Hayano}, {Takato}, {Morino}, {Kuhn}, {Serabyn}, {Norris},
  {Tuthill}, {Schworer}, {Stewart}, {Close}, {Huby}, {Perrin}, {Lacour},
  {Gauchet}, {Vievard}, {Murakami}, {Oshiyama}, {Baba}, {Matsuo}, {Nishikawa},
  {Tamura}, {Lai}, {Marchis}, {Duchene}, {Kotani}, \& {Woillez}}]{Jovanovic15}
{Jovanovic}, N., {Martinache}, F., {Guyon}, O., {et~al.} 2015, \pasp, 127, 890

\bibitem[{{Kasper} {et~al.}(2017){Kasper}, {Arsenault}, {K{\"a}ufl}, {Jakob},
  {Fuenteseca}, {Riquelme}, {Siebenmorgen}, {Sterzik}, {Zins}, {Ageorges},
  {Gutruf}, {Reutlinger}, {Kampf}, {Absil}, {Carlomagno}, {Guyon}, {Klupar},
  {Mawet}, {Ruane}, {Karlsson}, {Pantin}, \& {Dohlen}}]{Kasper17}
{Kasper}, M., {Arsenault}, R., {K{\"a}ufl}, H.~U., {et~al.} 2017, The
  Messenger, 169, 16

\bibitem[{{Kasper} {et~al.}(2010){Kasper}, {Beuzit}, {Verinaud}, {Gratton},
  {Kerber}, {Yaitskova}, {Boccaletti}, {Thatte}, {Schmid}, {Keller}, {Baudoz},
  {Abe}, {Aller-Carpentier}, {Antichi}, {Bonavita}, {Dohlen}, {Fedrigo},
  {Hanenburg}, {Hubin}, {Jager}, {Korkiakoski}, {Martinez}, {Mesa}, {Preis},
  {Rabou}, {Roelfsema}, {Salter}, {Tecza}, \& {Venema}}]{Kasper10}
{Kasper}, M., {Beuzit}, J.-L., {Verinaud}, C., {et~al.} 2010, in \procspie,
  Vol. 7735, Ground-based and Airborne Instrumentation for Astronomy III,
  77352E--77352E--9

\bibitem[{{K{\"a}ufl} {et~al.}(2018){K{\"a}ufl}, {Kasper}, {Arsenault},
  {Jakob}, {Leveratto}, {Zins}, {Fuenteseca}, {Riquelme}, {Siebenmorgen},
  {Sterzik}, {Ageorges}, {Gutruf}, {Kampf}, {Reutlinger}, {Absil},
  {Carlomagno}, {Guyon}, {Klupar}, {Mawet}, {Ruane}, {Karlsson}, {Pantin}, \&
  {Dohlen}}]{Kaufl18}
{K{\"a}ufl}, H.-U., {Kasper}, M., {Arsenault}, R., {et~al.} 2018, in Society of
  Photo-Optical Instrumentation Engineers (SPIE) Conference Series, Vol. 10702,
  \procspie, 107020D

\bibitem[{{Keller} {et~al.}(2010){Keller}, {Schmid}, {Venema}, {Hanenburg},
  {Jager}, {Kasper}, {Martinez}, {Rigal}, {Rodenhuis}, {Roelfsema}, {Snik},
  {Verinaud}, \& {Yaitskova}}]{Keller10}
{Keller}, C.~U., {Schmid}, H.~M., {Venema}, L.~B., {et~al.} 2010, in \procspie,
  Vol. 7735, Ground-based and Airborne Instrumentation for Astronomy III,
  77356G

\bibitem[{{Kemp} {et~al.}(1987){Kemp}, {Henson}, {Steiner}, \&
  {Powell}}]{Kemp87}
{Kemp}, J.~C., {Henson}, G.~D., {Steiner}, C.~T., \& {Powell}, E.~R. 1987,
  \nat, 326, 270

\bibitem[{{Keppler} {et~al.}(2018){Keppler}, {Benisty}, {M{\"u}ller},
  {Henning}, {van Boekel}, {Cantalloube}, {Ginski}, {van Holstein}, {Maire},
  {Pohl}, {Samland }, {Avenhaus}, {Baudino}, {Boccaletti}, {de Boer},
  {Bonnefoy}, {Chauvin}, {Desidera}, {Langlois}, {Lazzoni}, {Marleau},
  {Mordasini}, {Pawellek}, {Stolker}, {Vigan}, {Zurlo}, {Birnstiel},
  {Brandner}, {Feldt}, {Flock}, {Girard}, {Gratton}, {Hagelberg}, {Isella},
  {Janson}, {Juhasz}, {Kemmer}, {Kral}, {Lagrange}, {Launhardt}, {Matter},
  {M{\'e}nard}, {Milli}, {Molli{\`e}re}, {Olofsson}, {P{\'e}rez}, {Pinilla},
  {Pinte}, {Quanz}, {Schmidt}, {Udry}, {Wahhaj}, {Williams}, {Buenzli},
  {Cudel}, {Dominik}, {Galicher}, {Kasper}, {Lannier}, {Mesa}, {Mouillet},
  {Peretti}, {Perrot}, {Salter}, {Sissa}, {Wildi}, {Abe}, {Antichi},
  {Augereau}, {Baruffolo}, {Baudoz}, {Bazzon}, {Beuzit}, {Blanchard}, {Brems},
  {Buey}, {De Caprio}, {Carbillet}, {Carle}, {Cascone}, {Cheetham}, {Claudi},
  {Costille}, {Delboulb{\'e}}, {Dohlen}, {Fantinel}, {Feautrier}, {Fusco},
  {Giro}, {Gluck}, {Gry}, {Hubin}, {Hugot}, {Jaquet}, {Le Mignant}, {Llored},
  {Madec}, {Magnard}, {Martinez}, {Maurel}, {Meyer}, {M{\"o}ller-Nilsson},
  {Moulin}, {Mugnier}, {Orign{\'e}}, {Pavlov}, {Perret}, {Petit}, {Pragt},
  {Puget}, {Rabou}, {Ramos}, {Rigal}, {Rochat}, {Roelfsema}, {Rousset}, {Roux},
  {Salasnich}, {Sauvage}, {Sevin}, {Soenke}, {Stadler}, {Suarez}, {Turatto}, \&
  {Weber}}]{Keppler18}
{Keppler}, M., {Benisty}, M., {M{\"u}ller}, A., {et~al.} 2018, \aap, 617, A44

\bibitem[{{Kervella} {et~al.}(2019){Kervella}, {Arenou}, {Mignard}, \&
  {Th{\'e}venin}}]{Kervella19}
{Kervella}, P., {Arenou}, F., {Mignard}, F., \& {Th{\'e}venin}, F. 2019, \aap,
  623, A72

\bibitem[{{Kervella} {et~al.}(2016){Kervella}, {Mignard}, {M{\'e}rand}, \&
  {Th{\'e}venin}}]{Kervella16}
{Kervella}, P., {Mignard}, F., {M{\'e}rand}, A., \& {Th{\'e}venin}, F. 2016,
  \aap, 594, A107

\bibitem[{{Kervella} {et~al.}(2006){Kervella}, {Th{\'e}venin}, {Coud{\'e} du
  Foresto}, \& {Mignard}}]{Kervella06}
{Kervella}, P., {Th{\'e}venin}, F., {Coud{\'e} du Foresto}, V., \& {Mignard},
  F. 2006, \aap, 459, 669

\bibitem[{{Kraus} {et~al.}(2016){Kraus}, {Ireland}, {Huber}, {Mann}, \&
  {Dupuy}}]{Kraus16}
{Kraus}, A.~L., {Ireland}, M.~J., {Huber}, D., {Mann}, A.~W., \& {Dupuy}, T.~J.
  2016, \aj, 152, 8

\bibitem[{{Lagrange} {et~al.}(2006){Lagrange}, {Beust}, {Udry}, {Chauvin}, \&
  {Mayor}}]{Lagrange06}
{Lagrange}, A.~M., {Beust}, H., {Udry}, S., {Chauvin}, G., \& {Mayor}, M. 2006,
  \aap, 459, 955

\bibitem[{{Lagrange} {et~al.}(2009){Lagrange}, {Desort}, {Galland}, {Udry}, \&
  {Mayor}}]{Lagrange09}
{Lagrange}, A.-M., {Desort}, M., {Galland}, F., {Udry}, S., \& {Mayor}, M.
  2009, \aap, 495, 335

\bibitem[{{Lawler} {et~al.}(2014){Lawler}, {Di Francesco}, {Kennedy},
  {Sibthorpe}, {Booth}, {Vandenbussche}, {Matthews}, {Holland}, {Greaves},
  {Wilner}, {Tuomi}, {Blommaert}, {de Vries}, {Dominik}, {Fridlund}, {Gear},
  {Heras}, {Ivison}, \& {Olofsson}}]{Lawler14}
{Lawler}, S.~M., {Di Francesco}, J., {Kennedy}, G.~M., {et~al.} 2014, \mnras,
  444, 2665

\bibitem[{{Macintosh} {et~al.}(2015){Macintosh}, {Graham}, {Barman}, {De Rosa},
  {Konopacky}, {Marley}, {Marois}, {Nielsen}, {Pueyo}, {Rajan}, {Rameau},
  {Saumon}, {Wang}, {Patience}, {Ammons}, {Arriaga}, {Artigau}, {Beckwith},
  {Brewster}, {Bruzzone}, {Bulger}, {Burningham}, {Burrows}, {Chen}, {Chiang},
  {Chilcote}, {Dawson}, {Dong}, {Doyon}, {Draper}, {Duch{\^e}ne}, {Esposito},
  {Fabrycky}, {Fitzgerald}, {Follette}, {Fortney}, {Gerard}, {Goodsell},
  {Greenbaum}, {Hibon}, {Hinkley}, {Cotten}, {Hung}, {Ingraham},
  {Johnson-Groh}, {Kalas}, {Lafreniere}, {Larkin}, {Lee}, {Line}, {Long},
  {Maire}, {Marchis}, {Matthews}, {Max}, {Metchev}, {Millar-Blanchaer},
  {Mittal}, {Morley}, {Morzinski}, {Murray-Clay}, {Oppenheimer}, {Palmer},
  {Patel}, {Perrin}, {Poyneer}, {Rafikov}, {Rantakyr{\"o}}, {Rice}, {Rojo},
  {Rudy}, {Ruffio}, {Ruiz}, {Sadakuni}, {Saddlemyer}, {Salama}, {Savransky},
  {Schneider}, {Sivaramakrishnan}, {Song}, {Soummer}, {Thomas}, {Vasisht},
  {Wallace}, {Ward-Duong}, {Wiktorowicz}, {Wolff}, \&
  {Zuckerman}}]{Macintosh15}
{Macintosh}, B., {Graham}, J.~R., {Barman}, T., {et~al.} 2015, Science, 350, 64

\bibitem[{{Macintosh} {et~al.}(2014){Macintosh}, {Graham}, {Ingraham},
  {Konopacky}, {Marois}, {Perrin}, {Poyneer}, {Bauman}, {Barman}, {Burrows},
  {Cardwell}, {Chilcote}, {De Rosa}, {Dillon}, {Doyon}, {Dunn}, {Erikson},
  {Fitzgerald}, {Gavel}, {Goodsell}, {Hartung}, {Hibon}, {Kalas}, {Larkin},
  {Maire}, {Marchis}, {Marley}, {McBride}, {Millar-Blanchaer}, {Morzinski},
  {Norton}, {Oppenheimer}, {Palmer}, {Patience}, {Pueyo}, {Rantakyro},
  {Sadakuni}, {Saddlemyer}, {Savransky}, {Serio}, {Soummer},
  {Sivaramakrishnan}, {Song}, {Thomas}, {Wallace}, {Wiktorowicz}, \&
  {Wolff}}]{Macintosh14}
{Macintosh}, B., {Graham}, J.~R., {Ingraham}, P., {et~al.} 2014, Proceedings of
  the National Academy of Science, 111, 12661

\bibitem[{{Males} {et~al.}(2015){Males}, {Belikov}, \& {Bendek}}]{Males15}
{Males}, J.~R., {Belikov}, R., \& {Bendek}, E. 2015, in Society of
  Photo-Optical Instrumentation Engineers (SPIE) Conference Series, Vol. 9605,
  \procspie, 960518

\bibitem[{{Marois} {et~al.}(2008){Marois}, {Macintosh}, {Barman}, {Zuckerman},
  {Song}, {Patience}, {Lafreni{\`e}re}, \& {Doyon}}]{Marois08}
{Marois}, C., {Macintosh}, B., {Barman}, T., {et~al.} 2008, Science, 322, 1348

\bibitem[{{Mawet} {et~al.}(2019){Mawet}, {Hirsch}, {Lee}, {Ruffio}, {Bottom},
  {Fulton}, {Absil}, {Beichman}, {Bowler}, {Bryan}, {Choquet}, {Ciardi},
  {Christiaens}, {Defr{\`e}re}, {Gomez Gonzalez}, {Howard}, {Huby}, {Isaacson},
  {Jensen-Clem}, {Kosiarek}, {Marcy}, {Meshkat}, {Petigura}, {Reggiani},
  {Ruane}, {Serabyn}, {Sinukoff}, {Wang}, {Weiss}, \& {Ygouf}}]{Mawet19}
{Mawet}, D., {Hirsch}, L., {Lee}, E.~J., {et~al.} 2019, \aj, 157, 33

\bibitem[{{Mawet} {et~al.}(2014){Mawet}, {Milli}, {Wahhaj}, {Pelat}, {Absil},
  {Delacroix}, {Boccaletti}, {Kasper}, {Kenworthy}, {Marois}, {Mennesson}, \&
  {Pueyo}}]{2014ApJ...792...97M}
{Mawet}, D., {Milli}, J., {Wahhaj}, Z., {et~al.} 2014, \apj, 792, 97

\bibitem[{{McLean} {et~al.}(2017){McLean}, {Stam}, {Bagnulo}, {Borisov},
  {Devog{\`e}le}, {Cellino}, {Rivet}, {Bendjoya}, {Vernet}, {Paolini}, \&
  {Pollacco}}]{McLean17}
{McLean}, W., {Stam}, D.~M., {Bagnulo}, S., {et~al.} 2017, \aap, 601, A142

\bibitem[{{Milli} {et~al.}(2013){Milli}, {Mouillet}, {Mawet}, {Schmid},
  {Bazzon}, {Girard}, {Dohlen}, \& {Roelfsema}}]{Milli13}
{Milli}, J., {Mouillet}, D., {Mawet}, D., {et~al.} 2013, \aap, 556, A64

\bibitem[{{Mizuki} {et~al.}(2016){Mizuki}, {Yamada}, {Carson}, {Kuzuhara},
  {Nakagawa}, {Nishikawa}, {Sitko}, {Kudo}, {Kusakabe}, {Hashimoto}, {Abe},
  {Brander}, {Brandt}, {Egner}, {Feldt}, {Goto}, {Grady}, {Guyon}, {Hayano},
  {Hayashi}, {Hayashi}, {Henning}, {Hodapp}, {Ishii}, {Iye}, {Janson},
  {Kandori}, {Knapp}, {Kwon}, {Matsuo}, {McElwain}, {Miyama}, {Morino},
  {Moro-Martin}, {Nishimura}, {Pyo}, {Serabyn}, {Suenaga}, {Suto}, {Suzuki},
  {Takahashi}, {Takami}, {Takato}, {Terada}, {Thalmann}, {Turner}, {Watanabe},
  {Wisniewski}, {Takami}, {Usuda}, \& {Tamura}}]{Mizuki16}
{Mizuki}, T., {Yamada}, T., {Carson}, J.~C., {et~al.} 2016, \aap, 595, A79

\bibitem[{{Monnier} {et~al.}(2007){Monnier}, {Zhao}, {Pedretti}, {Thureau},
  {Ireland}, {Muirhead}, {Berger}, {Millan-Gabet}, {Van Belle}, {ten
  Brummelaar}, {McAlister}, {Ridgway}, {Turner}, {Sturmann}, {Sturmann}, \&
  {Berger}}]{Monnier07}
{Monnier}, J.~D., {Zhao}, M., {Pedretti}, E., {et~al.} 2007, Science, 317, 342

\bibitem[{{Neuh{\"a}user} {et~al.}(2007){Neuh{\"a}user}, {Mugrauer},
  {Fukagawa}, {Torres}, \& {Schmidt}}]{Neuhauser07}
{Neuh{\"a}user}, R., {Mugrauer}, M., {Fukagawa}, M., {Torres}, G., \&
  {Schmidt}, T. 2007, \aap, 462, 777

\bibitem[{{Nowak} {et~al.}(2018){Nowak}, {Le Coroller}, {Arnold}, {Dohlen},
  {Estevez}, {Fusco}, {Sauvage}, \& {Vigan}}]{Nowak18}
{Nowak}, M., {Le Coroller}, H., {Arnold}, L., {et~al.} 2018, \aap, 615, A144

\bibitem[{{Patapis} {et~al.}(2018){Patapis}, {K{\"u}hn}, \&
  {Schmid}}]{Patapis18}
{Patapis}, P., {K{\"u}hn}, J., \& {Schmid}, H.~M. 2018, in Society of
  Photo-Optical Instrumentation Engineers (SPIE) Conference Series, Vol. 10706,
  Advances in Optical and Mechanical Technologies for Telescopes and
  Instrumentation III, 107065J

\bibitem[{{Pavlov} {et~al.}(2008){Pavlov}, {M{\"o}ller-Nilsson}, {Feldt},
  {Henning}, {Beuzit}, \& {Mouillet}}]{Pavlov08}
{Pavlov}, A., {M{\"o}ller-Nilsson}, O., {Feldt}, M., {et~al.} 2008, in
  \procspie, Vol. 7019, Advanced Software and Control for Astronomy II, 701939

\bibitem[{{Quarles} \& {Lissauer}(2016)}]{Quarles16}
{Quarles}, B. \& {Lissauer}, J.~J. 2016, \aj, 151, 111

\bibitem[{{Queloz} {et~al.}(2000){Queloz}, {Mayor}, {Weber}, {Bl{\'e}cha},
  {Burnet}, {Confino}, {Naef}, {Pepe}, {Santos}, \& {Udry}}]{Queloz00}
{Queloz}, D., {Mayor}, M., {Weber}, L., {et~al.} 2000, \aap, 354, 99

\bibitem[{{Schmid} {et~al.}(2018){Schmid}, {Bazzon}, {Roelfsema}, {Mouillet},
  {Milli}, {Menard}, {Gisler}, {Hunziker}, {Pragt}, {Dominik}, {Boccaletti},
  {Ginski}, {Abe}, {Antoniucci}, {Avenhaus}, {Baruffolo}, {Baudoz}, {Beuzit},
  {Carbillet}, {Chauvin}, {Claudi}, {Costille}, {Daban}, {de Haan}, {Desidera},
  {Dohlen}, {Downing}, {Elswijk}, {Engler}, {Feldt}, {Fusco}, {Girard},
  {Gratton}, {Hanenburg}, {Henning}, {Hubin}, {Joos}, {Kasper}, {Keller},
  {Langlois}, {Lagadec}, {Martinez}, {Mulder}, {Pavlov}, {Podio}, {Puget},
  {Quanz}, {Rigal}, {Salasnich}, {Sauvage}, {Schuil}, {Siebenmorgen}, {Sissa},
  {Snik}, {Suarez}, {Thalmann}, {Turatto}, {Udry}, {van Duin}, {van Holstein},
  {Vigan}, \& {Wildi}}]{Schmid18}
{Schmid}, H.~M., {Bazzon}, A., {Roelfsema}, R., {et~al.} 2018, \aap, 619, A9

\bibitem[{{Schmid} {et~al.}(2006{\natexlab{a}}){Schmid}, {Beuzit}, {Feldt},
  {Gisler}, {Gratton}, {Henning}, {Joos}, {Kasper}, {Lenzen}, {Mouillet},
  {Moutou}, {Quirrenbach}, {Stam}, {Thalmann}, {Tinbergen}, {Verinaud},
  {Waters}, \& {Wolstencroft}}]{Schmid06a}
{Schmid}, H.~M., {Beuzit}, J.-L., {Feldt}, M., {et~al.} 2006{\natexlab{a}}, in
  IAU Colloq. 200: Direct Imaging of Exoplanets: Science \& Techniques, ed.
  C.~{Aime} \& F.~{Vakili}, 165--170

\bibitem[{{Schmid} {et~al.}(2012){Schmid}, {Downing}, {Roelfsema}, {Bazzon},
  {Gisler}, {Pragt}, {Cumani}, {Salasnich}, {Pavlov}, {Baruffolo}, {Beuzit},
  {Costille}, {Deiries}, {Dohlen}, {Dominik}, {Elswijk}, {Feldt}, {Kasper},
  {Mouillet}, {Thalmann}, \& {Wildi}}]{Schmid12}
{Schmid}, H.-M., {Downing}, M., {Roelfsema}, R., {et~al.} 2012, in Society of
  Photo-Optical Instrumentation Engineers (SPIE) Conference Series, Vol. 8446,
  Society of Photo-Optical Instrumentation Engineers (SPIE) Conference Series,
  8

\bibitem[{{Schmid} {et~al.}(2011){Schmid}, {Joos}, {Buenzli}, \&
  {Gisler}}]{Schmid11}
{Schmid}, H.~M., {Joos}, F., {Buenzli}, E., \& {Gisler}, D. 2011, \icarus, 212,
  701

\bibitem[{{Schmid} {et~al.}(2006{\natexlab{b}}){Schmid}, {Joos}, \&
  {Tschan}}]{Schmid06b}
{Schmid}, H.~M., {Joos}, F., \& {Tschan}, D. 2006{\natexlab{b}}, \aap, 452, 657

\bibitem[{{Schmidt} {et~al.}(2016){Schmidt}, {Neuh{\"a}user}, {Brice{\~n}o},
  {Vogt}, {Raetz}, {Seifahrt}, {Ginski}, {Mugrauer}, {Buder}, {Adam},
  {Hauschildt}, {Witte}, {Helling}, \& {Schmitt}}]{Schmidt16}
{Schmidt}, T.~O.~B., {Neuh{\"a}user}, R., {Brice{\~n}o}, C., {et~al.} 2016,
  \aap, 593, A75

\bibitem[{{Schroeder} {et~al.}(2000){Schroeder}, {Golimowski}, {Brukardt},
  {Burrows}, {Caldwell}, {Fastie}, {Ford}, {Hesman}, {Kletskin}, {Krist},
  {Royle}, \& {Zubrowski}}]{Schroeder00}
{Schroeder}, D.~J., {Golimowski}, D.~A., {Brukardt}, R.~A., {et~al.} 2000, \aj,
  119, 906

\bibitem[{{Schworer} \& {Tuthill}(2015)}]{Schworer15}
{Schworer}, G. \& {Tuthill}, P.~G. 2015, \aap, 578, A59

\bibitem[{{Seager} {et~al.}(2000){Seager}, {Whitney}, \& {Sasselov}}]{Seager00}
{Seager}, S., {Whitney}, B.~A., \& {Sasselov}, D.~D. 2000, \apj, 540, 504

\bibitem[{{Smith} \& {Tomasko}(1984)}]{Smith84}
{Smith}, P.~H. \& {Tomasko}, M.~G. 1984, \icarus, 58, 35

\bibitem[{{Spiegel} \& {Burrows}(2012)}]{Spiegel12}
{Spiegel}, D.~S. \& {Burrows}, A. 2012, \apj, 745, 174

\bibitem[{{Stam}(2008)}]{Stam08}
{Stam}, D.~M. 2008, \aap, 482, 989

\bibitem[{{Stam} {et~al.}(2004){Stam}, {Hovenier}, \& {Waters}}]{Stam04}
{Stam}, D.~M., {Hovenier}, J.~W., \& {Waters}, L.~B.~F.~M. 2004, \aap, 428, 663

\bibitem[{{Stolker} {et~al.}(2019){Stolker}, {Bonse}, {Quanz}, {Amara},
  {Cugno}, {Bohn}, \& {Boehle}}]{Stolker19}
{Stolker}, T., {Bonse}, M.~J., {Quanz}, S.~P., {et~al.} 2019, \aap, 621, A59

\bibitem[{{Stolker} {et~al.}(2017){Stolker}, {Min}, {Stam}, {Molli{\`e}re},
  {Dominik}, \& {Waters}}]{Stolker17}
{Stolker}, T., {Min}, M., {Stam}, D.~M., {et~al.} 2017, \aap, 607, A42

\bibitem[{{Sudarsky} {et~al.}(2003){Sudarsky}, {Burrows}, \&
  {Hubeny}}]{Sudarsky03}
{Sudarsky}, D., {Burrows}, A., \& {Hubeny}, I. 2003, \apj, 588, 1121

\bibitem[{{Swift} {et~al.}(2012){Swift}, {Eggert}, {Hicks}, {Hamel},
  {Caspersen}, {Schwegler}, {Collins}, {Nettelmann}, \& {Ackland}}]{Swift12}
{Swift}, D.~C., {Eggert}, J.~H., {Hicks}, D.~G., {et~al.} 2012, \apj, 744, 59

\bibitem[{{Thalmann} {et~al.}(2011){Thalmann}, {Janson}, {Buenzli}, {Brandt},
  {Wisniewski}, {Moro-Mart{\'{\i}}n}, {Usuda}, {Schneider}, {Carson},
  {McElwain}, {Grady}, {Goto}, {Abe}, {Brandner}, {Dominik}, {Egner}, {Feldt},
  {Fukue}, {Golota}, {Guyon}, {Hashimoto}, {Hayano}, {Hayashi}, {Hayashi},
  {Henning}, {Hodapp}, {Ishii}, {Iye}, {Kandori}, {Knapp}, {Kudo}, {Kusakabe},
  {Kuzuhara}, {Matsuo}, {Miyama}, {Morino}, {Nishimura}, {Pyo}, {Serabyn},
  {Suto}, {Suzuki}, {Takahashi}, {Takami}, {Takato}, {Terada}, {Tomono},
  {Turner}, {Watanabe}, {Yamada}, {Takami}, \& {Tamura}}]{Thalmann11}
{Thalmann}, C., {Janson}, M., {Buenzli}, E., {et~al.} 2011, \apjl, 743, L6

\bibitem[{{Thalmann} {et~al.}(2008){Thalmann}, {Schmid}, {Boccaletti},
  {Mouillet}, {Dohlen}, {Roelfsema}, {Carbillet}, {Gisler}, {Beuzit}, {Feldt},
  {Gratton}, {Joos}, {Keller}, {Kragt}, {Pragt}, {Puget}, {Rigal}, {Snik},
  {Waters}, \& {Wildi}}]{Thalmann08}
{Thalmann}, C., {Schmid}, H.~M., {Boccaletti}, A., {et~al.} 2008, in Society of
  Photo-Optical Instrumentation Engineers (SPIE) Conference Series, Vol. 7014,
  Society of Photo-Optical Instrumentation Engineers (SPIE) Conference Series,
  3

\bibitem[{{Tomasko} \& {Smith}(1982)}]{Tomasko82}
{Tomasko}, M.~G. \& {Smith}, P.~H. 1982, \icarus, 51, 65

\bibitem[{{Tuomi} {et~al.}(2013){Tuomi}, {Jones}, {Jenkins}, {Tinney},
  {Butler}, {Vogt}, {Barnes}, {Wittenmyer}, {O'Toole}, {Horner}, {Bailey},
  {Carter}, {Wright}, {Salter}, \& {Pinfield}}]{Tuomi13}
{Tuomi}, M., {Jones}, H.~R.~A., {Jenkins}, J.~S., {et~al.} 2013, \aap, 551, A79

\bibitem[{{van Dam} {et~al.}(2004){van Dam}, {Le Mignant}, \&
  {Macintosh}}]{vanDam04}
{van Dam}, M.~A., {Le Mignant}, D., \& {Macintosh}, B.~A. 2004, \ao, 43, 5458

\bibitem[{{van Holstein} {et~al.}(2017){van Holstein}, {Snik}, {Girard}, {de
  Boer}, {Ginski}, {Keller}, {Stam}, {Beuzit}, {Mouillet}, {Kasper},
  {Langlois}, {Zurlo}, {de Kok}, \& {Vigan}}]{vanHolstein17}
{van Holstein}, R.~G., {Snik}, F., {Girard}, J.~H., {et~al.} 2017, in Society
  of Photo-Optical Instrumentation Engineers (SPIE) Conference Series, Vol.
  10400, Society of Photo-Optical Instrumentation Engineers (SPIE) Conference
  Series, 1040015

\bibitem[{{Vigan} {et~al.}(2015){Vigan}, {Gry}, {Salter}, {Mesa}, {Homeier},
  {Moutou}, \& {Allard}}]{Vigan15}
{Vigan}, A., {Gry}, C., {Salter}, G., {et~al.} 2015, \mnras, 454, 129

\bibitem[{{Wiegert} \& {Holman}(1997)}]{Wiegert97}
{Wiegert}, P.~A. \& {Holman}, M.~J. 1997, \aj, 113, 1445

\bibitem[{{Zhao} {et~al.}(2018){Zhao}, {Fischer}, {Brewer}, {Giguere}, \&
  {Rojas-Ayala}}]{Zhao18}
{Zhao}, L., {Fischer}, D.~A., {Brewer}, J., {Giguere}, M., \& {Rojas-Ayala}, B.
  2018, \aj, 155, 24

\bibitem[{{Zucker} {et~al.}(2004){Zucker}, {Mazeh}, {Santos}, {Udry}, \&
  {Mayor}}]{Zucker04}
{Zucker}, S., {Mazeh}, T., {Santos}, N.~C., {Udry}, S., \& {Mayor}, M. 2004,
  \aap, 426, 695

\end{thebibliography}

\begin{appendix}

\section{Advanced data reduction steps}
\label{sec:Advanced data reduction steps}
\subsection{Frame transfer smearing correction}
ZIMPOL uses frame transfer CCDs which shift each frame at the end of an illumination 
from the image area to a covered read-out area of the detector. 
There the previously illuminated frame is read-out during integration of the next 
frame. The detector is also illuminated during the fast frame transfer, which
lasts 56~ms for fast polarimetry, and this causes a frame transfer
smearing of the frame in the column direction. The smearing amounts to a maximum 
of 5~\% for the shortest integration time of 1.1~s and less for longer integrations 
\citep[see][]{Schmid18, Schmid12}. The smearing corrects itself in the polarization signal $Q^Z$ but is for short exposures apparent in the intensity signal $I^Z$. We correct this frame transfer smearing in each intensity image with the subtraction of a correctly scaled mean row profile from every row 
in the bias subtracted image.

\subsection{Telescope polarization correction}
\label{sec:Telescope polarization correction}
\begin{figure*}
\centering
\includegraphics[width=\textwidth]{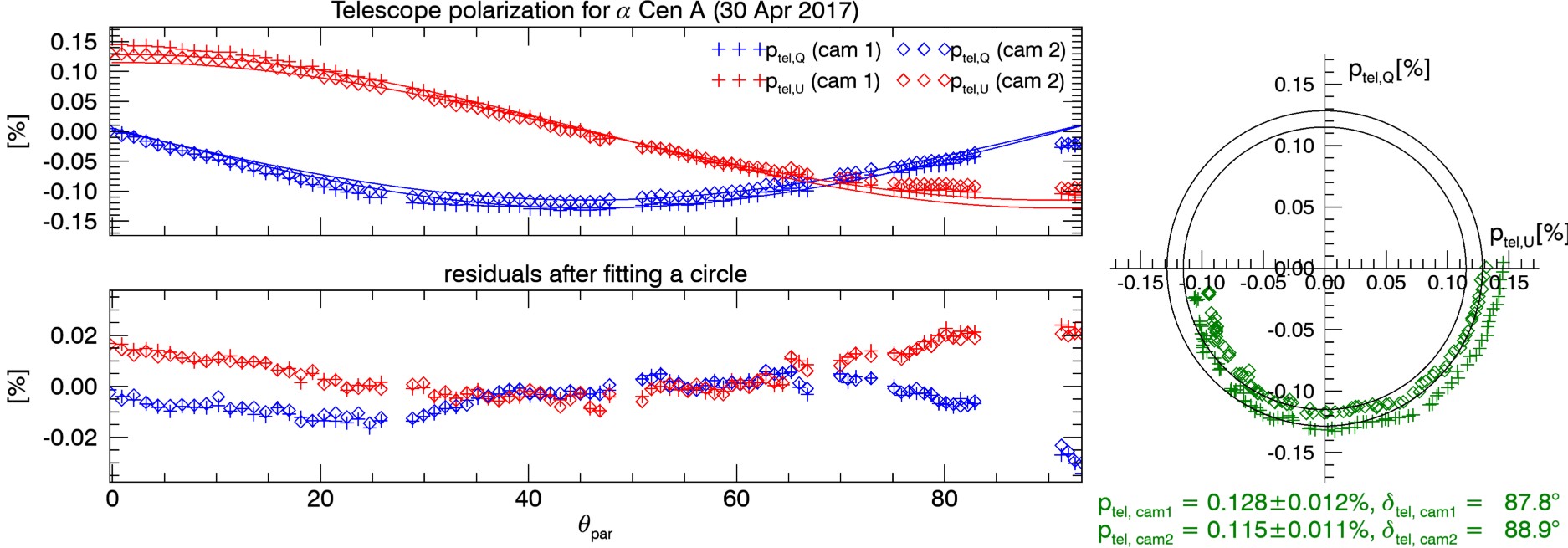}
\caption{Median telescope polarization for $\alpha$~Cen~A from 2017 in N\_R-band measured for both Stokes parameters -- Q and U -- as a function of the telescope parallactic angle $\theta_{\rm par}$. The plots show the measurements for camera 1 (+) and camera 2 ($\diamond$). The measurements have been fit with a model function (solid lines), resulting in a measurement of the telescope polarization amplitude $p_{\rm tel}$ and phase $\delta_{\rm tel}$ for both cameras.}
\label{fig:instr_pol}
\end{figure*}

\begin{figure}
\centering
\begin{tabular}{c}
\resizebox{\hsize}{!}{\includegraphics{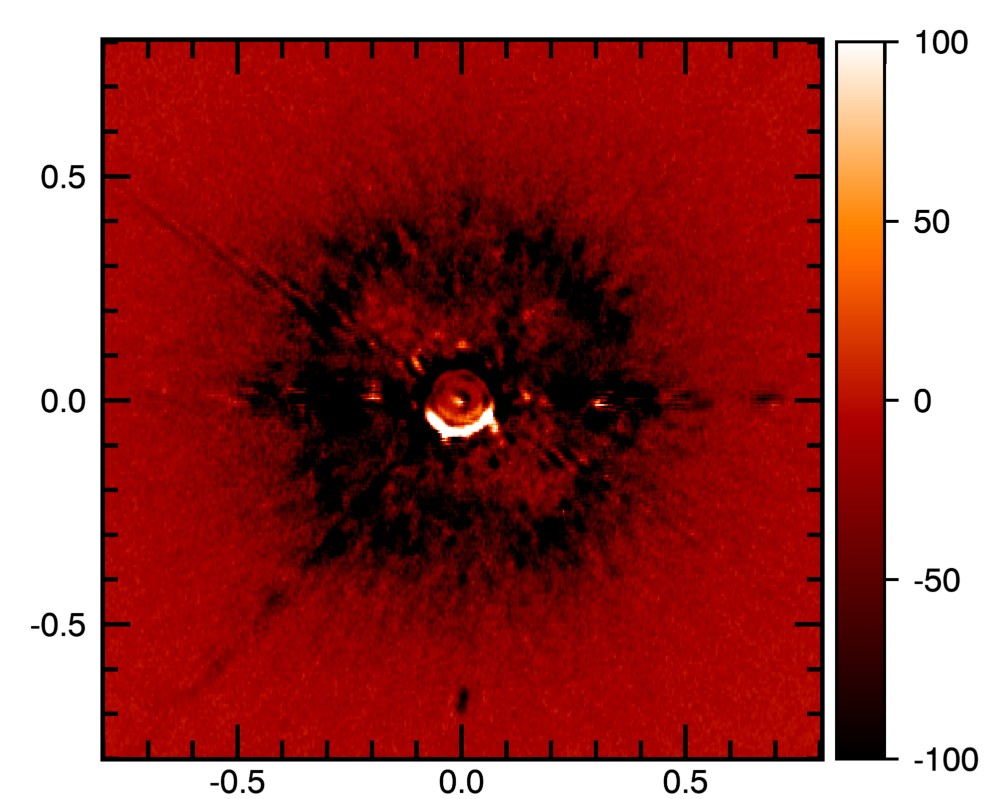}} \\
\resizebox{\hsize}{!}{\includegraphics{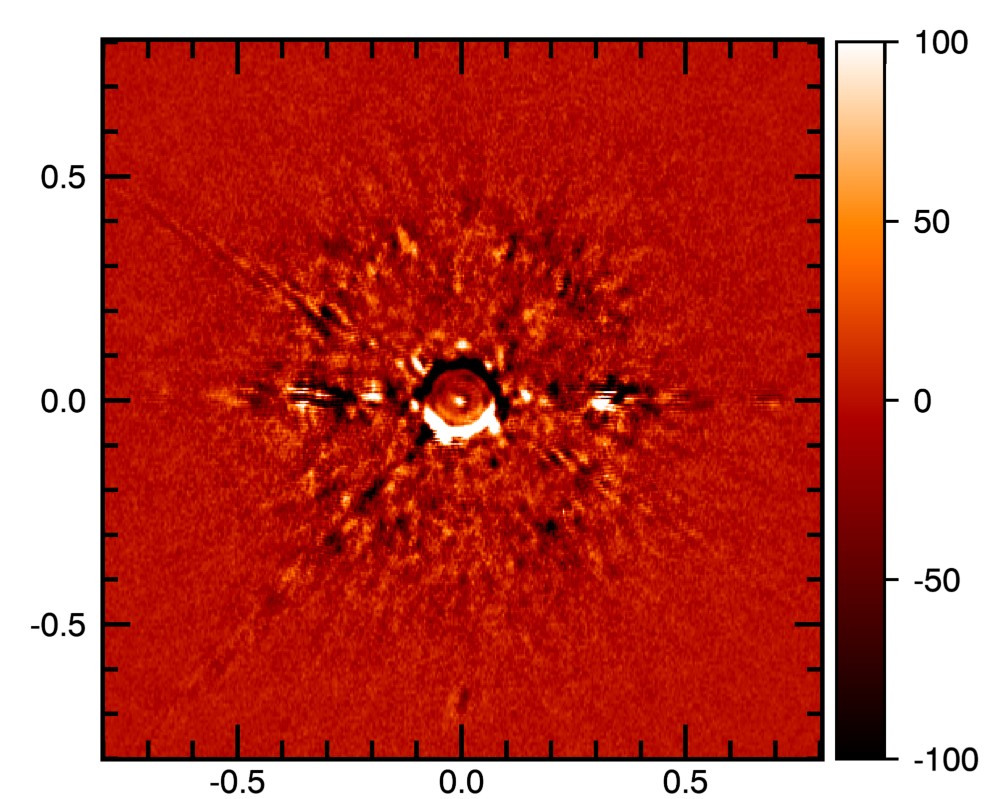}} \\
\end{tabular}
\caption{Comparison of a coronagraphic exposure before (top) and after (bottom) 
the subtraction of the telescope polarization $p_{\rm tel}$. The images show the polarized 
intensity Q of a combined pair of one single zero-phase and one $\pi$-phase 
($2 \times 1.2$~s) exposure of $\alpha$~Cen~A in the N\_R filter. The axes are in arc seconds and the color scale is in ADU.}
  \label{fig:ins_pol_corr_exmpl_ACenA30Apr}
\end{figure}

The SPHERE/ZIMPOL instrument uses a half wave plate (HWP) polarization switch to select opposite polarization modes Q$^+$ and Q$^-$, or U$^+$ and U$^-$ to compensate the instrumental polarization \citep{Bazzon12}. However, there is a remaining residual telescope polarization $p_{\rm tel}$ 
from the optical components located in front of the HWP switch. The magnitude of the residual telescope polarization $p_{\rm tel}^2 = p_{\rm tel,Q}^2(\theta_{par}) + p_{\rm tel,U}^2(\theta_{\rm par})$ is wavelength dependent but otherwise essentially constant. The orientation of $p_{\rm tel}$ rotates in the Stokes Q-U plane as a function of the parallactic angle of the telescope $\theta_{\rm par}$. The measured Stokes parameters $Q_{\rm obs}$ and $U_{\rm obs}$ are therefore a combination of the astrophysical polarization of the target $Q, U$ and the induced telescope polarization:
\begin{equation}
\begin{aligned}
Q_{\rm obs}(x,y) &= Q(x,y) + p_{{\rm tel},Q}(\theta_{\rm par}) \cdot I(x,y) \\
U_{\rm obs}(x,y) &= U(x,y) + p_{{\rm tel},U}(\theta_{\rm par}) \cdot I(x,y)
\end{aligned}
\label{equ:tel_pol_equ}
\end{equation}
To first order, the telescope polarization does not depend on the image
coordinates $(x, y)$. Therefore, we can estimate the fractional 
telescope polarization for unpolarized sky sources 
($Q(x,y)\approx 0$ and $U(x,y)\approx 0$) by measuring 
the mean\footnote{Average over pixels in a ring around the star with inner radius $r$ larger than the coronagraph and outer radius $R$: \[\langle f(x,y) \rangle = \frac{\sum\limits_{r^2 < x^2+y^2 < R^2} f(x,y)}{\sum\limits_{r^2 < x^2+y^2 < R^2} 1} \]} fractional polarization $Q_{\rm obs}/I$ and $U_{\rm obs}/I$:
\begin{equation}
\begin{aligned}
 p_{{\rm tel},Q} \approx \langle Q_{\rm obs}(x,y)\rangle / \langle I(x,y)\rangle  \\
 p_{{\rm tel},U} \approx \langle U_{\rm obs}(x,y)\rangle / \langle I(x,y)\rangle 
\end{aligned}
\label{equ:tel_pol_measure}
\end{equation}
\citet{Schmid18} used measurements of unpolarized standard stars for the determination of the amplitude $p_{\rm tel}$ in several different filters for SPHERE/ZIMPOL, which can be described by 
\begin{equation}
\begin{aligned}
p_{{\rm tel},Q}(\theta_{par}) &= p_{\rm tel} \cdot \cos \left( 2\left(\theta_{\rm par} + \delta_{\rm tel} \right) \right)\,, \\
p_{{\rm tel},U}(\theta_{par}) &= p_{\rm tel} \cdot \sin \left( 2\left(\theta_{\rm par} + \delta_{\rm tel} \right) \right)\,,
\end{aligned}
\label{equ:tel_pol_model}
\end{equation}
where $\delta_{\rm tel}$ is a wavelength dependent offset angle for the telescope polarization. 

Our targets are not zero-polarization standard stars, but it is expected that regular main-sequence stars are highly spherically symmetric and are hot enough to not have any clouds/hazes, therefore $\alpha$ Cen A and B, Sirius~A, $\epsilon$ Eri or $\tau$ Cet are expected to show only very little intrinsic integrated polarization in broad-band filters. For example for the Sun, an upper limit on the integrated linear polarization of $<10^{-6}$ was determined by \citet{Kemp87}. Altair could be an exception and it may show larger intrinsic linear polarization, because it is a rapidly rotating and therefore ellipsoidal A-star \citep{Monnier07}. However, existing polarimetry of Altair yield a very low polarization of $<10^{-5}$ \citep{Bailey10}. No interstellar polarization component is expected for our stars, because of the distance of only a few parsecs, as is confirmed by high precision polarimetry \citep{Bailey10, Cotton16}. 

Because our targets were observed over a large range of parallactic angles, they show the steady rotation of $p_{\rm tel}$ as illustrated in Fig.~\ref{fig:instr_pol} for the narrow R-band observations of $\alpha$~Cen~A taken in April 30, 2017. Each point in the plot shows the average telescope polarization measured in an annulus  -- extending from 0.13$\arcsec$ to 0.72$\arcsec$ -- in the stellar PSF halo centred on the star. The data can be fit well with the model in Eq.~(\ref{equ:tel_pol_model}) with a $p_{\rm tel}=0.12\%$ and $\delta_{\rm tel}=88.4^{\circ}$. The residuals indicate that the instrumental second order effects and the intrinsic polarization of $\alpha$ Cen A are of order $\lessapprox 0.02$\% and possibly a systematic shift of the polarization towards positive $U$ values in Fig.~\ref{fig:instr_pol}. However, fitting a model with a systematic shift as additional free parameter does not significantly improve the residuals, from this we conclude that the majority of the residuals are caused by instrumental effects. 

The instrument polarization produces in our $Q$ and $U$ differential polarization image a faint copy of the intensity images at the level of 0.1\% (see Fig.~\ref{fig:ins_pol_corr_exmpl_ACenA30Apr}, top). We correct this by measuring $p_{{\rm tel},Q}$ and $p_{{\rm tel},U}$ (Eq.~(\ref{equ:tel_pol_measure})) and subtracting the scaled intensity frames $p_{{\rm tel},Q} \cdot I$, $p_{{\rm tel},U} \cdot I$ from the corresponding $Q$ and $U$ frames. Fig. \ref{fig:ins_pol_corr_exmpl_ACenA30Apr} shows the polarized intensity before (top panel) and after (bottom panel) the subtraction.

The instrument polarization is not a dominating source of noise for the search of a localized point-source. Our analysis of the noise after different reduction and calibration steps (see Fig.~\ref{fig:data_redu_steps_single_img}) shows only small improvements for most separations. This is understandable if we separate the noise in a $Q_{\rm obs}$ frame with instrument polarization, as defined in Eq.~(\ref{equ:tel_pol_equ}), into its two main contributions $\delta Q$ and $p_{\rm tel} \cdot \delta I$. The noise of the polarized intensity $\delta Q$ originates mainly from residuals of the PDI speckle cancellation at smaller separations and photon noise at larger separations (see Fig. \ref{fig:data_redu_steps_single_img}, green line). The noise in the total intensity $\delta I$ is dominated by speckles at all separations (see Fig. \ref{fig:data_redu_steps_single_img}, black line). The small scale noise introduced by the instrument polarization is not significant as long as $\delta Q \gg p_{\rm tel} \cdot \delta I$. In our data this condition is usually satisfied, since $p_{\rm tel}$ is on the order of $10^{-3}$ while the noise ratio $\delta I / \delta Q$ is only on the order of $10$. Even though the improvement in point-source contrast is small, we still do the subtraction of the instrument polarization because we want to reach the best possible contrast and remove as many instrumental effects as possible.

For the considerations in this section we always assumed that the astrophysical polarization of the target $Q$ and $U$ are zero when averaged over large portion of the image. In Appendix~\ref{Appendix: instrmental polarization calculation} we investigate in more detail what happens to a non-zero astrophysical polarization signal of a faint companion if we apply the instrumental polarization correction as described above. We show that the process of subtracting the instrument polarization as described above has an insignificant effect on the signal of a faint point-source potentially present in the data.

\subsection{Differential polarimetric beam shift}
\begin{figure}
\centering
\begin{tabular}{c}
\resizebox{\hsize}{!}{\includegraphics{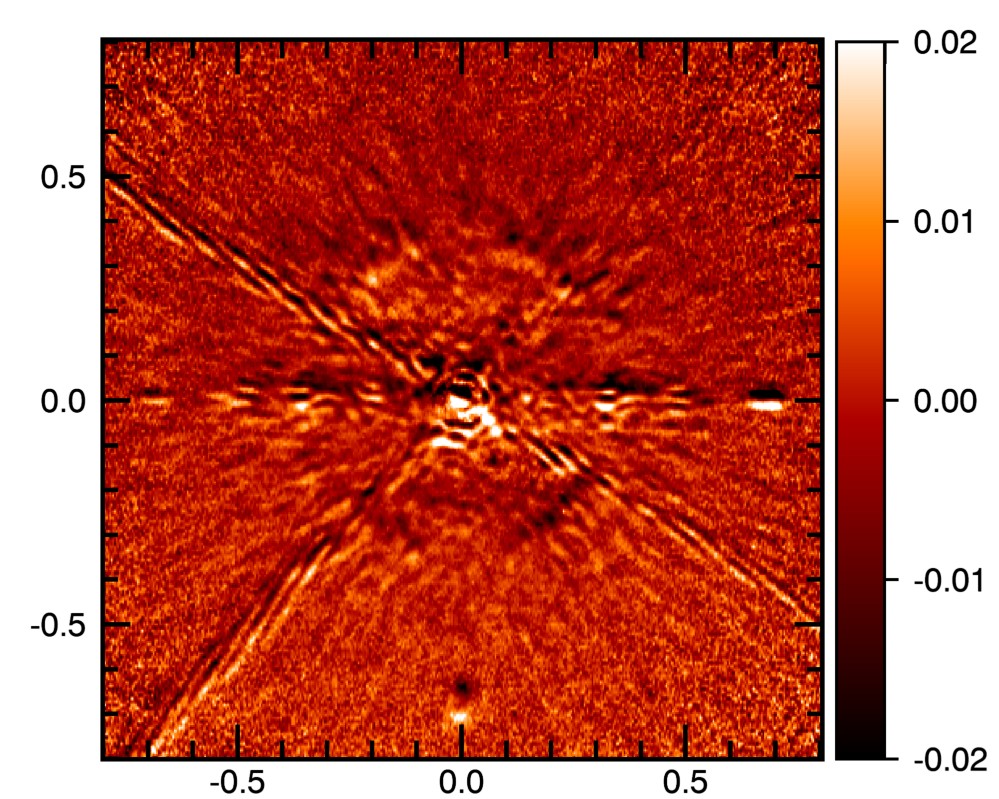}} \\
\resizebox{\hsize}{!}{\includegraphics{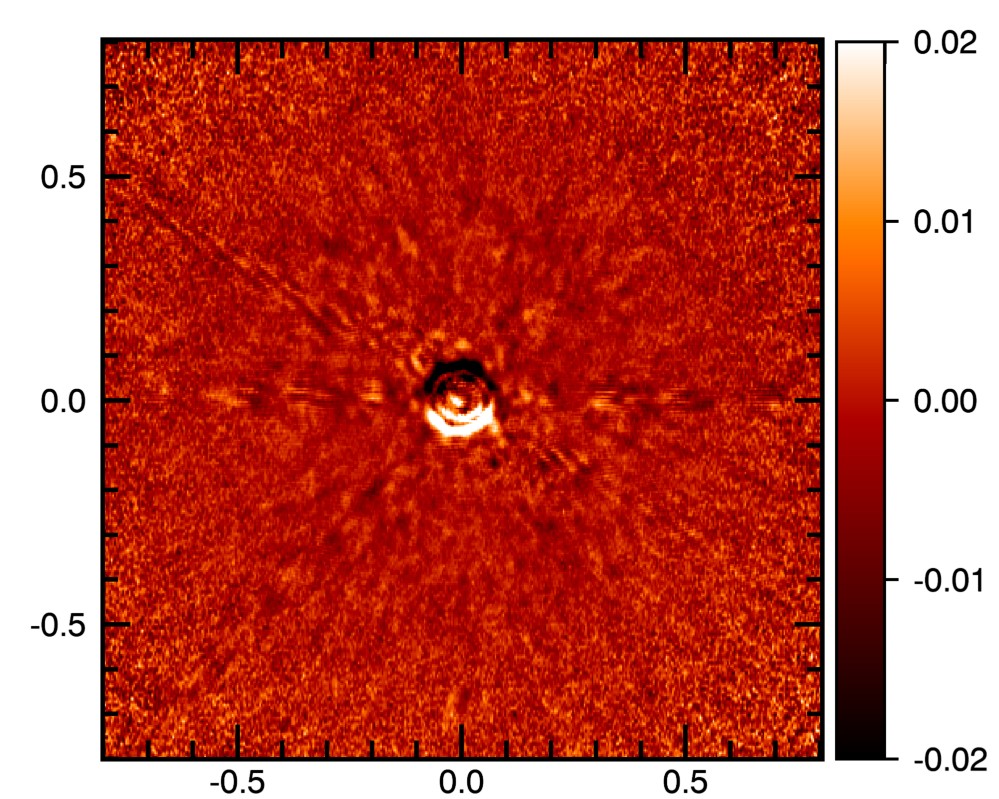}} \\
\end{tabular}
\caption{Comparison of a coronagraphic exposure before (top) and after (bottom) the beam shift correction. The images show the fractional polarization Q/I of a single 0-phase and $\pi$-phase combined $2 \times 1.2$ second exposure of $\alpha$~Cen~A in the N\_R filter. The axes are in arc seconds and the color scale is dimensionless.}
  \label{fig:bs_corr_exmpl_ACenA30Apr}
\end{figure}

The even and odd rows of the detectors in ZIMPOL measure the two opposite linear polarization states $I_{\parallel}$ and $I_{\perp}$ simultaneously and on the same detector pixels. This would in principle allow for a perfect speckle suppression in the final Q and U polarization frames. However, it was noticed that $I_{\parallel}$ and $I_{\perp}$ are not perfectly aligned on the detector even though they go through the same optical path in the instrument. This unexpectedly large differential polarization beam shift of up to 0.3 pixels ($\approx 1$~mas) is caused mostly by reflections on inclined mirrors. The effect is for ZIMPOL described in \citet{Schmid18} and one example is shown in Fig. \ref{fig:bs_corr_exmpl_ACenA30Apr}. For high-contrast applications the beam shift must be determined and corrected, otherwise the speckles will not cancel sufficiently in the PDI process and a pattern of positive and negative speckle residuals remains.

However, the beam shift and possibly also other differential aberration effects cannot be corrected perfectly. One reason for this is the wavelength dependence of the beam shift. This produces for broad-band observations radially elongated speckles, for which the innermost (shortest wavelength) part suffers a different beam shift than the outermost (longest wavelength) part. Tests have shown that the beam shift is similar in R and I-band, but it can be significantly different in V-band. Thus, the wavelength dependence is mainly a problem for observations with the VBB filter because it spans a large portion of the visual spectrum. Observations with any of the R- and I-band filters should not suffer as much. Despite this, it is important to apply a "mean" beam shift correction for any filter, because it can improve the high-contrast performance significantly.

Currently, there exists no comprehensive beam shift model for the SPHERE/ZIMPOL instrument. Therefore, the first step is a beam shift measurement, preferentially based on the science data which need to be corrected. This can be achieved with one of the following methods:

\begin{itemize}
\item For data taken with the semi-transparent coronagraph CLC-MT-WF or without coronagraph the beam shift can often be measured in individual images as offset of the PSF peak between even- and odd-row (or $I_\perp$ and $I_\parallel$) frames. This is only possible if the PSF peak is not saturated and well defined so that the relative PSF-offsets can be determined with high precision. The observing conditions influence the measuring precision and a careful selection of good beam shift data is essential. For PSF peaks observed through the semi-transparent coronagraph the determinations are difficult if the peak is close to the "edge" of the coronagraphic flux minimum. 

\item In some cases it is easier to derive the beam shift by cross-correlating the speckle pattern or numerically solving for the differential offset of $I_{\parallel}$ and $I_{\perp}$ that minimizes the residuals in $Q^Z=I_\perp-I_\parallel$. This method often works well because the beam shift is identical for the whole FOV as shown in Fig.~\ref{fig:bs_corr_exmpl_ACenA30Apr}. The minimization works better for shorter exposures and narrow filters because the individual speckles are more numerous and better defined. In long exposures the atmospheric turbulences smooth short lived speckles and broad-band observations extend the shape of the speckles strongly in radial direction, reducing their signal with respect to the background.
\end{itemize}

Both methods allow the determination of the beam shift with a precision of order 0.01 pixels ($\approx 0.04$~mas) in a single exposure for data with good observing conditions. For a proper beam shift correction the high precision is necessary because the effect is only of order 0.1 pixels, but the difference of the correction is noticeable even in a single exposure. Fig.\ref{fig:bs_corr_exmpl_ACenA30Apr} shows a short exposure of $\alpha$~Cen~A before and after applying the beam shift correction. Most of the residual speckle pattern can be suppressed, improving significantly the point-source contrast limit in the speckle-dominated region. In the combined final image the effect of the beam shift can not only be seen as additional speckles on a small scale but also as a disturbing feature on a larger scale. This is because the whole speckle halo of the stellar PSF is beam shifted. In our high-contrast images this artificially produces negatively and positively polarized large scale features that can limit the sensitivity for real large scale polarized signals.

Unfortunately, the beam shift correction also introduces some new systematic noise residuals. All intensity features originating from components located downstream of the inclined mirrors -- namely the M3 mirror, the pupil tip-tilt mirror and image de-rotator mirrors -- are not subject to the beam shift effect. Applying the beam shift correction to the polarimetric data, as described above, will introduce spurious residual patterns in the "corrected" polarimetric image. This concerns the intensity edges of the attenuating focal plane mask of the coronagraph, or intensity patterns from the dead actuators of the deformable mirror, as well as bad pixels, charge traps and dust on the micro-lens array of the ZIMPOL detector \citep{Schmid18}. An example in Fig.~\ref{fig:bs_corr_exmpl_ACenA30Apr} is the black and white pattern visible at the edge of the coronagraph because of the applied beam shift correction in vertical direction. This effect increases the effective IWA of the result and it cannot be corrected. The intensity patterns from dust on the micro-lens array can be efficiently removed by flat-fielding, and the pixel scale effects can be strongly reduced by masking, bad-pixel cleaning, dithering or angular differential imaging.

The beam shift changes continuously with the telescope pointing direction and depends on both the parallactic angle $\theta_{\rm par}$ and altitude angle $\theta_{\rm alt}$. Therefore it is advantageous for long observations if the beam shift can be accurately measured from the science data itself without requiring any additional overhead. However, for coronagraphic data or saturated data, the beam shift measurement may fail and therefore we regularly take non-coronagraphic PSF measurements.

\section{Subtraction of instrument polarization}
\label{Appendix: instrmental polarization calculation}
By measuring the residual polarization in our data for different parallactic angles $\theta_{\rm par}$ (see Sec.~\ref{sec:Telescope polarization correction}) we determined that all targets of our survey are only weakly polarized ($p_{\rm st} \lesssim 10^{-4}$). At this low level of polarization we cannot distinguish any more between the intrinsic polarization of the target and second order instrumental polarization effects. We have also shown that the residual telescope polarization $p_{\rm tel}$ is of order $10^{-3}$. In Sec.~\ref{sec:Telescope polarization correction} we explain how this polarization offset can be measured and removed from the data. This is a common way of removing the instrument polarization and it does not harm any polarized signals in the vicinity of the star as long as the stellar PSF is only polarized due to instrumental or interstellar polarization because both processes only add a constant fractional polarization offset to all sources in the FOV. In the following section we investigate the effect of the instrumental polarization correction on a polarized point-source if the star itself exhibits an intrinsic polarization. 

We analyse this problem with a model for the observed Stokes $I_{\rm obs}(x,y)$ and $Q_{\rm obs}(x,y)$ signal, with $(x,y)$ being the image coordinates (for $U_{\rm obs}(x,y)$ the analysis is equivalent). In our model the observed intensity distribution $I_{\rm obs}(x,y)$ consists of contributions from the PSF of the star itself $I_{\rm PSF,st}(x,y)$ and an offset PSF from the planet $I_{\rm PSF,pl}(x,y)$, scaled with the flux contrast $C_{\rm flux}(\alpha)$. The intensities $I_{\rm PSF,st}(x,y)$ and $I_{\rm PSF,pl}(x,y)$ are offset from each other but otherwise identical. The polarized intensity distribution $Q_{\rm obs}(x,y)$ consists of the intrinsic polarization of the star $p_{\rm st,Q} I_{\rm PSF,st}(x,y)$, the polarized signal from the planet $C_{\rm pol,Q}(\alpha) I_{\rm PSF,pl}(x,y)$ and a term that describes the polarization offset due to instrumental and/or interstellar polarization $p_{\rm tel,Q} (I_{\rm PSF,st}(x,y) + C_{\rm flux}(\alpha) I_{\rm PSF,pl}(x,y))$. For the sake of readability, we omit the dependencies on scattering angle $\alpha$ and image coordinates $(x,y)$ during the derivations.

\begin{equation}
\begin{aligned}
 I_{\rm obs} &= I_{\rm PSF,st} + C_{\rm flux} I_{\rm PSF,pl} \\
 Q_{\rm obs} &= p_{\rm st,Q} I_{\rm PSF,st} + C_{\rm pol,Q} I_{\rm PSF,pl} + p_{\rm tel,Q} (I_{\rm PSF,st} + C_{\rm flux} I_{\rm PSF,pl})
\end{aligned}
\label{equ:tel_pol_extended_model}
\end{equation}

From Eq.~(\ref{equ:tel_pol_extended_model}) we want to extract the signal of the polarized planet with all instrumental and (inter-)stellar contributions removed, so that the corrected polarized intensity $Q_{\rm obs}'(x,y)$ is:

\begin{equation}
Q_{\rm obs}' = C_{\rm pol,Q} I_{\rm PSF,pl} = p_{\rm pl,Q} C_{\rm flux} I_{\rm PSF,pl}
\label{equ:tel_pol_extended_model_desired}
\end{equation}

The contrasts $C_{\rm flux}(\alpha)$ and $C_{\rm pol,Q}(\alpha)$ are linked by the fractional polarization of the planet $p_{\rm pl,Q}(\alpha)=C_{\rm pol,Q}(\alpha)/C_{\rm flux}(\alpha)$. Fractional polarizations have an additional subscript $Q$ that can be either negative or positive because they are entries of the two dimensional vector for the linear polarization $p = (p_{Q}, p_{U})$ in the Stokes $Q-U$ plane. All of the derivations here are equivalent for the Stokes $U$ measurements.

Using this model, the flux weighted fractional polarization in a large aperture $\langle Q_{\rm obs}\rangle/\langle I_{\rm obs}\rangle$ is given by

\begin{equation}
\frac{\langle Q_{\rm obs}\rangle}{\langle I_{\rm obs}\rangle} = \frac{\langle p_{\rm st,Q} I_{\rm PSF,st} + C_{\rm pol,Q} I_{\rm PSF,pl} \rangle}{\langle I_{\rm PSF,st} + C_{\rm flux} I_{\rm PSF,pl} \rangle} + p_{\rm tel,Q}
\label{equ:tel_pol_extended_model_measure}
\end{equation}

We usually do not know the exact intrinsic polarization of the star, except that it is small, and contributions from the planet, except that they are very faint. Therefore, we correct the telescope polarization by just assuming that the two contributions can be neglected and subtract the scaled intensity $I_{\rm obs}$ like:

\begin{equation}
Q_{\rm obs,norm} = Q_{\rm obs} - \frac{\langle Q_{\rm obs}\rangle}{\langle I_{\rm obs}\rangle} I_{\rm obs}
\label{equ:tel_pol_extended_model_subtr}
\end{equation}

We then combine Eq.~(\ref{equ:tel_pol_extended_model_measure}) and (\ref{equ:tel_pol_extended_model_subtr}) and notice that all terms with $p_{\rm tel,Q}$ have successfully vanished:

\begin{equation}
\begin{aligned}
Q_{\rm obs,norm} =& p_{\rm st,Q} I_{\rm PSF,st} + C_{\rm pol,Q} I_{\rm PSF,pl} \\
&- \frac{p_{\rm st,Q} \langle I_{\rm PSF,st} \rangle + C_{\rm pol,Q} \langle I_{\rm PSF,pl} \rangle}{\langle I_{\rm PSF,st} \rangle + C_{\rm flux} \langle I_{\rm PSF,pl} \rangle} (I_{\rm PSF,st} + C_{\rm flux} I_{\rm PSF,pl})
\end{aligned}
\label{equ:tel_pol_extended_model_subtr2}
\end{equation}

We define $ f_{pl,st} = \langle I_{\rm PSF,pl} \rangle / \langle I_{\rm PSF,st} \rangle$ then simplify the result by factoring out the stellar and planetary PSFs:

\begin{equation}
\begin{aligned}
Q_{\rm obs,norm} = &- C_{\rm flux} I_{\rm PSF,st} f_{pl,st} \frac{p_{\rm pl,Q} - p_{\rm st,Q}}{1+C_{\rm flux} f_{pl,st}} \\
&+ C_{\rm flux} I_{\rm PSF,pl} \frac{p_{\rm pl,Q} - p_{\rm st,Q}}{1+C_{\rm flux} f_{pl,st}}
\end{aligned}
\label{equ:tel_pol_extended_model_subtr3}
\end{equation}



Now we can estimate the order of magnitude of the different terms for our particular case:

\begin{itemize}
\item $f_{pl,st}$ corresponds to the total number of counts in the planetary PSF divided by the number of counts in the stellar PSF measured in the ring of pixels around the star that was used to calculate $\langle ... \rangle$. We determined that the value of this parameter for the $\alpha$~Cen~A data is $\lesssim 3$ if the planetary PSF is inside the ring and it is always smaller than one if the planetary PSF is outside the ring.
\item The flux contrast of a reflecting planet $C_{\rm flux}$ is expected to be of order $10^{-7}$ or smaller.
\end{itemize}

Considering the simplifications above, we can approximate $(1+C_{\rm flux} f_{pl,st}) \approx 1$ and simplify Eq.~(\ref{equ:tel_pol_extended_model_subtr3}) to:

\begin{equation}
Q_{\rm obs,norm} \approx C_{\rm flux} \left( p_{\rm pl,Q} - p_{\rm st,Q} \right) I_{\rm PSF,pl} \left( 1 - f_{pl,s} \frac{I_{\rm PSF,st}}{I_{\rm PSF,pl}} \right)
\label{equ:tel_pol_extended_model_subtr4}
\end{equation}

Eq.~(\ref{equ:tel_pol_extended_model_subtr4}) still shows two additional terms compared to the desired result in Eq.~(\ref{equ:tel_pol_extended_model_desired}). The expression in the right hand bracket in Eq.~(\ref{equ:tel_pol_extended_model_subtr4}) mainly depends on the separation of the stellar and planetary PSFs because, for this term to have a small contribution to $Q_{\rm obs,norm}$, the value of $I_{\rm PSF,st}(x,y)/I_{\rm PSF,pl}(x,y)$ needs to be small at the $(x,y)$-position where the planet PSF peaks. The PSFs for the $\alpha$~Cen~A observation in Fig.~\ref{fig:contrast_methods_ACenA30Apr} show that for any separation larger than the IWA of about 0.15\arcsec this ratio is smaller than $\sim10^{-3}$ and the whole expression in the right hand bracket does not reduce the $Q$ signal of a planet by more than 0.3\%. If we neglect this small correction factor, we finally arrive at the expression

\begin{equation}
Q_{\rm obs,norm}(x,y) \approx C_{\rm flux} \left( p_{\rm pl,Q}(\alpha) - p_{\rm st,Q} \right) I_{\rm PSF,pl}(x,y)
\label{equ:tel_pol_extended_model_result}
\end{equation}

for the telescope polarization corrected (or normalized) $Q$ image, where we re-introduced the correct dependencies of the parameters on scattering angle $\alpha$ and image coordinates $(x,y)$. The result means that, for a high-contrast point-source, our process of removing the instrument polarization modifies the fractional polarization of the planet $p_{\rm pl,Q}(\alpha)$ with the intrinsic fractional polarization of the star $p_{\rm st,Q}$. However, given that we have shown (see Appendix~\ref{sec:Telescope polarization correction}) that the polarization of the star $p_{\rm st}$ is $\lessapprox 2 \cdot 10^{-4}$ for all our targets, we expect the change of the observed polarization of the planet to be $\lessapprox 0.002$~\%. This is insignificant compared to the expected $p_{\rm pl,Q}(\alpha) \approx 10\%$ of a reflecting planet, therefore we conclude that our process of removing the instrument polarization does not harm the polarized signal of a planet significantly in our specific case.

\section{Results and discussion for the additional targets}
\label{Appendix: Results and discussion for the additional targets}

\subsection{$\alpha$ Centauri B}
\begin{figure}
\resizebox{\hsize}{!}{\includegraphics{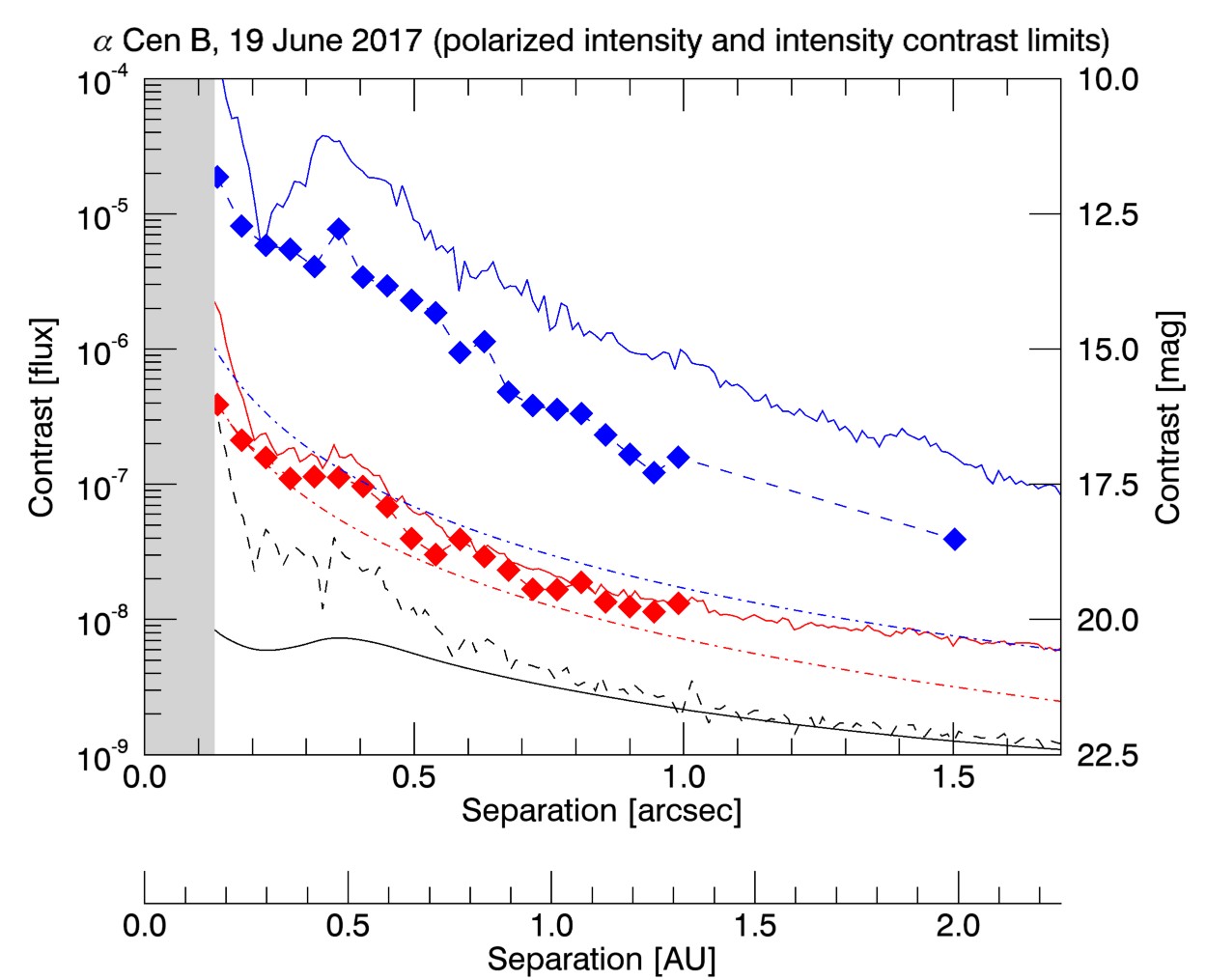}}
\caption{Radial contrast limits as a function of separation for the deepest ZIMPOL high-contrast dataset of $\alpha$~Cen~B in the VBB filter. The plot shows the 5$\sigma$ limits for the intensity and polarized intensity from the same observation as well as the 1$\sigma$ photon noise limit for the polarized intensity. The meaning of the colors and symbols is the same as explained in Fig.~\ref{fig:contrast_methods_ACenA30Apr}.}
  \label{fig:contrast_methods_ACenB19June17}
\end{figure}

In Fig.~\ref{fig:contrast_methods_ACenB19June17} we show the deepest contrast limits for $\alpha$~Cen~B, derived from a dataset with a total $t_{\rm exp}$ of 206.8~min in the VBB filter during a half-night with excellent observing conditions (see Table~\ref{table: Refplanets data}). $\alpha$~Cen~B is about four times fainter than the A component but using the VBB filter instead of the N\_R (or N\_I) compensates for this, resulting in similar count rates and contrast limits as for component A. The raw polarimetric contrast can be improved considerably with the use of PCA-ADI, even surpassing the limits for $\alpha$~Cen~A for some separations. Because of the excellent observing conditions, despite using the shortest possible DIT of 1.1~sec, some frames have saturated pixels just at the edge of the coronagraph located within the effective IWA of the data (highlighted by the grey bar in the contrast curve plots).

From a stability point-of-view, there is no reason why it should not be possible for $\alpha$~Cen~B to Harbor a Jupiter sized planet, just like for $\alpha$~Cen~A. However, for the B component of the system, the RV limits are much more stringent than for the A component. The radial velocity limits for $\alpha$~Cen B from Zhao (2018) exclude planets with $M sin(i) > 8.4~M_{\textrm{Earth}}$ for the classically defined habitable zone from about 0.7 to 1.3~AU, and even more stringent limits for smaller separations.

For $\alpha$~Cen~A we have discussed some arguments why the RV limits would still allow a giant planet to be in orbit around this star and that our deep contrast limits could allow us to observe such a giant planet. Due to the stringent RV limits, however, the arguments cannot be applied to $\alpha$~Cen~B. Orbital inclinations with $sin(i)\lesssim0.5$ are unlikely due to stability arguments. The resulting optimistic upper limit for the mass of a potential companion of $\sim20~M_{\textrm{Earth}}$, comparable to the mass of Neptune or Uranus, makes it unlikely that $\alpha$~Cen~B could Harbor a planet large enough to be detectable with the limits presented in Fig.~\ref{fig:contrast_methods_ACenB19June17} for a single half-night. The possibility of detecting a low mass planet around $\alpha$~Cen~B with ZIMPOL was discussed by \citet{Milli13} in the light of the former exoplanet candidate $\alpha$~Cen~Bb. They concluded that a detection of the reflected light should be possible. However, the study focused on very close separations and the use of a four-quadrant phase-mask coronagraph (not commissioned for ZIMPOL). Such small separations are mostly inaccessible with the Lyot coronagraph used in our survey.

A previous search around $\alpha$~Cen B was again done by \citet{Schroeder00} with HST. The contrast limits are comparable to $\alpha$~Cen~A, between 0.5\arcsec-1.5\arcsec the limits are between 7.5-8.5~mag at a wavelength of $\sim$1.02~$\mu$m. Our contrast limits (see Fig.~\ref{fig:contrast_methods_ACenB19June17} and summary in Table~\ref{table: Refplanets contrast}) push these limits for the R+I-band by a large amount. 

\subsection{Altair}
\begin{figure}
\resizebox{\hsize}{!}{\includegraphics{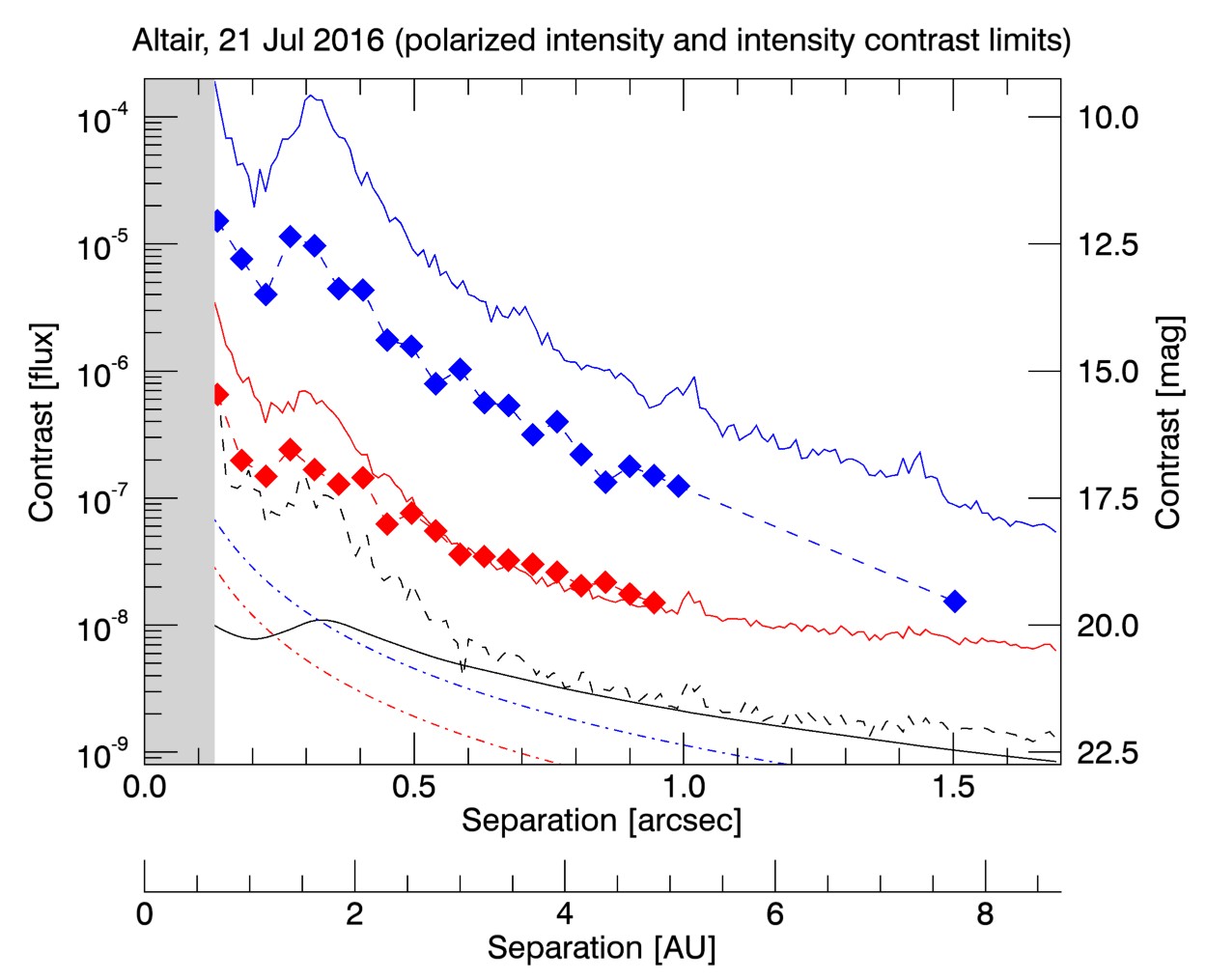}}
\caption{Radial contrast limits as a function of separation for the deepest ZIMPOL high-contrast dataset of Altair in the R\_PRIM filter. The plot shows the 5$\sigma$ limits for the intensity and polarized intensity from the same observation as well as the 1$\sigma$ photon noise limit for the polarized intensity. The meaning of the colors and symbols is the same as explained in Fig.~\ref{fig:contrast_methods_ACenA30Apr}.}
  \label{fig:contrast_methods_altair22Jul16}
\end{figure}

In Fig.~\ref{fig:contrast_methods_altair22Jul16} we show the deepest contrast limits for Altair derived from a dataset with a total $t_{\rm exp}$ of 151.3~min in the R\_PRIM filter. The filter is $\sim$2.6 times broader than the narrow R-band filter used for $\alpha$~Cen A, but Altair is fainter by about the same factor. This results in similar numbers for the captured photons per second and similar contrast limits. The main difference is the larger distance, and therefore the lower expected signal for given planet parameters and angular separation, as shown for our model planet in Fig.~\ref{fig:contrast_methods_altair22Jul16}.

Altair is an active star and a fast rotator, this makes it difficult to use the RV method to determine precise upper mass limits for possible companions. For example, the survey from \citet{Lagrange09} shows that the RV limits for planets around a fast rotating early type star like Altair allow only the detection of high-mass and short-period exoplanets. Therefore, a direct imaging search is competitive and complementary with respect to the RV studies.

The deepest contrast limits for Altair were derived in the HST survey of \citet{Schroeder00} conducted with the Hubble Space Telescope (HST) who achieved for separations 0.5\arcsec-1.5\arcsec about 7.5-8.5~mag for a wavelength of $\sim$1.02~$\mu$m. Our limits in intensity are about 12.8-17.7~mag and in polarized intensity 17.9-20.3~mag but with an effective IWA of only $\sim$0.13\arcsec. The limits are deep in terms of contrast, however, Altair is with 5~pc the most distant object in our survey. As a result of that, the polarimetric contrast of the reflected light from a Jupiter sized planet with our model atmosphere would be lower than 10$^{-8}$ for all separations larger than 0.22\arcsec. Reaching such contrast levels is not possible with only a few hours of observation.

\subsection{Sirius~A}

\begin{figure*}
\centering
\begin{tabular}{cc}
\includegraphics[totalheight=3.0in]{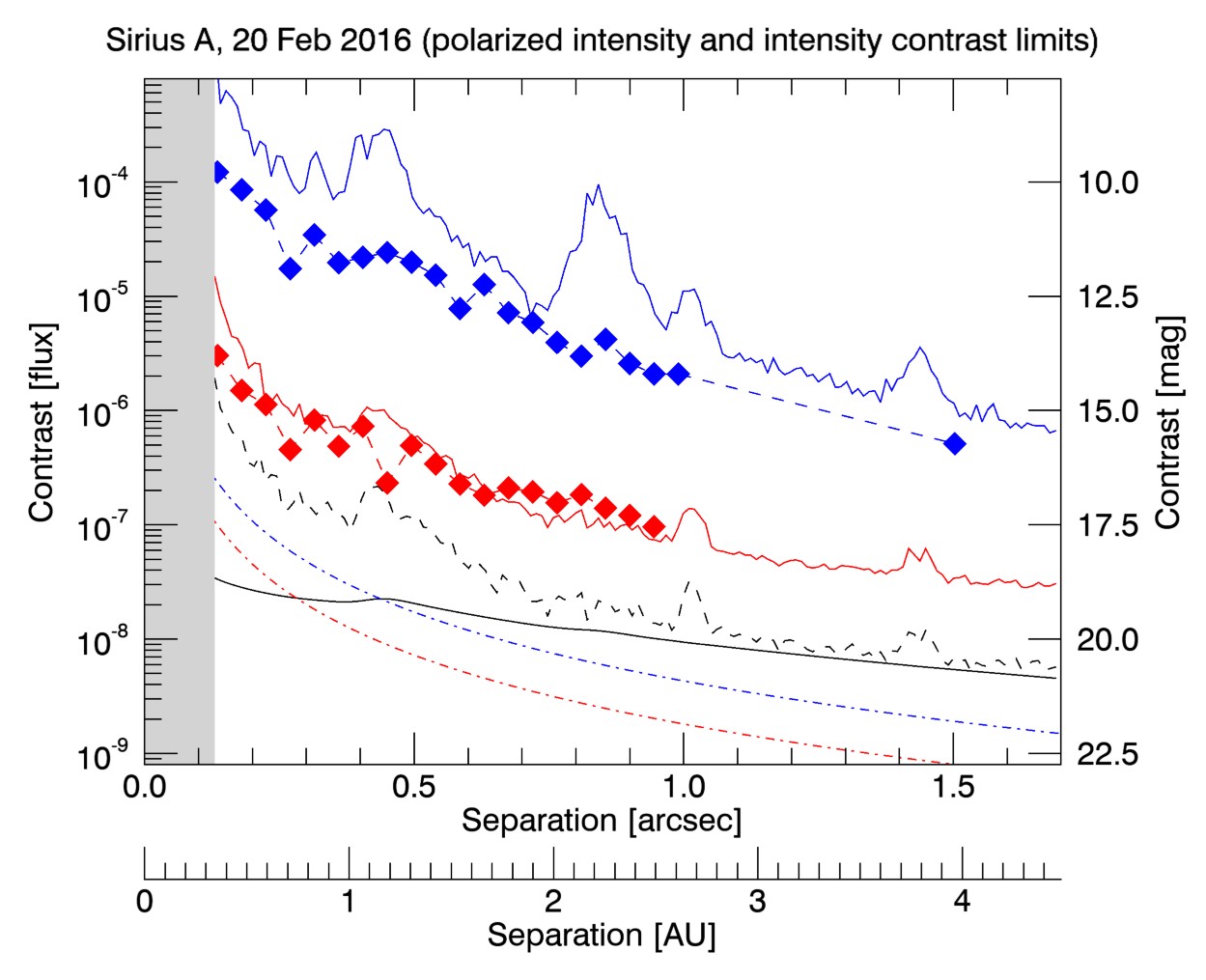} & \includegraphics[totalheight=3.0in]{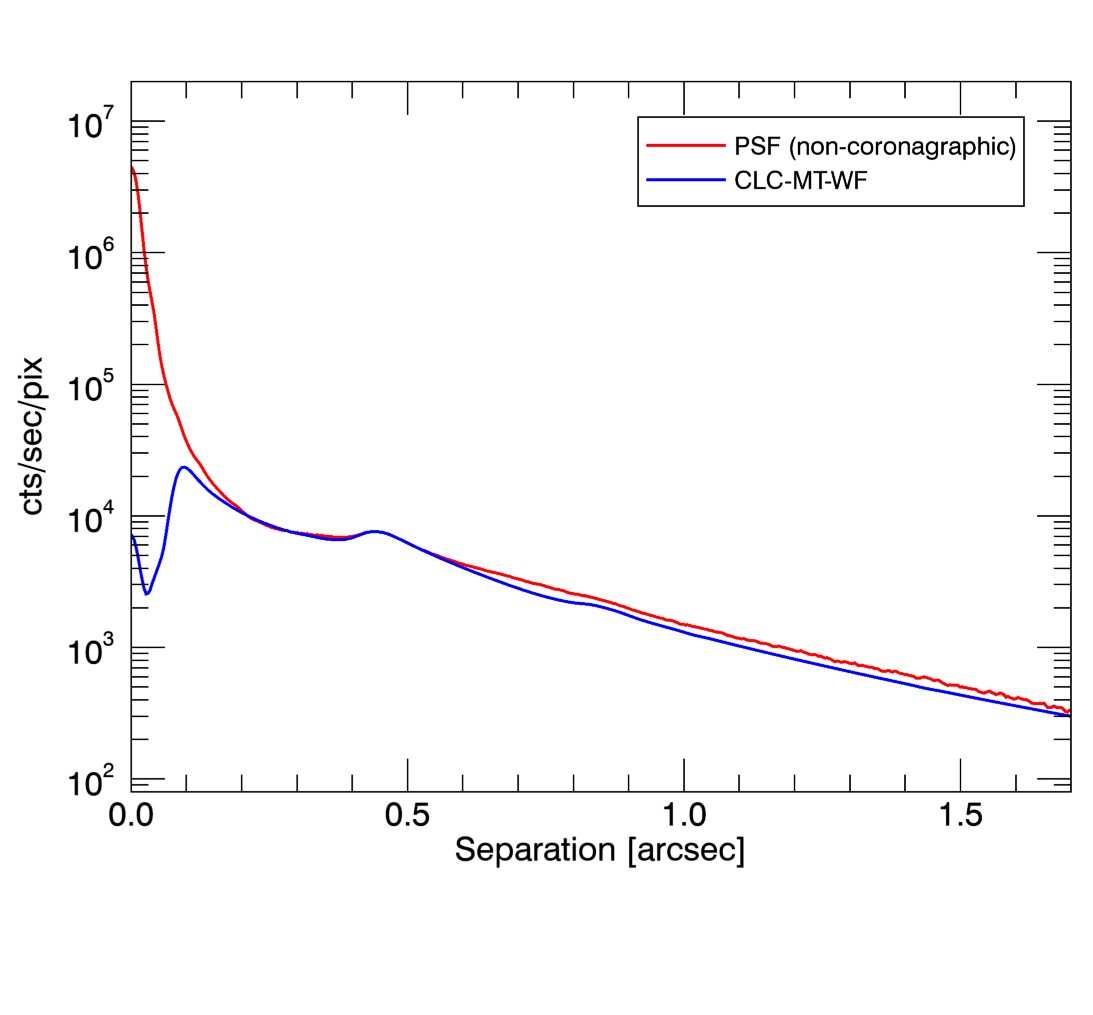} \\
\end{tabular}
\caption{Left: Radial contrast limits as a function of separation for the deepest ZIMPOL high-contrast dataset of Sirius~A in the N\_I filter. The plot shows the 5$\sigma$ limits for the intensity and polarized intensity from the same observation as well as the 1$\sigma$ photon noise limit for the polarized intensity. The meaning of the colors and symbols is the same as explained in Fig.~\ref{fig:contrast_methods_ACenA30Apr}. Right: The coronagraphic PSF compared to the non-coronagraphic PSF of Sirius~A in the N\_I filter}
  \label{fig:contrast_methods_SiriusA21Feb16}
\end{figure*}

In Fig. \ref{fig:contrast_methods_SiriusA21Feb16} we show the deepest contrast limits for Sirius~A based on a total $t_{\rm exp}$ of 175.2~min. The N\_I filter was used to avoid saturation of this very bright, blue star in coronagraphic mode with the shortest possible DIT. Unfortunately, the atmospheric conditions for these observations were poor (see Table~\ref{table: Refplanets data}), resulting in a degradation of the resolution and the contrast, which makes it more difficult to perform some of the data reduction steps (e.g. centering, beam shift correction, ...). The resulting contrast limits are much worse than what would be possible for such a bright target. The coronagraphic PSF in the right panel in Fig.~\ref{fig:contrast_methods_SiriusA21Feb16} shows well that the level of the PSF halo is enhanced and the non-coronagraphic PSF peak is significantly lowered when compared to the PSFs from the observation of $\alpha$~Cen~A shown in Fig.~\ref{fig:psf_profiles}. Both effects have a negative impact on the SNR of a point-source and the contrast limit of the data.

The radial velocity mass/separation limits for low mass objects are loose for Sirius~A because it is an intermediate mass ($\sim$2~M$_{\odot}$) A1V star with strong intrinsic RV variation \citep[e.g.][]{Lagrange09}. The possibility for stable planetary orbits around Sirius~A was investigated by \citet{Holman99} and \citet{Bond17}, suggesting that stable orbits with periods up to 2.24~yr are possible. This corresponds to a semimajor axis of 2.2~AU or 0.83\arcsec in angular separation. \citet{Bond17} used precise HST astrometry and could not exclude the presence of a third body in the system with a mass smaller than $\sim$15-25~M$_{\textrm{Jupiter}}$.

There have been attempts to find massive companions to Sirius~A in the infrared by \citet{Schroeder00} using HST and reaching a contrast limit of about 7.5-8.5~mag between 0.5\arcsec-1.5\arcsec for a wavelength of $\sim$1.02~$\mu$m. \citet{Thalmann11} used Subaru IRCS and AO188 in the $4.05~\mu $m narrow-band Br~$\alpha$ filter. At an IWA of 0.7\arcsec they were able to achieve a contrast of about 11~mag, and about 14~mag at a separation of 1.5\arcsec. The deepest limits were obtained with the IRDIS and IFS instrument of SPHERE/VLT in the near infrared from 0.95 to $2.3~\mu $m. \citep{Vigan15} using SDI in combination with ADI with an IWA of only 0.2\arcsec. They report contrasts up to 14.3~mag at 0.2\arcsec and $\sim$16.3~mag in the 0.4-1.0\arcsec range. With our combination of PDI and ADI we achieved slightly better contrasts of about 14.7~mag at 0.2\arcsec and $\sim$17.1~mag in the 0.4-1.0\arcsec range in I-band ($\lambda=817~\textrm{nm} $) with a smaller effective IWA of $\sim$0.13\arcsec. However, our observation suffers from poor observing conditions and contrast limits like for the other bright targets in our survey (e.g. $\alpha$~Cen A/B) should be possible for Sirius~A as well under good seeing conditions. This would improve our polarized intensity contrast limits by about 3~mag at all separations.

Our contrast limits are relatively far away from detecting the Jupiter sized reference model planet when comparing to $\alpha$~Cen~A/B. This is partially due to the poor observing conditions and partially due to the distance of 2.64~pc which is about twice as far as $\alpha$~Cen. Because the contrast of a companion scales for a given angular separation like $L[pc]^{-2}$ with distance $L$. Thus, the reflected light contrast for a reference planet at the same angular separation to its host star is four times more demanding for Sirius~A compared to $\alpha$~Cen. For bright stars $<1.5^m$, the contrast efficiency of ZIMPOL is limited by the frame rate, or the ability to collect as many photons as possible without saturating the coronagraphic images. Therefore, the $C\propto L^{-2}$ is increasing the required $t_{\rm exp}$ for a detection of a planet in reflected light around Sirius~A by factor of 16 when compared to $\alpha$~Cen~A/B. However, for smaller separations $\lessapprox 0.3 \arcsec$ the reflected light contrast of a possible Jupiter-sized planet around Sirius~A increases to values above $2 \cdot 10^{-8}$ which could be in reach for ZIMPOL within a few consecutive observing nights, assuming a $t_{\rm exp}^{-1/2}$ noise scaling.

\subsection{$\tau$~Ceti}
\begin{figure}
\resizebox{\hsize}{!}{\includegraphics{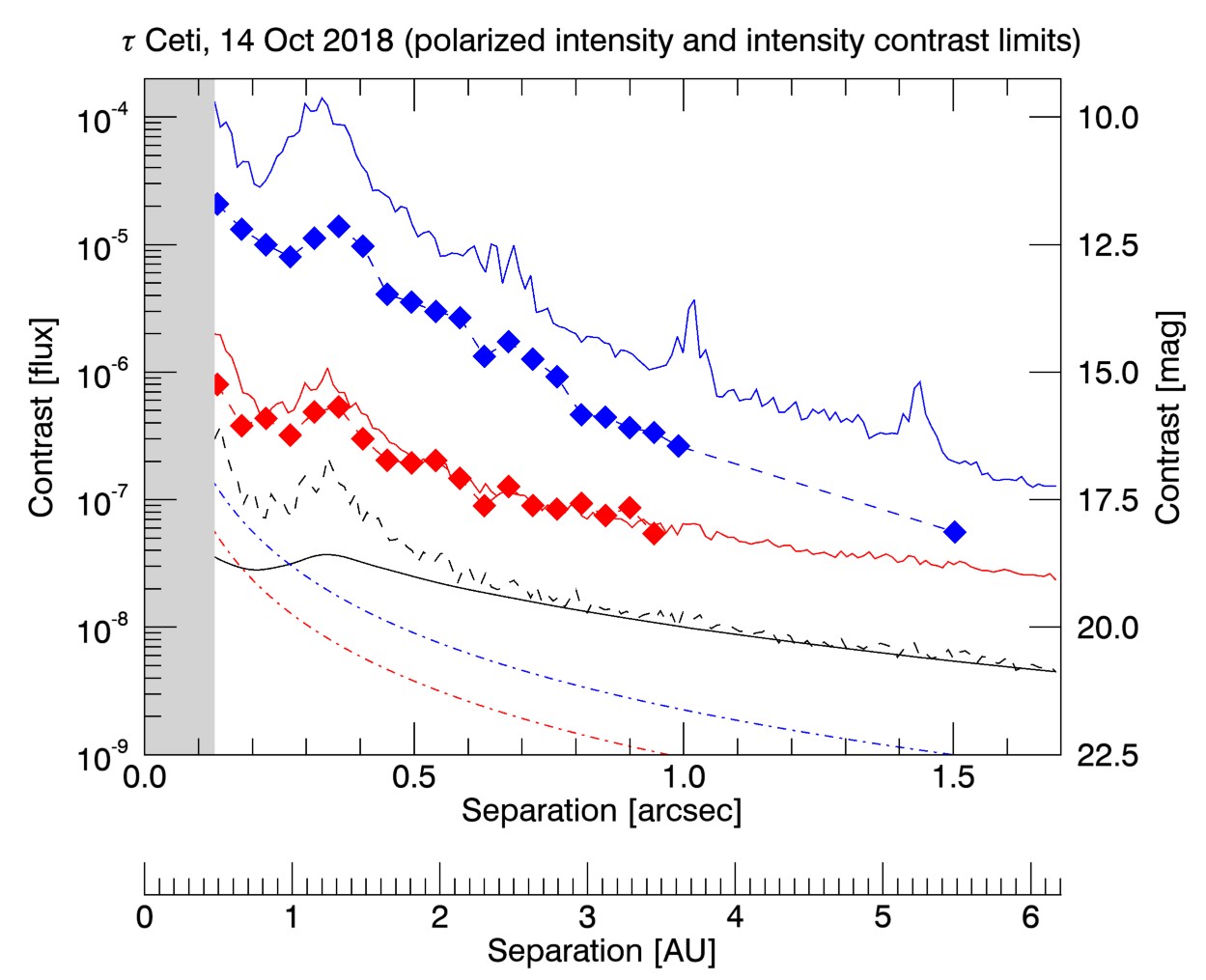}}
\caption{Radial contrast limits as a function of separation for the deepest ZIMPOL high-contrast dataset of $\tau$~Ceti in the R\_PRIM filter. The plot shows the 5$\sigma$ limits for the intensity and polarized intensity from the same observation as well as the 1$\sigma$ photon noise limit for the polarized intensity. The meaning of the colors and symbols is the same as in Fig.~\ref{fig:contrast_methods_ACenA30Apr}, except that we only subtracted 10~PCs during the PCA-ADI step for the polarized intensity. For this dataset the resulting contrast limits were significantly better with only 10 instead of the 20~PCs that we used for all other dataset.}
  \label{fig:contrast_methods_tauCeti14Oct18}
\end{figure}

In Fig.~\ref{fig:contrast_methods_tauCeti14Oct18} we show the deepest contrast limits for a half-night of observing $\tau$~Ceti during a time with excellent observing conditions (see Table~\ref{table: Refplanets data}) derived from a total $t_{\rm exp}$ of 168~min in the R\_PRIM filter. We used a long $t_{DIT}=14$~sec per exposure to ensure that the contrast is photon noise limited in the whole ZIMPOL FOV. A problem with long exposures in P1 mode is rotational smearing. The field rotation can be quite fast because $\tau$~Ceti passes close to the zenith. We could have used the broader VBB filter for this observation to maximize the number of collected photons, however we selected R\_PRIM, because some of the instrumental effects (e.g. instrument polarization, beam shift) can be corrected more accurately in the data reduction for the narrower filters because of strongly wavelength dependent effects which increase with filter width. This strategy seems to be beneficial for the planet search at small separation $<0.4\arcsec$ where the speckle noise dominates, while it is less favourable at larger separation in the photon noise limited region. The resulting contrast limits for one half-night for $\tau$~Ceti are not as deep as for most other targets in our survey because of the resulting photon counts are about 10 times lower than for our brighter targets.

The presence of RV planets around $\tau$~Ceti has been proposed by \citet{Tuomi13} and \citet{Feng17}. However, the measured signals indicate masses $M sin(i)\lesssim 6.6~$M$_{\textrm{Earth}}$ and such planets would be too faint to be observable in our data. High-mass planets >3~M$_{\textrm{Jupiter}}$ are excluded by a separate study based on Gaia and Hipparcos astrometry for the separation range 3--30~AU or $\approx1-10\arcsec$ \citep{Kervella19}.

Deep direct imaging contrast limits for $\tau$~Ceti are given by \citet{Schroeder00} who achieved between 9.0-11.5~mag in the separation range from 0.5\arcsec-1.5\arcsec with HST at a wavelength of $\sim$1.02~$\mu$m. At longer wavelengths, \citet{Boehle19} report a limiting contrast of about 11.0-12.0~mag in L'-band with NACO at the VLT in the separation range from 1.0\arcsec-1.5\arcsec. With one excellent night we were able to achieve contrast limits in intensity of about 12.3-16.9~mag and in polarized intensity 16.7-18.8~mag for the separation range from 0.5\arcsec-1.5\arcsec but with an IWA down to $\sim$0.13\arcsec. The contrast limits are deep but -- as shown in Fig. \ref{fig:contrast_methods_tauCeti14Oct18} -- they are still far above what we calculate for our model Jupiter sized planet, both in intensity and polarized intensity. This is because $\tau$~Ceti is, with 3.65~pc, one of the more distant targets and with m$_R=2.9$ also one of the fainter targets of our survey. As a result, the polarimetric contrast of the reflected light from our Jupiter sized reference planet model would be lower than 10$^{-8}$ for any separations larger than 0.3\arcsec.

It is known that $\tau$~Ceti hosts a large debris disk and \citet{Lawler14} measured with Herschel an inner edge between 1 and 10~AU, an outer edge at about $\sim$55~AU and an inclination of $35^{\circ}\pm10^{\circ}$ from face-on. The total mass of the disk is estimated to be only $\sim$1~$M_{\textrm{Earth}}$ \citep{Greaves04} and it is extended over a large range of separations. We did not detect the signal of an extended source around $\tau$~Ceti in our data, therefore it is either too faint to be seen directly in our data or the inner edge is located outside of our FOV of about 1.7\arcsec or about 6~AU at this distance.

\subsection{$\epsilon$~Eridani}
\begin{figure}
\resizebox{\hsize}{!}{\includegraphics{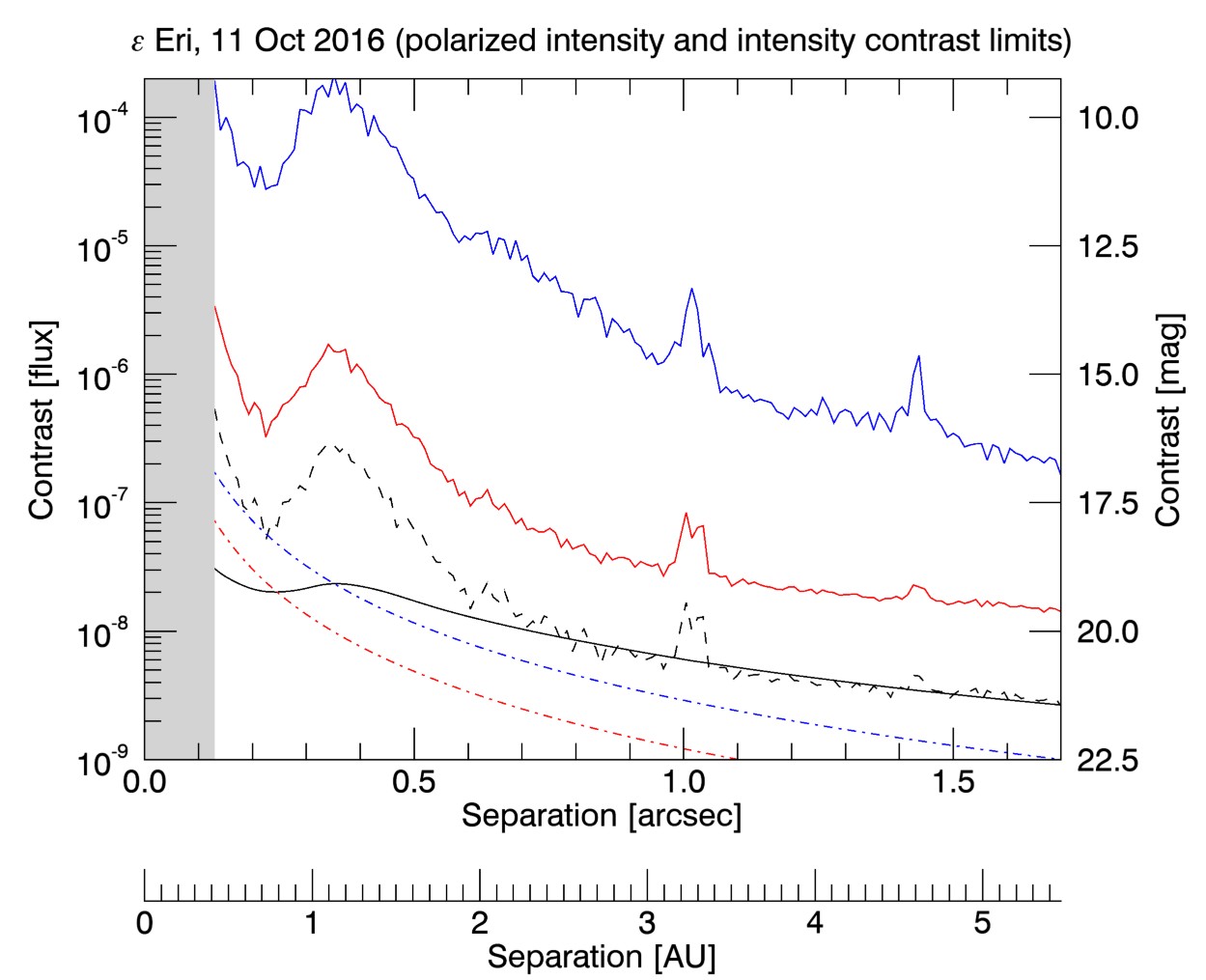}}
\caption{Radial contrast limits as a function of separation for the deepest ZIMPOL high-contrast dataset of $\epsilon$~Eri in the VBB filter. The plot shows the 5$\sigma$ limits for the intensity and polarized intensity from the same observation as well as the 1$\sigma$ photon noise limit for the polarized intensity. The meaning of the colors and symbols is the same as in Fig.~\ref{fig:contrast_methods_ACenA30Apr}}
  \label{fig:contrast_methods_epsEri11Oct16}
\end{figure}

In Fig.~\ref{fig:contrast_methods_epsEri11Oct16} we show the deepest contrast limits of $\epsilon$~Eri for a single half-night with good observing conditions (see Table~\ref{table: Refplanets data}) and a total $t_{\rm exp}$ of 192~min in the VBB filter. This is the only target observed in the field stabilized P2 polarimetry mode. Without field rotation, we cannot apply ADI to this dataset. The targets $\epsilon$~Eri and $\tau$~Ceti are almost identical in brightness and therefore it is interesting to compare the contrast limits of the non-ADI $\epsilon$~Eri data with the ADI $\tau$~Ceti data and the different filters used. The achieved contrast for $\epsilon$~Eri is significantly deeper at larger separations because of the longer total $t_{\rm exp}$ and the broader filter, and therefore increased photon counts by a factor of $\sim$2. At closer separations in the speckle-dominated regime the contrast limits for $\tau$~Ceti are better, despite the smaller amount of collected photons. The ADI data of $\tau$~Ceti profit from reduced quasi-static aberrations because the speckles are averaged and significantly reduced by the PSF subtraction. From this comparison we estimate that field rotation would improve the contrast limits for $\epsilon$~Eri inside the speckle ring by up to a factor of 5 for the polarization up to 10 for the intensity, but at the time of the observation the addition of ADI was not yet considered as an option.

The deepest direct imaging and radial velocity limits for $\epsilon$~Eri are both presented in \citet{Mawet19}. They also present the strongest evidence so far for the existence of $\epsilon$~Eri~b. A giant planet with a mass of $\sim$1.2~$M_{\textrm{Jupiter}}$ for an orbital inclination of $34^{\circ}\pm2^{\circ}$ when assumed to be coplanar with the outer debris disk. The star is also monitored by Gaia but the measured astrometric trends are not yet precise enough to confirm the planet \citep{Kervella19}. The planet's separation of $\sim$3.5~AU is well within the FOV of ZIMPOL and in the photon noise dominated regime of the contrast curve (see Fig. \ref{fig:contrast_methods_epsEri11Oct16}). However, we were not able to detect the planet because a Jupiter sized reference model planet would produce at this orbital separation, a polarization contrast below $1.2 \cdot 10^{-9}$ (22.3~mag). Our 5$\sigma$ contrast limit with one half-night of observation at this separation is $2.2 \cdot 10^{-8}$ (19.1~mag). We were also not able to spot any extended polarized emission from the disk around $\epsilon$~Eri. The well known part of the disk around $\epsilon$~Eri as seen by Herschel and ALMA is located between 11--13~AU \citep{Greaves14, Booth17}, which is outside of our FOV with ZIMPOL. 

The deepest high spatial resolution imaging limits for the thermal emission of $\epsilon$~Eri~b were obtained with VLT/NACO in Lp-band \citep{Mizuki16} and Keck/NIRC2 in Ms-band \citep{Mawet19}. The 5$\sigma$ contrast curves presented in \citet{Mawet19} range from 0.3\arcsec to 1.5\arcsec with contrast limits in the range from $\sim$9.0--11.8~mag and $\sim$10.3--13.5~mag for Lp-band and Ms-band, respectively. With our best half-night we were able to achieve contrast limits of about 10.0--16.3~mag in reflected intensity and 15.0--19.6~mag in polarized intensity for the same range of separations with an effective IWA of $\sim$0.13\arcsec. Due to the broad passband of the VBB filter the contrast limits for $\epsilon$~Eri are deep at the separation where $\epsilon$~Eri~b is expected to be orbiting but this is still a factor of $\sim$20 or $\sim$3.3~mag away from the expected signal for planet~b.

\end{appendix}

\end{document}